\documentclass[a4paper, 12 pt, reqno]{article}
\usepackage{comment}

\usepackage{mathrsfs,enumerate,latexsym,graphicx}
\usepackage{xcolor}
\usepackage{amssymb,amsmath}
\usepackage{amsfonts}
\usepackage{amsthm}
\usepackage[hidelinks]{hyperref}
\usepackage{graphicx}
\usepackage{subcaption}
\usepackage{caption}
\usepackage{booktabs}

\usepackage[margin=1.2in]{geometry}
\usepackage{setspace}
\usepackage{bbm}
\usepackage{calrsfs}
\usepackage{txfonts}

\usepackage{pdflscape}

\usepackage{float}
\restylefloat{figure}

\usepackage[]{appendix}
\usepackage{doi}
\usepackage{enumitem}

\usepackage[comma]{natbib}

\expandafter\def\expandafter\normalsize\expandafter{%
	\normalsize
	\setlength\abovedisplayskip{6pt}
	\setlength\belowdisplayskip{6pt}
	\setlength\abovedisplayshortskip{4pt}
	\setlength\belowdisplayshortskip{4pt}
}

\usepackage{float}
\restylefloat{figure}

\usepackage{calrsfs}
\usepackage{lipsum}

\newtheoremstyle{break}
 {\topsep}{\topsep}%
 {\itshape}{}%
 {\bfseries}{}%
 {\newline}{}%
 \theoremstyle{break}

\newtheorem{corollary} {Corollary}

\newtheorem{theorem}{Theorem}

\DeclareMathOperator*{\E}{\mathbb{E}}
\DeclareMathOperator*{\V}{\mathbb{V}}

\DeclareMathOperator*{\prob}{\mathbb{P}}

\newcommand{\Real}{\mathbb{R}}

\newcommand{\obs}{\mathbf{x}}
\newcommand{\iobs}{\mathbf{z}}

\newcommand{\con}{\mathbf{c}}

\newcommand{\Logit}{\Lambda}

\newcommand{\gucon}{\varphi(\obs_i)}
\newcommand{\guconhat}{\hat{\varphi}(\obs_i)}

\newcommand{\guconz}{\varphi(\obs_i, \iobs_i)}

\newcommand{\gcon}{\psi_{\con_i}(\obs_i)}
\newcommand{\gconhat}{\hat{\psi}_{\con_i}(\obs_i)}

\renewcommand{\vec}[1]{\textbf{#1}}
\renewcommand{\vec}[1]{\boldsymbol{#1}}

\newcommand{\kh}[1]{{\color{red}{#1}}}

\newcommand*\diff{\mathop{}\!\mathrm{d}}

\usepackage{array}
\newcolumntype{H}{>{\setbox0=\hbox\bgroup}c<{\egroup}@{}}
\newcolumntype{Z}{>{\setbox0=\hbox\bgroup}c<{\egroup}@{\hspace*{-\tabcolsep}}}

\begin{document}

\title{Promotion through Connections: \linebreak
Favors or Information?}	
\date{June 22, 2026}
\author{Yann Bramoull\'{e} and Kenan Huremovi\'{c}\thanks{	
		We thank participants in seminars and conferences and Olivier
		Chanel, Anushka Chawla, Russell Davidson, Rahul Deb, Joseph Engelberg, Mathieu Faure, Bernard Fortin, Habiba Djebbari, Bruno Decreuse, Ludovic Renou, Mark Rosenzweig, Marc Sangnier, Christian Schluter, Adam Szeidl, Jin Wang, Jeffrey Wooldridge, Paolo Zacchia, and Natalia Zinovyeva for helpful comments, and four anonymous referees and the editor for their constructive feedback. For financial support, Yann Bramoull\'{e} thanks the European Research Council (Consolidator Grant 616442), the French government under the “France 2030” investment plan managed by the French National Research Agency (ANR-17-EURE-0020) and the Excellence Initiative of Aix-Marseille University - A*MIDEX. Bramoull\'{e}: Aix-Marseille University, CNRS, Aix-Marseille School of Economics, CEPR, yann.bramoulle@univ-amu.fr; Huremovi\'{c}: IMT School for Advanced Studies Lucca,  kenan.huremovic@imtlucca.it.}
	}
\maketitle

\begin{abstract}
    \singlespacing
Connections appear to be helpful in many contexts such as obtaining a job, a promotion, a grant, a loan, or publishing a paper. This may be due to favoritism or to information conveyed by connections. Building on earlier work on discrimination, we propose a new method that identifies these channels using data observed at the time of promotion. The method exploits distinct implications of the two effects on the relationship between observables and success. We show that extra information on connected candidates generates excess variance in latent errors while favors yield different promotion thresholds. We characterize the conditions under which both effects are identified and operationalize these ideas econometrically within a semiparametric framework. We also derive testable restrictions of the model and show how to account for connection endogeneity. We reanalyze data on academic promotions in Spain and Italy and political promotions in China. We detect evidence of favoritism for all types of candidates and of information effects for candidates applying to junior positions. We find strong support for the model’s testable restrictions.
\end{abstract}

\thispagestyle{empty}
 \newpage
\setcounter{page}{1}
\renewcommand{\kh}[1]{{\textcolor{red}{#1}}}

\section{Introduction} \label{sec:Introduction}

Connections appear to be helpful in many contexts such as obtaining a job, a promotion, a grant, a loan, or publishing a paper.\footnote{The literature on jobs and connections is large and expanding, see e.g. \cite{beaman2012gets}, \cite{brown2016informal}, \cite{hensvik2016social}, \cite{pallais2016referential}. On promotions, see \cite{combes2008publish}, \cite{zinovyeva2015role}, \cite{jia2015political}, \cite{bagues2017does}. On grants, see \cite{li2017expertise}. On loans, see \cite{engelberg2012friends}, \cite{schoenherr2019political}. On publications, see  \cite{laband1994favoritism},  \cite{brogaard2014networks}, \cite{colussi2015social}, \cite{fisman2018social}, \cite{ductor2022coauthor}.} Two main mechanisms help explain these
wide-ranging effects. On the one hand, connections may convey information
on candidates, projects, and papers that helps recruiters, juries and editors make better decisions. On the other hand, decision-makers may unduly favor connected candidates, leading to worse decisions.\footnote{Favor exchange within a group may increase the group's welfare to the detriment of society, see \cite{bramoulle2016favoritism}. In this paper, we consider the immediate implications of favoritism.} These two mechanisms have opposite welfare implications and empirical researchers have been trying to tease out the different forces behind the impacts of connections. Existing studies generally do so by building measures of candidates' \textquotedblleft true\textquotedblright\ quality. Researchers then compare the quality of connected and unconnected successful candidates. Information effects are likely to dominate if connected successful candidates have higher quality; favors are likely to dominate if connected successful candidates have lower quality.\footnote{For instance, articles published
in top economics and finance journals by authors connected to editors tend
to receive more citations, a sign that editors use their connections to
identify better papers (\cite{brogaard2014networks}). By contrast,
Full Professors in Spain who were connected to members of their promotion
jury publish less after promotion (\cite{zinovyeva2015role}), consistent
with favoritism.}

This empirical strategy carries some important limitations. First, building a measure of true quality may not be easy or feasible. Looking at researchers' publications or articles' citations requires a long enough time lag following promotion or publication. Second, identification is only valid if the impact of success on measured quality is the same for connected and unconnected successful candidates, see \citet[p.283]{zinovyeva2015role}. Third, connections may both convey
information and attract favors, and this strategy does not allow researchers to estimate the relative strengths of the two effects.

We develop a new method to identify why connections matter, building on earlier work on discrimination. Our method allows researchers to estimate the magnitudes of the two effects, without relying on measures of true quality. Our method exploits data collected at the time of application: observable characteristics of candidates and whether they were successful.\footnote{As with any analysis of the reasons behind the effect of connections, our method requires some exogenous shocks on connections. In other words, we assume that the problem of identifying whether connections matter
has been solved and focus on the question of identifying why they matter.} It looks for evidence of the two effects on the relationship between
candidates' observables and success.

Consider candidates applying for promotion and being evaluated by a jury. Some candidates are connected to jury members. Assume for now that the connections are conditionally random. When connections convey information, the jury has an extra signal regarding connected candidates' ability. This signal is unobserved by the econometrician and could be positive or negative. To the econometrician, then, the promotion decision looks more random for connected candidates.\footnote{A similar idea underlies Theorem 4 in \cite{lu2016random}; we discuss this relation in
more detail below.} We show how the strength of the information channel can
be recovered, under appropriate assumptions, from this \textit{excess
variance} in the latent error of connected candidates. Favors can then be
recovered by estimating and comparing the \textit{promotion thresholds} faced by connected and unconnected candidates. Favors lead to systematic biases in evaluation and the difference between promotion thresholds measures the magnitude of the underlying favoritism effect.

We develop our analysis in three stages. We first provide formal results on the identification of favors and information effects from connections, based on data on applications and promotion decisions. We show that the promotion bonus from favors and the distribution of latent errors of connected candidates are both identified under two assumptions. First, researchers must adopt an anchoring parametric assumption on the latent error distribution of unconnected candidates. Second, favors and information effects must be independent of some observable characteristics and these characteristics must induce sufficient variation in conditional promotion probabilities. Under these two assumptions, researchers can recover favors and information effects by contrasting the conditional promotion probabilities of connected and unconnected candidates. Moreover, identification holds without imposing any restriction on the \textit{grade function}, capturing how jury's evaluation depends on observable characteristics. This grade function is non-parametrically identified under the same assumptions. We further derive testable implications of these identifying assumptions. We notably show that the conditional promotion probability of connected candidates must be an increasing function of the conditional promotion probability of unconnected candidates, a strong testable restriction.

In the second stage, we implement this identification strategy and analyze the impact of connections on promotion in three different settings: academic promotion in Spain and in Italy and political promotion in China. We develop a common econometric framework based on a logistic anchor and a flexible estimation of the grade function. We consider two variants of the framework: a parametric variant, where the latent error distribution of connected candidates is also assumed to be logistic, with unknown variance, and a semiparametric variant, where this latent error distribution is left unrestricted. The estimation of the parametric model combines heteroscedastic logit estimations with increasingly complex polynomial approximations of the grade function. The estimation of the semiparametric model relies on a machine learning regularization technique (LASSO).\footnote{This technique enables us to approximate the grade function and the distribution of the error of connected candidates at near ``oracle rate'', see \cite{belloni2014inference}. More traditional methods, such as those proposed in \cite{klein1993efficient}, \cite{horowitz1992smoothed} or \cite{manski1985semiparametric} are not adapted to our context, due to the curse of dimensionality arising in the applications.}

We reanalyze three datasets: on academic promotions in Spain between 2002 and 2006 from \cite{zinovyeva2015role}, on academic promotions in Italy between 2012 and 2014 from \cite{bagues2017does}, and on political promotions in China between 1993 and 2009 from \cite{jia2015political}. The Spanish and Italian contexts notably differ in their levels of competitiveness. Only $11\%$ of the candidates get promoted in the Spanish data, while this number rises to $37\%$ in the Italian data. From data on applications and promotion decisions, the original articles assess the overall impact of connections on promotion.\footnote{\cite{zinovyeva2015role} analyze the impact of connections on academic promotions in Spain and \cite{jia2015political} the impact of connections on political promotions in China. While \cite{bagues2017does} focus on gender effects in academic promotions in Spain and Italy, they also look at the impact of connections, see p.1229.} We apply our proposed identification strategy and estimate the relative magnitudes of favors and information effects in the impact of connections on the same data.

The empirical results depict a coherent picture. We detect evidence of positive information effects for candidates to Associate Professor positions in both Spain and Italy. This is consistent with the fact that uncertainty on candidates' academic abilities is still high at this career stage. By contrast, we do not detect evidence of information effects from connections for candidates to Full Professor positions in Spain or Italy nor in the political promotion of Chinese provincial leaders. We detect strong evidence of favors from connections for all types of candidates: to Associate and Full Professor positions in Spain and Italy and to political promotion in China. We also find strong support for the monotonicity relationship implied by the assumption that favors and information effects are independent of observable characteristics.

In the third stage, we extend our framework to account for connection endogeneity. We adapt a control function approach proposed by \cite{wooldridge2014quasi}, building on insights from \cite{heckman1976common} and \cite{blundell2004endogeneity}. We consider a parametric model of connection formation. We clarify the conditions under which adding the generalized residuals from this regression to the observable characteristics allows researchers to identify favors and information effects in the impact of connections. We apply this extended framework to reestimate the impacts of connections in the Italian data. One concern in the Italian context is that participation to the jury is not mandatory, and about $8\%$ of the randomly selected evaluators decided not to participate in the jury. Consistently with this modest number, our estimates of favors and information effects are essentially unchanged when accounting for the endogeneity of connections.

Our analysis contributes to a growing empirical literature on connections. We develop the first empirical method to identify favors and information effects from data on candidates' characteristics and promotion decisions. This method could be applied in many other contexts, and could notably be used to cross-validate results obtained from
quality measures.

Our analysis builds on ideas first identified in the literature on discrimination.\footnote{For a comprehensive survey of the literature on discrimination and recent developments, see \citet{onuchic2022recent}.} \cite{siegelman1993urban} and \cite{heckman1998detecting} clarify key implications of differences in the variances of unobservables across groups. They show that variance differences invalidate the use of standard models of binary outcomes to detect discrimination. \cite{neumark2012detecting} shows how probit models with heteroscedasticity can help address this issue. He reanalyzes the data from \cite{bertrand2004} and finds stronger evidence for race discrimination than in the original study, once differences in variance across racial groups are accounted for.

We adapt and extend these ideas to the analysis of connections. We show that differences in variance help identify the informational content of connections, an idea consistent
with Theorem 4 in \cite{lu2016random}. \cite{lu2016random} provides a theoretical analysis of random choice under private information. He shows that better private information generates choices that look more dispersed to the econometrician. To our knowledge, we are the first to implement this insight in an empirical context.

We obtain novel identification results, and extend \cite{neumark2012detecting}'s identification arguments well beyond the probit case. We show that favors and information effects are identified in a semiparametric framework, with an unknown grade function and an unknown distribution of latent errors of connected candidates. Identification relies on a parametric anchor for the latent error distribution of unconnected candidates and on independence between (some) observable characteristics and favors and information effects. Identification fails to hold if the unconnected error distribution is unknown, or if favors and information effects vary with all observable characteristics. By contrast, we show that identification extends to setups where favors and information effects may depend on the number and types of connections and on some observable characteristics, like gender. 

We then demonstrate the empirical interest and feasibility of this novel identification strategy, by reanalyzing promotion data in three different contexts. In this empirical implementation, we showcase a new way for utilizing machine learning techniques to estimate causal effects within a semistructural econometric framework.

The paper proceeds as follows. Section \ref{sec:benchmark} analyzes identification in a benchmark model. Section \ref{sec:ExtendedModel} extends the model and identification results along several dimensions. Section \ref{sec:empirical_applications} describes the three datasets and empirical contexts. Section \ref{sec:EmpricalImplementation} presents our econometric framework and discusses key features of the empirical implementation. Section \ref{sec:EmpiricalAnalysis} presents the empirical estimates of favors and information effects. Section \ref{sec:endogeneous_connections} extends our framework to account for connection endogeneity. Section \ref{sec:discussion} provides a concluding discussion.

\section{Benchmark model}
\label{sec:benchmark}

We now introduce a benchmark model of promotion. This model relies on three assumptions: (1) connections to the jury are conditionally random; (2) favors and information gains from connections are homogeneous, and (3) only depend on the presence of a connection to the jury. We develop a semiparametric framework, and show that favors and information effects can be separately identified. In short, favors induce a left-translation of the promotion probability (first-order stochastic dominance shift) while information gains lead to an overall flattening (second-order stochastic dominance shift). We relax each of these three assumptions in what follows. We relax (2) and (3) in Section \ref{sec:ExtendedModel}, where we show that our identification strategy extends to situations where favors and information effects may vary with candidates' observable characteristics and with the number and types of connections. We relax (1) in Section \ref{sec:endogeneous_connections}, and consider a setup with endogenous connections.

In our framework, candidates apply for promotion and are evaluated by a jury making promotion decisions. The econometrician has data on the entire pool of candidates and on their promotion outcomes. We assume that the jury grades candidates and that candidates with higher grades are promoted. The econometrician, however, has no information on the jury's evaluations.\footnote{Access to such information would, of course, expand identification possibilities, see \cite{li2017expertise} on a framework where grades are observed} Grades depend on the jury's belief about candidates' ability and on connections to jury members.

We assume that candidate $i$'s ability, $a_i$ can be decomposed into three parts:
\begin{equation}
\label{eq:ability}
a_{i}=\varphi(\mathbf{x}_{i}) +u_{i}+v_{i}  
\end{equation}
where $\mathbf{x}_{i}\in \Real^{m}$ denotes a vector of $m$ characteristics observed by the jury and by the econometrician. Function $\varphi:\Real^{m} \rightarrow \Real$ is referred to as the \textit{grade function} and is unknown to the econometrician.  Variable $u_i$ is unobserved by the jury and the econometrician, and $v_i$ is observed by the jury but not the econometrician. We assume that $\E(u_i)=\E(v_i)=0$, $v_{i} \Perp \mathbf{x}_{i}$ and $u_i \Perp (\mathbf{x}_i,v_i)$.

Candidates can be socially connected to jury members. In this benchmark model, we assume that grades may be affected by the presence of a connection to the jury, but not by their number or types. Let $c_i \in \{0,1\}$ describe whether candidate $i$ has at least one connection to the jury ($c_i=1$) or none ($c_i=0$). We also assume that connections to the jury are conditionally random. This notably holds when jury members are chosen at random within a pool, and candidates' connections to the members of this pool are observable. From these connections, we can compute $\E(c_i)$, where the expectation is taken over all possible choices of jury members in the pool. Then, $c_i$ is purely random conditional on $\E(c_i)$. We assume that the jury can observe $\E(c_i)$, and we therefore incorporate it into $\obs_i$ in the following discussion. When connections are conditionally random, $c_i$ is conditionally independent from $u_i$ and $v_i$, $c_i \Perp  (u_i, v_i) | \mathbf{x}_i$. Denoting by $\mathbb{D}(u)$ the distribution of $u$, this implies that $\mathbb{D}(u_i|c_i=0, \mathbf{x}_i)=\mathbb{D}(u_i|c_i=1, \mathbf{x}_i)$ and $\mathbb{D}(v_i|c_i=0, \mathbf{x}_i)=\mathbb{D}(v_i|c_i=1, \mathbf{x}_i)$. Unobservables of connected and unconnected candidates then have the same conditional distributions.

Let $y_i \in \{0,1\}$ denote candidate $i$'s outcome, where $y_i=1$ if $i$ is promoted and $0$ otherwise. To think about identification, we can neglect issues related to sampling, see \cite{hsieh1985estimation, manski1985semiparametric}. We can assume that the econometrician recovers the conditional promotion probabilities for connected and unconnected candidates from the data: $\mathbb{P}(y_i=1|c_i=0,\mathbf{x}_i)$ and $\mathbb{P}(y_i=1|c_i=1,\mathbf{x}_i)$. The key question is, then, what can be learned from contrasting these two promotion probabilities? We next describe natural assumptions on how connections affect juries' grades and under which favors and information effects can be separately identified.

Consider unconnected candidates first. We assume that their grade is equal to the jury's expected belief about their ability. Since $\E(u_i|\mathbf{x}_i,v_i)=0$, this grade is equal to $\E(a_i|\mathbf{x}_i,v_i)=\varphi(\mathbf{x}_{i}) +v_{i}$. This implicitly assumes that the jury is risk-neutral. Accounting for risk aversion is a potentially important and challenging direction for future research, which we further discuss in Section \ref{sec:discussion}.

We assume that the distribution of $v_i$ is known by the econometrician. Denote by $\Lambda_0$ the cumulative density function of $-v_i$. We will consider a logistic distribution in our empirical analysis below, i.e., $\Lambda_0(\varphi)=\Lambda(\varphi)=\frac{e^\varphi}{1+e^\varphi}$.\footnote{The logistic distribution offers theoretical and computational advantages over alternatives, see Section 5.3 below.} This parametric assumption provides a necessary anchor to the overall model; we will show below that identification fails to hold when $\Lambda_0$ is unknown.

We normalize the threshold promotion grade to zero. We can do this by adding a constant term and exam shifters to the function $\varphi$. Unconnected candidate $i$ is thus promoted when $\varphi(\mathbf{x}_{i}) +v_{i} \geq 0$ which yields:

\begin{equation*}
    \mathbb{P}(y_i=1|c_i=0,\mathbf{x}_i)=\Lambda_0(\varphi(\mathbf{x}_{i})).
\end{equation*}

Now consider connected candidates. We assume that being connected to the jury can have two different effects on grades. First,  the jury may want to unfairly favor candidates with connections. Favors take the form of a homogeneous bonus grade, equal to $B$. When $B>0$, connected candidates, in fact, face a lower promotion bar.

Second, the jury obtains extra information on the candidate's ability. Formally, denote by $\theta_i$ the extra signal obtained by the jury about the connected candidate $i$'s ability. The jury's updated belief about the candidate's unobserved ability is then equal to $\E(u_i|\theta_i,\obs_i,v_i)$. By the law of iterated expectations and since $\E(u_i)=0$, this updated belief also has zero mean.

We consider two nested assumptions on the information structure of this signal: \textit{independence from observables}: $(u_i,\theta_i,v_i) \Perp \obs_i$ and \textit{independence}: $(u_i,\theta_i) \Perp (\obs_i,v_i)$. Under independence from observables, the jury's extra information about the connected candidate does not depend on $\obs_i$ but could depend on $v_i$, i.e., on characteristics observed by the jury but not by the econometrician. Under independence, the jury's extra information does not depend on $\obs_i$ nor on $v_i$.

The grade given by the jury to the connected candidate $i$ is then equal to
\begin{equation}\label{eq:grade_by_jury}
B+\E(a_i|\mathbf{x}_i,v_i,\theta_i)=B+\varphi(\mathbf{x}_{i})+v_{i}+\E(u_i|\theta_i,\obs_i,v_i). 
\end{equation}

Denote by $w_i=v_{i}+\E(u_i|\theta_i,\obs_i,v_i)$ and note that $\mathbb{E}(w_i)=0$. This variable captures what is observed by the jury, but not by the econometrician, about the ability of connected candidates. Independence from observables implies that $\E(u_i|\theta_i,\obs_i,v_i)=\E(u_i|\theta_i,v_i)$ while under independence, $\E(u_i|\theta_i,\obs_i,v_i)=\E(u_i|\theta_i)$. In either case, $w_i \Perp \mathbf{x}_i$. Denote by $\Lambda_1$ the cumulative density function of $-w_i$. In the baseline version of our framework, we make no parametric assumption on $\Lambda_1$. To anchor non-parametric estimation, we will adopt a classical zero median assumption, $\Lambda_1(0)=1/2$ (see, e.g., \cite{manski1975maximum, manski1988identification, KHAN2013168}). However, the zero mean property of $w_i$ is sufficient for our identification results below.

Under either independence assumption, the conditional promotion probabilities for connected and unconnected candidates can be written as follows:
\begin{align}\label{eq:SimpleModel_General}
	\begin{split}
    &\mathbb{P}_0(\mathbf{x}_{i})=\mathbb{P}(y_{i}=1|c_{i}=0,\mathbf{x}_{i})=\Lambda_0\left(\varphi( \mathbf{x}_{i})\right) \\
		&\mathbb{P}_1(\mathbf{x}_{i})=\mathbb{P}(y_i=1|c_i=1, \mathbf{x}_{i}) = \Lambda_1 \left( \varphi( \mathbf{x}_{i})+B\right)
	\end{split}
\end{align}
We next derive testable restrictions and analyze identification. A key feature of this benchmark model is that $\mathbb{P}_0$ and $\mathbb{P}_1$ both depend on multidimensional observables $\mathbf{x}_i$ through a common single-valued index $\varphi$, the grade function. This feature is associated with strong testable restrictions. It implies, in particular, that $\mathbb{P}_1$ is a monotonically increasing function of $\mathbb{P}_0$. To see this, denote by $\Lambda_0^{-1}$ the inverse function of $v_i$'s cumulative density function. A direct implication of Equation (\ref{eq:SimpleModel_General}) is that
\begin{equation}\label{eq:P1(P0)}
    \mathbb{P}_1(\mathbf{x}_{i}) =\Lambda_1((\Lambda_0)^{-1}( \mathbb{P}_0(\mathbf{x}_{i}))+B)
\end{equation}
\noindent and we will test this monotonic relationship in Section \ref{subsec:testing_restrictions}.

Identification can be derived in three steps. First, the grade function, $\varphi$, is identified from the unconnected promotion probability $\mathbb{P}_0$ since $\Lambda_0$ is known. Second, the bonus from favors, $B$, can be identified from the connected promotion probability $\mathbb{P}_1$ and from the knowledge of $\varphi$ and the property of zero mean or zero median of $w_i$. Third, the cumulative density function of $w_i$ can be fully recovered from $\mathbb{P}_1$ once $B$ and $\varphi$ are known.

We next state this result formally and provide a detailed proof. We say that full support holds when the econometrician can recover $\mathbb{P}_0(\mathbf{x}_i)$ and $\mathbb{P}_1(\mathbf{x}_i)$ for all probabilities in $(0,1)$. Full support only holds when the grade function $\varphi$ spans all values in $\mathbb{R}$.\footnote{For instance with one continuous characteristic $x_i \in \mathbb{R}$ and a linear grade function $\varphi=\alpha+\beta x_i$, full support holds when $\beta \neq 0$ and the underlying distribution of $x$'s in the population is everywhere positive.} 

\begin{theorem}\label{th:IdentificationBenchmark}
Consider the benchmark model of promotion. Let $v_i$ and $w_i=v_i+\E(u_i|\theta_i,\obs_i,v_i)$ denote what is observed by the jury but not by the econometrician about the ability of candidate $i$ when $i$ is unconnected ($v_i$) or connected ($w_i$) to the jury.

Assume full support, that the bonus grade from favors is homogeneous and equal to $B$, and that the jury's extra information about connected candidates is independent from observables $(u_i,\theta_i,v_i) \Perp \mathbf{x}_{i}$.

If the distribution of $v_i$ is known, then $B$ and the distribution of $w_i$ are identified. Moreover, when the jury's extra information is independent, $(u_i,\theta_i) \Perp (\mathbf{x}_{i},v_i)$, then $w_i$ represents a mean-preserving spread of $v_i$.
\end{theorem}

\begin{proof}[Proof of Theorem \ref{th:IdentificationBenchmark}]
The grade function $\varphi$ is identified from $\mathbb{P}_0$ since $\varphi(\mathbf{x}_{i})=(\Lambda_0)^{-1}(\mathbb{P}_0(\mathbf{x}_{i}))$.

Next, we show that $B$ can be identified either from the zero median or from the zero mean condition. The zero median condition says that $\Lambda(0)=\frac{1}{2}$. Consider any observables $\hat{\mathbf{x}}_i$ such that $\mathbb{P}_1(\hat{\mathbf{x}}_i)=\frac{1}{2}$. These observables exist by full support. Since $\mathbb{P}_1(\hat{\mathbf{x}}_i)=\Lambda_1(\varphi(\hat{\mathbf{x}}_{i})+B)=\frac{1}{2}$, then  
$B=-\varphi(\hat{\mathbf{x}}_{i})$. The zero mean condition says that $\mathbb{E}(w_i)=0$. This is equivalent to  $\int w d\Lambda_1(w)=0$. Denote by $P_1(\varphi)=\Lambda_1(\varphi+B)$ and set $\varphi=w-B$. This yields $\int (\varphi+B) dP_1(\varphi)=0$ and hence $B=-\int \varphi dP_1(\varphi)$. This integral can be computed since, by full support, $P_1(\varphi)$ is known for all values of $\varphi$.

Once $\varphi$ and $B$ are known, the cumulative distribution function of $w_i$ is given by $\Lambda_1(w)=\mathbb{P}_1(\mathbf{x}_i)$ for observables $\mathbf{x}_i$ such that $\varphi(\mathbf{x}_{i})=w-B$. These observables exist by full support.

Finally, note that when $(u_i,\theta_i) \Perp (\mathbf{x}_{i},v_i)$, then  $w_i=v_i+\E(u_i|\theta_i)$ with $\E(\E(u_i|\theta_i)|v_i)=0$. By \cite{rothschild1970increasing}, $w_i$ represents a mean-preserving spread of $v_i$.
\end{proof}

Let us discuss and illustrate the key identification ideas underlying Theorem \ref{th:IdentificationBenchmark}. In the presence of favors only,  $\mathbb{P}_1(\mathbf{x}_{i})=\Lambda_0(\varphi(\mathbf{x}_{i})+B)$. When $B>0$, this corresponds to a left-shift of the promotion probability. In this case, $\mathbb{P}_1$ first-order stochastically dominates $\mathbb{P}_0$. Favors increase the likelihood of being promoted at all levels. In the presence of information effects only, $\mathbb{P}_1(\mathbf{x}_{i})=\Lambda_1(\varphi(\mathbf{x}_{i}))$. Under independence, this corresponds to a flattening of the promotion probability, and $\mathbb{P}_0$ second-order stochastically dominates $\mathbb{P}_1$. Since the jury has more private information on candidates, the observed probability of being promoted varies less with observable characteristics. The promotion probability thus tends to be higher at low levels and lower at high levels. For good candidates, connections apparently have a negative impact. This is due to the asymmetry in the effects of good and bad news on good candidates' chances of being promoted. While good news does not improve already good chances by much, bad news reduces these chances significantly. On average, more accurate information conveyed by connections thus reduces the observed probability of being promoted for good candidates.

\begin{figure}[!h]
	\caption{Effects of a connection on promotion probability}
	\label{fig:IdentificationPlot}%
	\includegraphics[scale=0.9]{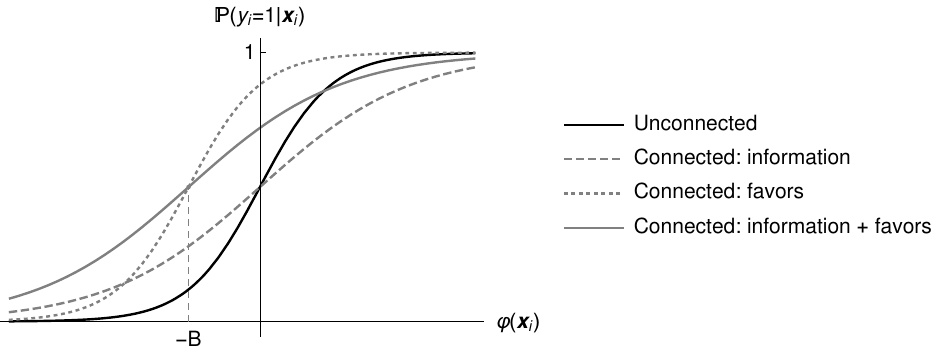}
\end{figure}

In general, both effects may be at play, leading to a combination of a left-shift and a flattening, illustrated in Figure \ref{fig:IdentificationPlot}. The graph depicts the probability of being promoted as a function of the normalized grade $\varphi = \varphi(\obs_i)$. The plain black curve depicts the promotion probability of an unconnected candidate. Probability is $\frac{1}{2}$ when $\varphi=0$. The short-dashed curve depicts the promotion probability of a connected candidate when only favors are involved. The curve is translated to the left, inducing a first-order stochastic dominance shift. Probability is $\frac{1}{2}$ when $\varphi=-B$. The long-dashed curve depicts the promotion probability in the presence of information effects only. The whole curve is less steep but has the same intercept, probability is $\frac{1}{2}$ when $\varphi=0$. Finally, the grey curve depicts the promotion probability of connected candidates in the presence of the two effects.

We would like to highlight three features of this approach. Observe, first, that Theorem \ref{th:IdentificationBenchmark} holds even when $B<0$ and when the extra signal from connections is correlated with $v_i$. A negative $B$ would correspond to a right-shift of the promotion probability and would mean that connected candidates are disfavored relative to unconnected ones. And negative correlation between the extra signal and $v_i$ could lead to a promotion probability that varies more with observable characteristics, not less. Theorem \ref{th:IdentificationBenchmark} tells us that $B$ and the distribution of $w_i$ are identified in these cases as well. 

Second, the model is not identified if we do not assume that the distribution of $v_i$ is known. To see why, consider the case where $B=0$ and $\mathbb{P}_0(\mathbf{x}_{i})=\Lambda_0\left(\varphi( \mathbf{x}_{i})\right)$ and $\mathbb{P}_1(\mathbf{x}_{i})=\Lambda_1\left(\varphi( \mathbf{x}_{i})\right)$. Consider any strictly increasing transform $f$ such that the stochastic variables $f(-v_i)$ and $f(-w_i)$ have zero mean. Then, the model with grade function $f(\varphi)$ and cdf's $\Lambda_0(f^{-1})$ and $\Lambda_1(f^{-1})$ yields the same conditional promotion probabilities. 

Third, Theorem \ref{th:IdentificationBenchmark} shows that no parametric restriction is needed to identify the distribution of $w_i$, conditional on assuming that the distribution of $v_i$ is known. Favors are then measured by a single number, while information effects are characterized by an entire distribution. In our empirical applications below, we will also consider a parametric version of our model, where both favors and information effects are measured by a single number. In this version, we assume that $w_i$ follows a logistic distribution with variance $\mathbb{V}(w_i)=\sigma^2\mathbb{V}(v_i)$.\footnote{This is consistent with independence and $w_i=v_i+\mathbb{E}(u_i|\theta_i)$ when $\mathbb{E}(u_i|\theta_i)$ is distributed according to the $GL(1/\sigma)$ distribution described in \cite{cardell1997variance}.} In this case, $\mathbb{P}_1(\mathbf{x}_{i})=\Lambda(\frac{\varphi(\mathbf{x}_{i})+B}{\sigma})$. Under independence, $\sigma$ should be greater than $1$, and higher values of $\sigma$ correspond to more accurate information obtained through connections.

\section{Extended Model}\label{sec:ExtendedModel}

We now show that identification of favors and information effects may hold even when these two effects depend on the number and types of connections to the jury and on candidates' observable characteristics. The presence of an exclusion restriction is a key necessary condition, however, i.e., the existence of an observable characteristic that does not affect favors and information effects. Identification fails when favors and information effects may vary with all observable characteristics. 

Formally, let $\mathbf{c}_i$ be a vector representing the number and types of connections that candidate $i$ has with the jury. For instance if candidate $i$ has $n^S_i$ strong ties and $n^W_i$ weak ties with jury members, then $\mathbf{c}_i=(n^S_i,n^W_i)$. Unconnected candidates are such that $\mathbf{c}_i=\mathbf{0}$. We now distinguish between two kinds of observable characteristics, excluded variables $\mathbf{x}_i$ and included ones $\mathbf{z}_i$. Bias and information effects do not depend on $\mathbf{x}_i$ but may depend on $\mathbf{z}_i$. Gender has to be included in $\mathbf{z}_i$, for example, to study whether the impacts of connections differ for male and female candidates.

We maintain and extend the assumptions of the benchmark model. To think about identification, we can assume that the econometrician recovers conditional promotion probability $\mathbb{P}(y_i=1|\mathbf{c}_i,\mathbf{x}_i,\mathbf{z}_i)$. Connections are assumed to be random conditional on a subset of observable characteristics. Importantly, we assume that the relevant conditioning variables do not affect the impact of connections - they belong to $\mathbf{x}_i$ not to $\mathbf{z}_i$, $\mathbf{c}_i \Perp (u_i,v_i) | \mathbf{x}_i$. 
This means that, conditional on excluded variables $\mathbf{x}_i$, unobservables have the same distributions for all possible realizations of $\mathbf{c}_i$.

Candidate $i$'s ability is the sum of three terms, $a_i=\varphi(\mathbf{x}_i,\mathbf{z}_i)+u_i+v_i$ with $\mathbb{E}(u_i|\mathbf{z}_i)=\mathbb{E}(v_i|\mathbf{z}_i)=0$, $v_i \Perp \obs_i |\mathbf{z}_i$ and $u_i \Perp (\obs_i,v_i)|\mathbf{z}_i$. Potential differences in expected abilities as a function of observables $\mathbf{x}_i$ or $\mathbf{z}_i$ are directly included in function $\varphi$. The probability distribution of $-v_i$ may now depend on $\mathbf{z}_i$ and is denoted by $\Lambda_{0,\mathbf{z}_i}$.\footnote{This distribution does not depend on $\mathbf{x}_i$ by assumption, nor on $\mathbf{c}_i$ since connections are conditionally random.} Thus, the conditional promotion probability of unconnected candidates is equal to
\begin{equation}\label{eq:ProbaPromotionUnconExtended}
    \mathbb{P}(y_i=1|\mathbf{c}_i=\mathbf{0},\mathbf{x}_i,\mathbf{z}_i)=\Lambda_{0,\mathbf{z}_i}(\varphi(\mathbf{x}_i,\mathbf{z}_i)).
\end{equation}

Next, consider connected candidates with connections $\mathbf{c}_i \neq \mathbf{0}$. The jury gives them a bonus grade from favors equal to $B(\mathbf{c}_i,\mathbf{z}_i)$ and obtains a signal $\theta_i$ whose distribution might also depend on $\mathbf{c}_i$ and $\mathbf{z}_i$. Dependence on $\mathbf{c}_i$ can capture situations where the extent of favors and the accuracy of the extra signal from connections are higher for candidates with more, or stronger, connections to the jury. Similarly, dependence on, say, gender arises when a jury gives different favors to men and to women and receives signals from connections of different accuracies for men and for women.

In this extended model, we consider the two following nested assumptions on the structure of the signal: independence from excluded observables,  $(u_i,\theta_i,v_i) \Perp \obs_i | (\mathbf{c}_i,\mathbf{z}_i)$ and independence, $(u_i,\theta_i) \Perp (\obs_i,v_i) | (\mathbf{c}_i,\mathbf{z}_i)$. Let $w_i=v_i+\mathbb{E}(u_i|\theta_i,\obs_i,v_i,\mathbf{z}_i,\mathbf{c}_i)$ denote what is observed by the jury but not by the econometrician on the ability of candidate $i$. As in the benchmark model, $w_i$ has zero mean and, under either independence assumption, $w_i \Perp \obs_i | (\mathbf{c}_i,\mathbf{z}_i)$. Denote by $\Lambda_{\mathbf{c}_i,\mathbf{z}_i}$ the distribution of $w_i$, conditional on connections $\mathbf{c}_i$ and included observables $\mathbf{z}_i$.

The conditional promotion probability of candidates with connections $\mathbf{c}_i$ is thus equal to:
\begin{equation}\label{eq:ProbaPromotionConExtended}
    \mathbb{P}(y_i=1|\mathbf{c}_i,\mathbf{x}_i,\mathbf{z}_i)=\Lambda_{\mathbf{c}_i,\mathbf{z}_i}(\varphi(\mathbf{x}_i,\mathbf{z}_i)+B(\mathbf{c}_i,\mathbf{z}_i)).
\end{equation}
\noindent We extend the full support assumption and say that full support holds conditionally on $\mathbf{c}_i,\mathbf{z}_i$ if for every possible value of $\mathbf{c}_i,\mathbf{z}_i$, the econometrician can recover $\mathbb{P}(y_i=1|\mathbf{c}_i,\mathbf{x}_i,\mathbf{z}_i)$ for all probabilities in $(0,1)$. This holds when the conditional grade function $\varphi(\cdot,\mathbf{z}_i)$ spans all values in $\mathbb{R}$. 

\begin{theorem}\label{th:IdentificationExtended}
    Consider the extended model of promotion, where favors and information effects from connections may depend on number and types of connections $\mathbf{c}_i$ and on observables $\mathbf{z}_i$.

    Assume that there exist excluded observables $\mathbf{x}_i$ such that full support holds conditionally on $(\mathbf{c}_i,\mathbf{z}_i)$, and that the bonus grade from favors and the jury's extra information about connected candidates do not depend on $\mathbf{x}_i$, $(u_i,\theta_i,v_i) \Perp \obs_i | (\mathbf{c}_i,\mathbf{z}_i)$. 

    If the conditional distribution of $v_i$, $\Lambda_{0,\mathbf{z}_i}$, is known, then for $\mathbf{c}_i \neq \mathbf{0}$ the conditional bonus grade from favors, $B(\mathbf{c}_i,\mathbf{z}_i)$, and the conditional distribution of $w_i$, $\Lambda_{\mathbf{c}_i,\mathbf{z}_i}$, are identified. 

    Moreover, under independence, $(u_i,\theta_i) \Perp (\obs_i,v_i) | (\mathbf{c}_i,\mathbf{z}_i)$, $w_i|(\mathbf{c}_i,\mathbf{z}_i)$ represents a mean-preserving spread of $v_i|\mathbf{z}_i$.
\end{theorem}

\begin{proof}[Proof of Theorem \ref{th:IdentificationExtended}]
    The proof is analogous to the proof of Theorem \ref{th:IdentificationBenchmark}.
    For any fixed value of  $\iobs_i$ the grade $\varphi\left( \obs_i,\iobs_i \right)$ is identified as $\varphi\left( \obs_i,\iobs_i \right) = \Logit_{0,\iobs_i}^{-1} \left(\prob(\obs_i, \iobs_i) \right)$.  Since this holds for any $\iobs_i$, the grade function $\varphi$ is identified.  

    Next, we show that, for any fixed $(\mathbf{c}_i, \iobs_i)$,  $B(\mathbf{c}_i, \iobs_i)$ can be identified either from zero median or from the zero mean condition. 
    Suppose first that zero median condition holds for all $\Logit_{\mathbf{c}_i, \iobs_i}$. Choose any connection profile $\mathbf{c}_i$ and $\iobs_i$. Now, consider any excluded observable $\hat{\obs}_i = \obs_i(\mathbf{c}_i, \iobs_i)$ such that $\prob(y_i=1|\hat{\obs}_i, \iobs_i, \mathbf{c}_i) =0$. Under the full support assumption such excluded observable exists. Then $B(\mathbf{c}_i, \iobs_i) = -\varphi(\hat{\obs}_i, \iobs_i).$
    Under the zero-mean condition $\int w \diff \Logit_{\mathbf{c}_i, \iobs_i} = 0$. For a given ($\mathbf{c}_i, \iobs_i$) define $P_{\mathbf{c}_i, \iobs_i}(\varphi) = \Logit_{\mathbf{c}_i, \iobs_i}(\varphi+B(\mathbf{c}_i, \iobs_i))$ where $\varphi = \varphi(\cdot, \mathbf{c}_i, \iobs_i)$, then $B(\mathbf{c}_i, \iobs_i) = -\int \varphi \diff P_{\mathbf{c}_i, \iobs_i}(\varphi)$. 
    
    Since we can identify $B(\mathbf{c}_i, \iobs_i)$ for any given value of $B(\mathbf{c}_i, \iobs_i)$, function $B$ is identified.

     The  conditional cumulative distribution function $w_i|\mathbf{c}_{i}, \iobs_i$  is then given by \linebreak $\Logit_{\mathbf{c}_i, \iobs_i}(w_i)= $  $\prob\left(y_i=1|\obs_i,\iobs_i, \mathbf{c}_i\right)$ for $\obs_i$ such that $\guconz = w_i - B(\mathbf{c}_i, \iobs_i)$.

     Under independence, $w_i|(\mathbf{c}_i,\mathbf{z}_i)=v_i|\mathbf{z}_i+\E(u_i|\theta_i)|(\mathbf{c}_i,\mathbf{z}_i)$ with $\E((\E(u_i|\theta_i)|(\mathbf{c}_i,\mathbf{z}_i))|(v_i|\mathbf{z}_i))=0$. By \cite{rothschild1970increasing}, $w_i|(\mathbf{c}_i,\mathbf{z}_i)$ represents a mean-preserving spread of $v_i|\mathbf{z}_i$.
\end{proof}

As in the benchmark model, identification is obtained by contrasting the promotion probability of connected candidates with connections $\mathbf{c}_i$ and included variables $\mathbf{z}_i$ and the promotion probability of unconnected candidates with the same included variables. The exclusion restriction, combined with the support assumption, implies that these two promotion probabilities span the full $[0,1]$ range. We can then apply the identification strategy from the benchmark model, and recover the bias from favors from the zero mean or zero median property and the distribution of the extra signal associated with connections.

As in the benchmark model, identification fails to hold when $\Lambda_{0,\mathbf{z}_i}$ is unknown. Identification also fails to hold in the absence of excluded variables. In that case, the conditional promotion probability $\mathbb{P}(y_i=1|\mathbf{c}_i,\mathbf{z}_i)$ is simply a scalar, from which researchers cannot separately identify one scalar $B(\mathbf{c}_i,\mathbf{z}_i)$ and a cumulative density function $\Lambda_{\mathbf{c}_i,\mathbf{z}_i}$.

In our empirical applications below, we will also consider a parametric version of the extended model, where conditional favors and information effects are both captured by a single parameter. 
In this parametric version, $\mathbb{P}(y_i=1|\mathbf{c}_i,\mathbf{x}_i,\mathbf{z}_i)=\Lambda\left(\frac{\varphi(\mathbf{x}_{i},\mathbf{z}_{i})+B(\vec{c}_i, \iobs_i)}{\sigma(\vec{c}_i, \iobs_i)}\right)$ with $B(\mathbf{0}, \iobs_i)=0$ and $\sigma(\mathbf{0},\iobs_i)=1$.

\section{Empirical Applications}\label{sec:empirical_applications}

When connections are conditionally random, researchers can simply estimate the causal impact of connections on promotion by computing the difference in average promotion probabilities between connected and unconnected candidates. In Theorems \ref{th:IdentificationBenchmark} and \ref{th:IdentificationExtended}, we provide conditions under which researchers can, with the same data, go beyond the estimation of the causal impact and identify the mechanisms behind the effect of connections.

We now apply our identification strategy to three different contexts: academic promotions in Spain from 2002 to 2006, academic promotions in Italy from 2012 to 2014, and promotions to the Politburo in China from 1993 to 2009. We consider publicly available data, which have been assembled by researchers to study promotion and, in particular, to estimate the causal impact of connections on promotion. Data on Spain come from \cite{zinovyeva2015role} and \cite{bagues2017does}, on Italy from \cite{bagues2017does}, and on China from \cite{jia2015political}. We next provide brief descriptions of the three datasets and institutional contexts, and refer the reader to the original articles for details.

Note that our objective is not to replicate or to evaluate the robustness of the original analyses. Rather, we want to implement our identification strategy on real data and to assess its performance in explaining actual promotion patterns. We build upon the original analyses, modifying them only when necessary to accommodate the specific requirements of our approach. In particular, we account for the same set of control variables and analyze the types of connections considered in the original papers.

\subsection{Academic promotions in Spain}
From 2002 to 2006, academics in Spain seeking promotion to Associate Professor (\textit{profesor titular}) or Full Professor (\textit{catedr\'{a}tico}) first had to
qualify in a national exam (\textit{habilita\'{c}ion}). Candidates were evaluated by an exam-specific jury composed of seven members, responsible for allocating a predetermined number of positions. These exams were highly competitive, and obtaining the national qualification essentially ensured promotion. A central feature of this system was that jury members were picked at random from a pool of eligible evaluators after the application deadline. Participation in these juries was mandatory, and only 2\% of the initially selected evaluators ended up not participating.

The data contain information on all candidates for academic promotion during that period, their connections to eligible evaluators and to jury members, and their success or failure in the national exam. Overall, there are $31,243$ applications to $967$ exams: $17,799$
applications to $465$ exams for Associate Professor (AP) positions and $%
13,444$ to $502$ exams for Full Professor (FP) positions.  On average, there are about 32 candidates per exam and only 11\% of the candidates obtained the qualification. We have
information on candidates' demographics and academic outcomes at the time of application. Observable characteristics include gender, age, whether the candidate obtained his PhD in Spain, number of publications, number of publications weighted by journal quality, number of PhD students supervised, number of PhD committees the candidate was on, and number of previous attempts at promotion. Table \ref{tbl:DescriptiveObservables} provides
descriptive statistics. As in \cite{zinovyeva2015role}, we normalize research indicators to have mean $0$ and variance $1$ within exams.\footnote{We also normalize age and past experience to have a mean of $0$ within each exam.} %

The data also contain information on six types of links between candidates and evaluators. We adopt \cite{zinovyeva2015role}'s classification of these links into strong and weak ties.\footnote{The data also contains information on indirect connections, for instance, when a candidate and an evaluator had a common member on their PhD committees.\ \cite{zinovyeva2015role} do not find any effect of indirect connections and we do not include them in our analysis.} Candidates are
said to have strong ties to their PhD advisor(s), to their coauthors and to their colleagues. They have weak ties to members of their PhD committee, to members of the PhD committees of their PhD students and to fellow PhD committee members.
Overall, 32\% of candidates have at least one strong connection with a member of their jury, and 19\% have at least one weak connection. Table \ref%
{tbl:DescriptiveConnections} provides further information on connections. In our empirical analysis, we restrict attention to exams with at least one connected and at least one unconnected candidate, reducing the number of observations by 0.8\%, to 31,000.

\subsection{Academic promotions in Italy}

In 2012, Italy put into place a national qualification system (\textit{abilitazione scientifica nazionale}) similar to the Spanish one. Our analysis focuses on data gathered during the initial phase of this system, from 2012 to 2014. As in Spain, any academic seeking promotion to Associate Professor (\textit{Professore di seconda fascia}) or Full Professor (\textit{Professore di prima fascia}) during this period first had to obtain this qualification. There are some important differences with the Spanish context, however. In Italy, the number of available positions was not predetermined, and obtaining a qualification did not guarantee promotion. There are more candidates per exam -- 188 in Italy, 32 in Spain -- and the average success rate is also much higher -- 37\% in Italy, 11\% in Spain. All candidates in a discipline  were evaluated by a common jury composed of five members (184 disciplines (\textit{settore concorsuale}) in total). Jury members were also picked at random in a pool of eligible evaluators. Unlike in Spain, however, participation was not mandatory and selected jury members could choose not to participate. Among initially selected evaluators, 8\% ended up not participating in the jury. This is a small, but perhaps not insignificant number, which raises potential concerns of endogeneity.

Overall, the data includes information on 69,020 applications to 368 exams: 47,426 to 184 exams for Associate Professor positions and 21,594 to 184 exams for Full Professor positions. We have information on candidates’ demographics and academic outcomes at the time of application, including gender, age,  length of CV in pages, number of published research articles, number of publications of different kinds, number of \textit{class A} journal articles,  number of publications weighted by journal quality, average number of coauthors per publication,  number of publication where the candidate's name is first and last in the list of authors. Table \ref{tbl:DescriptiveObservablesItaly} provides descriptive statistics. A main difference with the Spanish data is that we do not have information on PhD defenses and committees, and hence on connections inferred from this information.

We have information on two kinds of strong ties between applicants and jury members: coauthors and colleagues. Overall, 16\% of the candidates have at least one strong tie to a jury member, with coauthorship connections being more common. Table \ref{tbl:DescriptiveConnectionsItaly} provides descriptive statistics.\footnote{In Italy each exam has at least one connected and at least one unconnected  candidate.}

\subsection{Political promotions in China}
 
Our third application is about political promotion in China, based on data assembled by \cite{jia2015political}. China is organized into 31 provinces governed by two political leaders holding the offices of provincial secretary and provincial governor. The Politburo Standing Committee of the Communist Party of China decides which provincial leaders to promote to a member of the Politburo or a higher rank within the central government. Provincial leaders are relatively independent in governing the provinces, and the real GDP growth is a measure of performance used by the Politburo in assessing the quality of the leaders (\cite{maskin2000incentives}). Provincial leaders may be connected with members of the Politburo, which potentially affects their chances of being promoted.  

The dataset contains information on 258 provincial leadership spells\footnote{The Communist Party makes major personnel decisions every five years. The average duration of a spell is 4.3 years. There are only two years in our sample when no promotion is observed (1995 and 2008).} for 187 different officials in the period 1993 -- 2009, which includes information on the province-level annual real GDP growth.  The unconditional probability of promotion of a provincial leader in any given year is 7\%, and the probability of being promoted during a leadership spell is 26\%.   A provincial leader is said to be connected to a member of the Politburo if they used to work in the same branch of the Party or of the government at the same time in the past.  Table \ref{tbl:DescriptiveObservablesChina} provides descriptive statistics.

\section{Empirical implementation}\label{sec:EmpricalImplementation}

We now discuss three main features of the empirical implementation of our identification strategy: the conditionally random connections, the exam-specific promotion thresholds, and the econometric framework.

\subsection{Conditionally random connections}

Our identification results, Theorems \ref{th:IdentificationBenchmark} and \ref{th:IdentificationExtended}, rely on the assumption that connections are conditionally random. This property is also key for the estimation of the causal impact of connections, and is thoroughly discussed and assessed in the original papers. 

In Spain, jury members are picked at random among eligible evaluators, and jury participation is mandatory. This implies that actual connections to the jury are random, conditional on expected connections to the jury. That is, candidates, of course, have different networks, and candidates with larger networks are more likely to end up being connected to the jury. To control for this in practice, we follow \cite{zinovyeva2015role} and use the expected number of connections of a candidate to the jury. These numbers are computed in \cite{zinovyeva2015role} from the number of eligible evaluators and the number of weak and strong ties to these eligible evaluators, with the expectation taken over all possible choices of jury members. We present the corresponding balance tests in Table \ref{tbl:BalanceTest}.\footnote{To be consistent with our main regressions, we run balance tests conditioning directly on the expected numbers of connections. By contrast, \cite{zinovyeva2015role} control for expected connections through an extensive set of dummies, see their Table 4 p.278. Incorporating these dummies raises computational issues in our non-linear setup. Results from Table \ref{tbl:BalanceTest} show that even in a simple linear formulation, actual connections are conditionally uncorrelated with observable characteristics.} When controlling for candidates' expected numbers of connections, we do not detect significant residual correlations between observable characteristics and the actual number of connections. Note that the expected numbers of connections represent measures of social capital, built from information available to the jury, therefore we include them in the set of candidates' observable characteristics in the empirical analysis.

In Italy, jury members are also picked at random among eligible evaluators. Since participation is not mandatory, however, initially selected evaluators can resign, which raises potential selection issues. We use expected numbers of connections to the jury, as computed in \cite{bagues2017does},  and perform balance tests, described in Table \ref{tbl:RandomAssignementItaly}. We see that conditioning on the expected number of connections removes most, but not all, of the correlation between the actual number of connections and observable characteristics. In Section \ref{sec:endogeneous_connections} below, we develop an extension of our framework to account for connection endogeneity using a control function approach, and we apply this extension to the Italian data. We do not detect any significant effect of connection endogeneity. This is consistent with the fact that evaluator resignation is infrequent. In our main empirical analysis in Section \ref{sec:EmpiricalAnalysis}, we thus analyze Italian data by neglecting this residual endogeneity and by assuming conditionally random connections.

In China, two party officials are connected if they used to work in the same branch of the Party or of the government at the same time in the past. \cite{jia2015political} perform a number of balance tests to show that connections appear to be as good as random. In particular, they find that previous growth of provinces (the main indicator used for promotion) is not significantly different when ruled by connected versus unconnected officials once province and year fixed effects are accounted for. They do not find significant differences in government support to provinces ruled by connected or unconnected leaders, as measured by explicit transfer to provinces and credit market interventions. \footnote{We verify that this result holds when controlling for province and year fixed effects.} We thus develop our analysis by assuming that connections are random conditionally on province and year fixed effects.

\subsection{Exam-specific promotion thresholds}\label{subsec:ExamSpecificHeterogeneity}

In our approach, the bias from favors is identified from differences in promotion thresholds between connected and unconnected candidates. In our three applications, we observe promotion decisions across many exams and years. To implement our identification strategy, we must, therefore, incorporate exam-specific promotion thresholds.

In the two applications on academic promotion, we include either exam fixed effects or exam controls in the econometric specifications. While exam fixed effects impose fewer restrictions, their inclusion may raise practical issues. They may not be identified for exams with a small number of candidates because of full predictability. And they may raise incidental parameter problems (\cite{wooldridge2010econometric}). These issues arise in the Spanish application, where the number of candidates per exam and the likelihood of being promoted are both quite low. In our regressions on the Spanish data, we thus include exam controls in the econometric specifications.\footnote{We control for the number of candidates, the number of positions available, and the number of filled positions, including powers and interaction terms. When the model with fixed effects is estimable in practice, we test one model against another. We consistently find that the model with exam fixed effects does not explain the data better than the model with exam controls, see Table \ref{tbl:RobustnessFE_ParametricSpain} in the  Online Appendix B.} In the Italian context, by contrast, the exams have more candidates and are less competitive. Exam fixed effects are always identified and, at an average of 188 candidates per exam, the incidental parameter problem is very likely negligible. We therefore include exam fixed effects in our regressions on the Italian data.

In the application on political promotions in China, some promotion decisions are made every year, and there are no specific exams like in the cases of academic promotions in Spain and Italy. By controlling for year fixed effect, we control for the year-specific threshold. 

\subsection{Econometric framework}\label{subsec:econometric_framework}

We now present the econometric models and the estimation procedures used in our empirical analysis. We focus on specifications where favors and information effects only depend on connections to the jury. An important feature of our implementation is a flexible estimation of the grade function $\varphi$ via a polynomial approximation. This function plays a central role, and its flexible estimation helps reduce the risks of misspecification.

Consider, first, the version of our model where unobservables have a logistic distribution. Promotion probability for candidates with observed characteristics $\mathbf{x}_{i}$ and connections $\con_i$ is equal to

\begin{equation}
    \label{eq:knownmodel}
    \prob(y_i=1|\obs_i, \con_i)= \Logit \left( \frac{\varphi(\mathbf{x}_{i}) + B(\con_i)}{\sigma(\con_i)} \right)
\end{equation}

\noindent with $B(\mathbf{0})=0$ and $\sigma(\mathbf{0})=1$. We estimate two variants of this model. In the first variant, the effects of connections only depend on the presence of a connection to the jury, $B(\con_i)=B$ and $\sigma(\con_i)=\sigma=\exp(\delta)$ if $\con_i \neq \mathbf{0}$. In the second variant, favors and information effects depend on the number and type of connections. We adopt a linear formulation for the bias and a log-linear formulation for the variance, $B(\con_i) = \vec{\gamma}\con_i$ and $\sigma(\con_i)= \exp(\vec{\delta} \con_i)$. Potential non-linearities could easily be explored by relaxing these assumptions.

If the grade function $\varphi$ has a known functional form, model (\ref{eq:knownmodel}) corresponds to a standard heteroskedastic logit, and can be estimated via maximum likelihood. Since the grade function is unknown, however, we consider increasingly complex polynomial approximations of $\varphi$ and select the best approximation. We start with a linear specification and then introduce terms of successively higher order. We follow \cite{hansen2014nonparametric}, and choose the specification which has the best out-of-sample prediction based on the mean squared error criterion.\footnote{We evaluate the out-of-sample prediction using cross-validation.}

Next, we relax the assumption that the distribution of unobservables of connected candidates is known. Promotion probability of unconnected candidates is equal to $ \prob(y_i=1|\obs_i, \con_i=\mathbf{0})= \Logit_0 \left(\varphi(\mathbf{x}_{i}) \right)$ where we still assume that  $\Lambda_0 \equiv \Logit$ is the logit cdf. Promotion probability of connected candidates is now equal to  $\prob(y_i=1|\obs_i, \con_i)= \Logit_{\con_i} \left(\varphi(\mathbf{x}_{i}) + B(\con_i) \right)$ where $\Logit_{\con_i}$ is unknown. To estimate this model, we reformulate it in a unified logit framework as follows:
\begin{equation}\label{eq:FirstStage_General}
\prob(y_i =1|\obs_i, \con_i) = \Logit_0 \left[\varphi( \obs_{i}) + \mathbbm{1}_{\con_i \neq \mathbf{0}} \psi_{\con_i}(\obs_i)\right],
\end{equation}
\noindent where $\psi_{\con_i}(\obs_i)=\Logit_0^{-1}(\Logit_{\con_i}(\varphi(\obs_i)+B(\con_i))-\varphi(\obs_i)$. Since $\varphi$ and $\psi_{\con_i}$ are unknown, we use lasso logit to estimate model (\ref{eq:FirstStage_General}) with polynomial approximations of $\varphi( \obs_{i})$ and $\psi_{\con_i}(\obs_i)$. Lasso regularization allows us to approximate these two functions well while avoiding overfitting, in a context where a fully non-parametric estimation is infeasible, see \cite{belloni2014inference}. Importantly, we do not impose any additional restriction on the estimation of $\psi_{\con_i}$ in this estimation. This will allow us to test implications of the framework. 

Overall, we see that our models involve a large number of regressors, leading to computationally demanding non-linear estimations. In this context, the assumption of a logistic anchor offers a number of advantages over alternatives, such as normal distributions (probit) or complementary log log. Theoretically, the logit link is the canonical link for the binomial family in the generalized linear model framework. This yields tractable likelihood equations and better numerical stability (see, for instance, \cite{mccullagh1989,greene2008}).

From the estimation of model (\ref{eq:FirstStage_General}), we then recover the bias from favors $B(\con_i)$ and the distribution of unobservables for connected candidates $\Lambda_{\con_i}(\cdot)$ as follows. Note that
\begin{equation}\label{eq:SecondStage_General}
\Logit_0 \left[\varphi( \obs_{i}) + \mathbbm{1}_{\con_i \neq \mathbf{0}} \psi_{\con_i}(\obs_i)\right] = \Logit_{\con_i}(\varphi( \obs_{i})+B(\con_i))=F_{\con_i}(\varphi( \obs_{i}))
\end{equation}
\noindent and denote by $\guconhat$ and $\gconhat$ the estimated values of $\gucon$ and $\gcon$. For every value of $\con_i \neq \mathbf{0}$, we then non-parametrically estimate $F_{\con_i}$, using a penalized smoothing spline regression (e.g. \cite{hastie1990generalized}) of $\Logit_0 \left[\guconhat + \mathbbm{1}_{\con_i \neq \mathbf{0}} \gconhat\right]$ on $\guconhat$.\footnote{The results are almost identical when we use instead the semiparametric estimator proposed in \cite{ichimura1993semiparametric}. Penalized spline regression is computationally more efficient, however, an important factor when standard errors are computed via bootstrap.} From this estimate of $F_{\con_i}$ and the zero median condition, $\Lambda_{\con_i}(0)=F_{\con_i}(-B(\con_i))=\frac{1}{2}$, we recover the estimate of the bias from favors,

\begin{equation}\label{eq:RecoveringBSemiparametric}
B(\vec{c}_i) = - F_{\mathbf{c}_i}^{-1}\left(\frac{1}{2}\right),
\end{equation}

\noindent and then $\Lambda_{\con_i}(w)=F_{\con_i}(w-B(\con_i))$. 

This semiparametric model is less restrictive than the parametric model, and hence more demanding in terms of data. Observations for which the conditional probability to be promoted is close to $\frac{1}{2}$ notably play an important role in its estimation. Lack of support in the data around this range would yield imprecise estimates of the bonus from favors and hence of information effects. In our empirical analysis below, this is an issue for the data on academic promotion in Spain. We thus present estimates of the semiparametric model on data on academic promotion in Italy and on political promotion in China, and discuss the Spanish data in the Online Appendix C.\footnote{We also focus on the binary variant of the model where the impact of connections only depends on the presence of some connection to the jury, $B(\con_i)=B$ and $\Lambda_{\con_i}=\Lambda$ if $\con_i \neq \mathbf{0}$. Our estimation methodology extends directly to a setup where favors $B(\con_i)$ and information effects $\Lambda_{\con_i}$ may vary with the number and types of connections. In practice, however, estimates are informative only when the number of candidates with connections exactly equal to $\con_i$ is large enough.}

From the estimation of model (\ref{eq:knownmodel}) or (\ref{eq:FirstStage_General}), we can then assess, quantitatively, the causal impact of connections to the jury and the relative contributions of the two mechanisms -- favors and information -- to this causal impact. For candidates with observables $\obs_i$, the impact of having connections $\con_i$ rather than no connection on the probability to be promoted is equal to $\Lambda_{\con_i}(\varphi(\obs_{i})+B(\con_i))-\Lambda_0(\varphi( \obs_{i}))$. This impact can be decomposed into the effects of information $\Lambda_{\con_i}(\varphi(\obs_{i}))-\Lambda_0(\varphi( \obs_{i}))$ and of favors $\Lambda_{\con_i}(\varphi(\obs_{i})+B(\con_i))-\Lambda_{\con_i}(\varphi( \obs_{i}))$. These effects vary non-linearly with $\obs_i$, and in a way which is fundamentally different for favors and information. Effects of favors are always positive when $B(\con_i)>0$. By contrast, effects of information are generally positive for candidates with low chances, when $\Lambda_0(\varphi( \obs_{i}))<\frac{1}{2}$, and negative for candidates with high chances, when $\Lambda_0(\varphi( \obs_{i}))>\frac{1}{2}$.

In our regressions below, we report estimates of the average effect of connections $TE(\con_i)$ and its two components $IE(\con_i)$ and $FE(\con_i)$ such that $TE(\con_i)=IE(\con_i)+FE(\con_i)$ and

\begin{align*}
\begin{split}
   & TE(\con_i)=\frac{1}{n} \sum_{i=1}^n \Lambda_{\con_i}(\varphi(\obs_{i})+B(\con_i))-\Lambda_0(\varphi( \obs_{i})), \\
   & IE(\con_i)=\frac{1}{n} \sum_{i=1}^n \Lambda_{\con_i}(\varphi(\obs_{i})-\Lambda_0(\varphi( \obs_{i})), \\
    & FE(\con_i)=\frac{1}{n} \sum_{i=1}^n \Lambda_{\con_i}(\varphi(\obs_{i})+B(\con_i))-\Lambda_{\con_i}(\varphi( \obs_{i})).
    \end{split}
\end{align*}

We report these three estimates on the overall sample, and estimates of information effects on subsamples of candidates with low and high chances. In the applications on academic promotion in Spain and Italy, we focus on candidates with some connections to eligible evaluators, since only these candidates had some chance to end up connected to the jury. 

An important observation is that both parametric model (\ref{eq:knownmodel}) and semiparametric model (\ref{eq:FirstStage_General}) have testable restrictions. And since the first model is nested in the second, we can test how well the parametric restrictions hold within the semiparametric framework. We now describe how we implement these tests below. Consider the binary variant, where favors and information effects only depend on the presence of some connection to the jury. We find it helpful to rewrite the model implications in terms of log odds ratios. Denote by $e_1(\obs_i)$ and $e_0(\obs_i)$ the conditional log odds ratio of connected and unconnected candidates, respectively, defined as follows 

\begin{align}\label{eq:LogOddsDefinition}
\begin{split}
    e_1(\obs_i) \equiv ln \left( \frac{\prob(y_i=1|c_i=1, \obs_i)}{1-\prob(y_i=1|c_i=1, \obs_i)}\right)\\
    e_0(\obs_i) \equiv ln \left( \frac{\prob(y_i=1|c_i=0, \obs_i)}{1-\prob(y_i=1|c_i=0, \obs_i)}\right)
    \end{split}
\end{align}
\noindent and note that under the assumption that $v_i$ follows a logistic distribution, $e_0$ is precisely equal to the grade function, $e_0(\obs_i)=\varphi(\obs_{i})$.

Within the semiparametric framework, a testable restriction of the assumption of independence on observables, $(u_i,\theta_i,v_i) \Perp \obs_i$, is that $e_1$ is an increasing function of $e_0$. The stronger, testable restriction of the parametric model (\ref{eq:knownmodel}) is that this increasing function is linear when $v_i$ and $w_i$ follow a logistic distribution. Namely, $e_1 =  \frac{1}{\sigma} e_0 + \frac{B}{\sigma}$. Both claims follow directly from \eqref{eq:P1(P0)}, the definitions of $e_1$ and $e_0$ in \eqref{eq:LogOddsDefinition}, and specifications of empirical models in \eqref{eq:knownmodel} and \eqref{eq:FirstStage_General}. We collect both implications in the following Corollary, and test them in the empirical analysis.

\begin{corollary}
In semiparametric model (\ref{eq:FirstStage_General}), when the extra information on connected candidates is independent of observables, $e_1(\obs_i)$ is an increasing function of $e_0(\obs_i)$. In parametric model (\ref{eq:knownmodel}) where $v_i$ and $w_i$ follow a logistic distribution, $e_1(\obs_i)$ is a linear function of $e_0(\obs_i)$.
\end{corollary}

\section{Empirical analysis}\label{sec:EmpiricalAnalysis}

We develop our empirical analysis in three stages. First, we estimate the binary variant of the parametric model on the three datasets. We recover estimates of bias from favors $B$ and information effects $\sigma$ and of the contributions of both effects to the impact of connections. We also compare the magnitude of the effects across contexts. Second, we maintain the assumption of a binary impact of connections and estimate the semiparametric model. We test whether restrictions of the semiparametric and parametric models are satisfied. Third, and as a form of robustness check, we estimate a variant of the parametric model where the types and numbers of connections may matter.

Overall, we detect significant information effects for candidates to Associate Professor positions in Spain and Italy, but not for candidates to Full Professor positions, nor for Chinese politicians vying for positions in the Politburo. By contrast, the estimated bias from favors is positive and significant in all cases. These results are consistent with the fact that AP candidates are junior academics, at an early career stage where a CV provides an imperfect signal of quality. In contrast, promotions to Full Professor and higher ranks within the Politburo typically occur later in a candidate’s career, reducing uncertainty about their personality and ability.

\subsection{Binary impact of connections, parametric model}

We now estimate model (\ref{eq:knownmodel}) on the three datasets, using a polynomial approximation of the grade function and assuming that $B(\con_i)=B$ and $\sigma(\con_i)=\sigma=e^{\delta}$ if $\con_i \neq \mathbf{0}$. The presence of information effects under independence implies that $\delta>0$. We choose the approximation of function $\varphi$  by comparing different specifications based on their out-of-sample performance. In the Online Appendix B, we provide a description of these estimations and the polynomial approximation that provides the best fit. Importantly, since we estimate models with the same structure and a common distributional anchor, we can quantitatively compare the estimates across datasets and subsamples.

\begin{footnotesize}
\begin{table}[!htb]
    \caption{Parametric model, binary connections}\label{tbl:HetLogitBinaryParametricJoint}
    \vspace{-10pt}
     \begin{center}
    \begin{subtable}{.40\linewidth}
      \centering
        \caption*{\hspace{150pt} Spain}
        \small
        \begin{tabular}{lccc} 
        \hline \hline
& (All) & (AP) & (FP)\\ 
\hline \\[-1.8ex] 

 Bias ($B$) & 0.532$^{***}$ & 0.497$^{***}$ & 0.684$^{***}$ \\ 
  & (0.100) & (0.131) & (0.171) \\ 
  Information ($\delta$)& 0.100$^{**}$ & 0.179$^{***}$ & $-$0.030 \\ 
  & (0.051) & (0.067) & (0.085) \\ 
\\
Observations & 31,000 & 17,784 & 13,216\\
 \hline 
\end{tabular} 
    \end{subtable}%
\quad \quad \quad \; \;
    \begin{subtable}{.40\linewidth}
      \centering
        \caption*{Italy}
        \small
        \begin{tabular}{ccc} 
    \hline \hline
 (All) & (AP) & (FP)\\ 
\hline \\[-1.8ex] 
 0.219$^{***}$ & 0.230$^{***}$ & 0.233$^{***}$ \\ 
   (0.038) & (0.044) & (0.057) \\ 
   0.109$^{***}$ & 0.137$^{***}$ & 0.027 \\ 
   (0.040) & (0.051) & (0.052) \\ 
  \\
  69,020 & 47,426 &  21,594 \\
 \hline 
\end{tabular} 
    \end{subtable}%
    \kern-0.6em 
\begin{subtable}{.10\linewidth}
      \centering
        \caption*{China}
        \small
        \begin{tabular}{c}
    \hline \hline
    (All)\\
    \hline \\[-1.8ex] 
 1.460$^{**}$ \\ 
   (0.696) \\ 
  $-$0.164 \\ 
   (0.355) \\ 
 \\
966\\
 \hline 
\end{tabular} 
    \end{subtable} 
 \end{center}   
\vspace*{-10pt}
\begin{spacing}{0.7}
{\footnotesize\textit{Notes}:  Heteroscedastic logit estimates. The column name indicates the sample. Standard errors, in parentheses, are clustered at the committee level for Spain and Italy, and the province level for China. We include exam-specific controls for Spain, exam fixed effects for Italy, and year, term, and office$\times$province fixed effects for China. $^{*}$p$<$0.1;$^{**}$p$<$0.05; $^{***}$p$<$0.01. }
\end{spacing} 
\end{table}
\end{footnotesize}

Results are reported in Table \ref{tbl:HetLogitBinaryParametricJoint}. We detect evidence of favors and information effects from connections in academic promotions in Spain and Italy, and of favors in political promotions in China. In both Spain and Italy, information effects appear to be present for AP candidates, but not for FP candidates. The precision of the estimates, of course, varies across the datasets, consistently with the different sample sizes. It is high on the Italian data, fairly high on the Spanish data, and fairly low on the Chinese data. Estimated favors are largest for political promotion in China, and larger for academic promotions in Spain than in Italy. Information effects for AP candidates also appear to be larger in Spain than in Italy. The larger estimated impacts of connections in Spain may reflect the fact that we have information on more types of connections for Spain.

Are these effects quantitatively significant? How much do connections help? And how much does each mechanism contribute to the overall impact? We next provide estimates of the average effect of connections and of its two components for the three datasets. We also depict how these effects vary with candidates' observable chances of promotion, as measured by the estimated promotion probability if unconnected $\Lambda_0(\guconhat)$.

\begin{figure}[!ht]
\captionsetup{position=above}
    \caption{Marginal effects, parametric model, binary connections, Spain}\label{fig:MarginalEffects_Binary_Sieve_Spain}
     \begin{center}
    \includegraphics[scale=0.36]{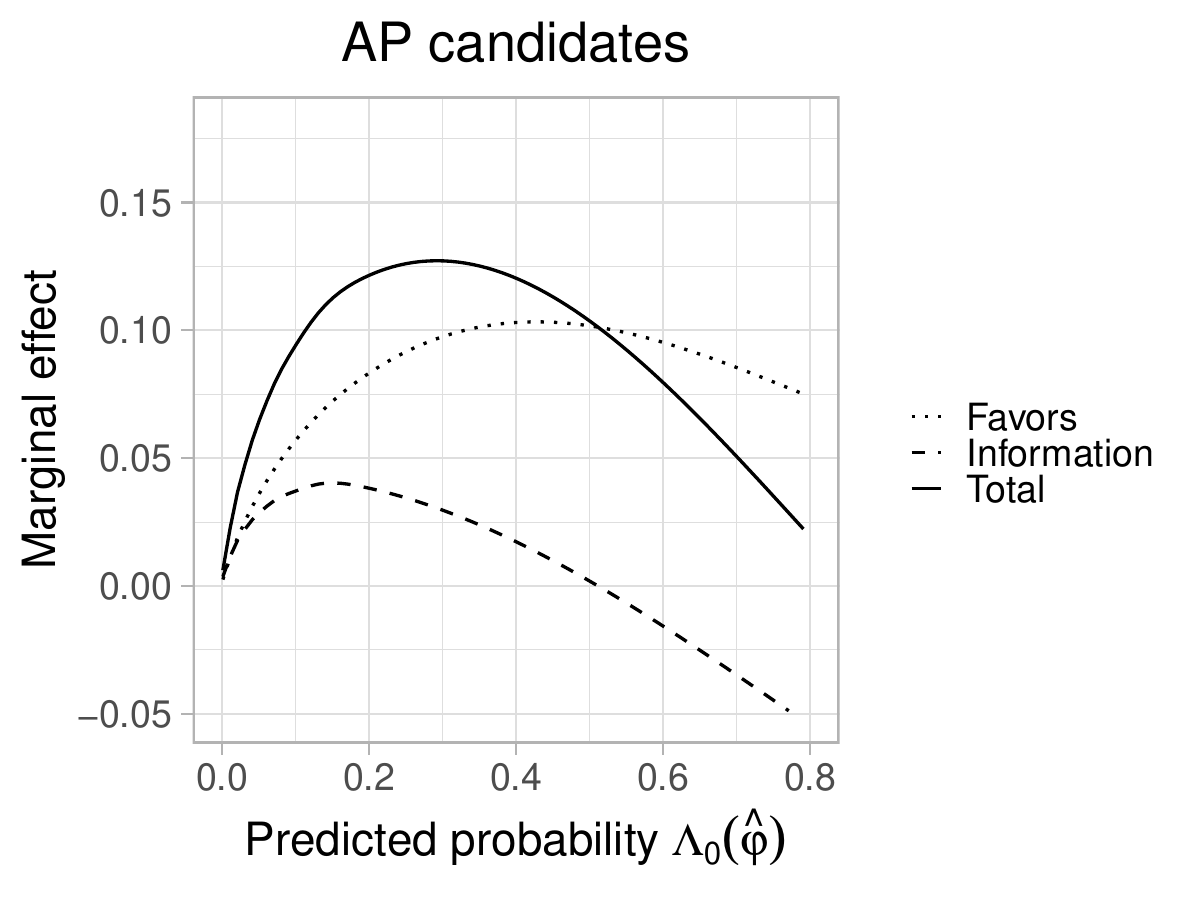}
    \includegraphics[scale=0.36]{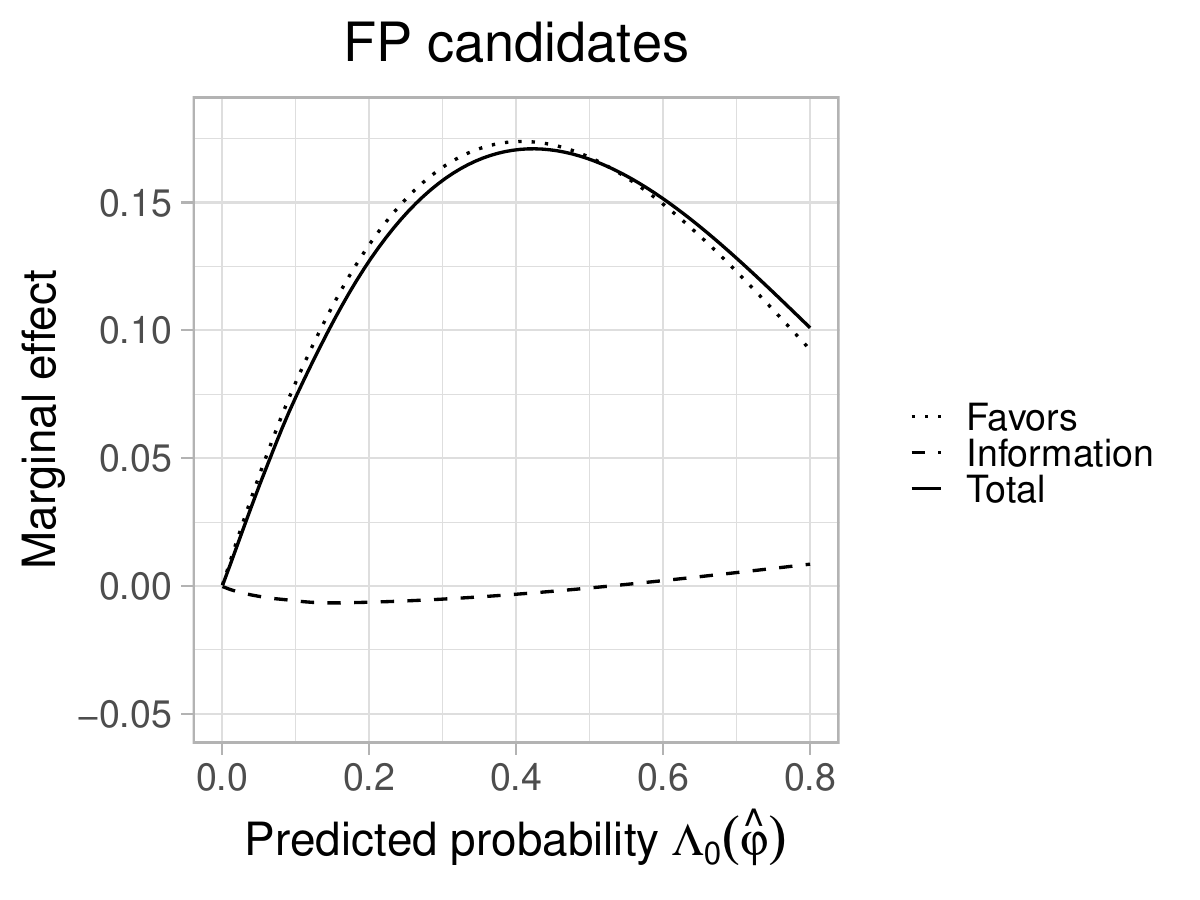}
\end{center}
    \par
    \begin{spacing}{0.7}
        {\footnotesize \textit{Notes}: The plots are constructed using the estimates from Model \eqref{eq:knownmodel} reported in Table \ref{tbl:HetLogitBinaryParametricJoint}. Subsamples are indicated above each plot.}
    \end{spacing}
\end{figure}

Table \ref{tbl:AvgEffects_Binary_Parametric_Spain} and Figure \ref{fig:MarginalEffects_Binary_Sieve_Spain} report results for the data on Spain. In Table \ref{tbl:AvgEffects_Binary_Parametric_Spain}, the first column provides the average estimated promotion probability if unconnected. We see that candidates with no connection have less than 1 in 10 chance to be promoted. The second column provides the average impact of connections. Being connected, by chance, to the jury has a large impact. It leads to an $82\%$ increase in promotion probability for AP candidates and to a $69 \%$ increase for FP candidates. For AP candidates, slightly more than $3/5$  of the effect of connections is due to favors and  about $2/5$ (39\%) due to information effects. Almost all candidates have a low chance of being promoted.

\begin{table}[!ht]
	\begin{center}
		\caption{Marginal effects, parametric model, binary connections, Spain} 
		\label{tbl:AvgEffects_Binary_Parametric_Spain}
		\small
  	\vspace{-10pt}
		\begin{tabular}{rccccccc}
			\hline
			\hline\\
			[-1.8ex]
			&\multicolumn{1}{c}{Baseline}& &\multicolumn{5}{c}{Marginal effects}\\
			\cline{2-2} \cline{4-8}\\
			[-1.8ex]
			&Predicted& &Total & Favors & Information &Information &Information \\
            && &&  & &Low chances& High chances\\
			\hline\\
			[-1.5ex]

AP & 0.092$^{***}$ & & 0.075$^{***}$ & 0.047$^{***}$ & 0.029$^{**}$ & 0.029$^{**}$ & -0.013$^{*}$ \\ 
   & (0.004) && (0.007) & (0.016) & (0.014) & (0.014) & (0.007) \\ 
  Observations & 15,492 && 15,492 & 15,492 & 15,492 & 15,440 & 52 \\ \\
  FP & 0.074$^{***}$ & &0.051$^{***}$ & 0.056$^{**}$ & -0.004 & -0.004 & 0.002 \\ 
   & (0.005) & &(0.007) & (0.021) & (0.017) & (0.017) & (0.012) \\ 
  Observations & 12717 && 12717 & 12717 & 12717 & 12702 & 15 \\ 



			\hline
		\end{tabular}
	\end{center}
	\vspace*{-10pt}
	\begin{spacing}{0.7}
	{\footnotesize 	\textit{Notes:} Average marginal effect of being connected calculated for candidates with at least one connection to eligible evaluators based on the estimates reported in Table \ref{tbl:HetLogitBinaryParametricJoint}. Bootstrapped standard errors clustered at the committee level are in parentheses. $^{*}$p$<$0.1;$^{**}$p$<$0.05; $^{***}$p$<$0.01.} 
\end{spacing}
\end{table}

Figure \ref{fig:MarginalEffects_Binary_Sieve_Spain} describes how these effects vary with observable characteristics. We see that the overall and favoritism effects on both samples and the information effect for AP candidates all display an increasing, then decreasing pattern. This inverted U shape is, of course, related to the binary, non-linear nature of promotion. Favoritism effects are always positive and tend to peak at probability $0.46$. Information effects for AP candidates become negative for candidates with high chances and tend to peak at probability $0.15$. Combining the two, the overall impact of connections is highest for AP candidates at a probability of $0.28$ and for FP candidates at a probability of $0.43$. 

\begin{figure}[!ht]
 	\caption{Marginal effects, parametric model, binary connections, Italy}
 	\label{fig:MarginalEffects_Binary_Sieve_Italy}
 	\begin{center}
 		\includegraphics[scale=0.36]{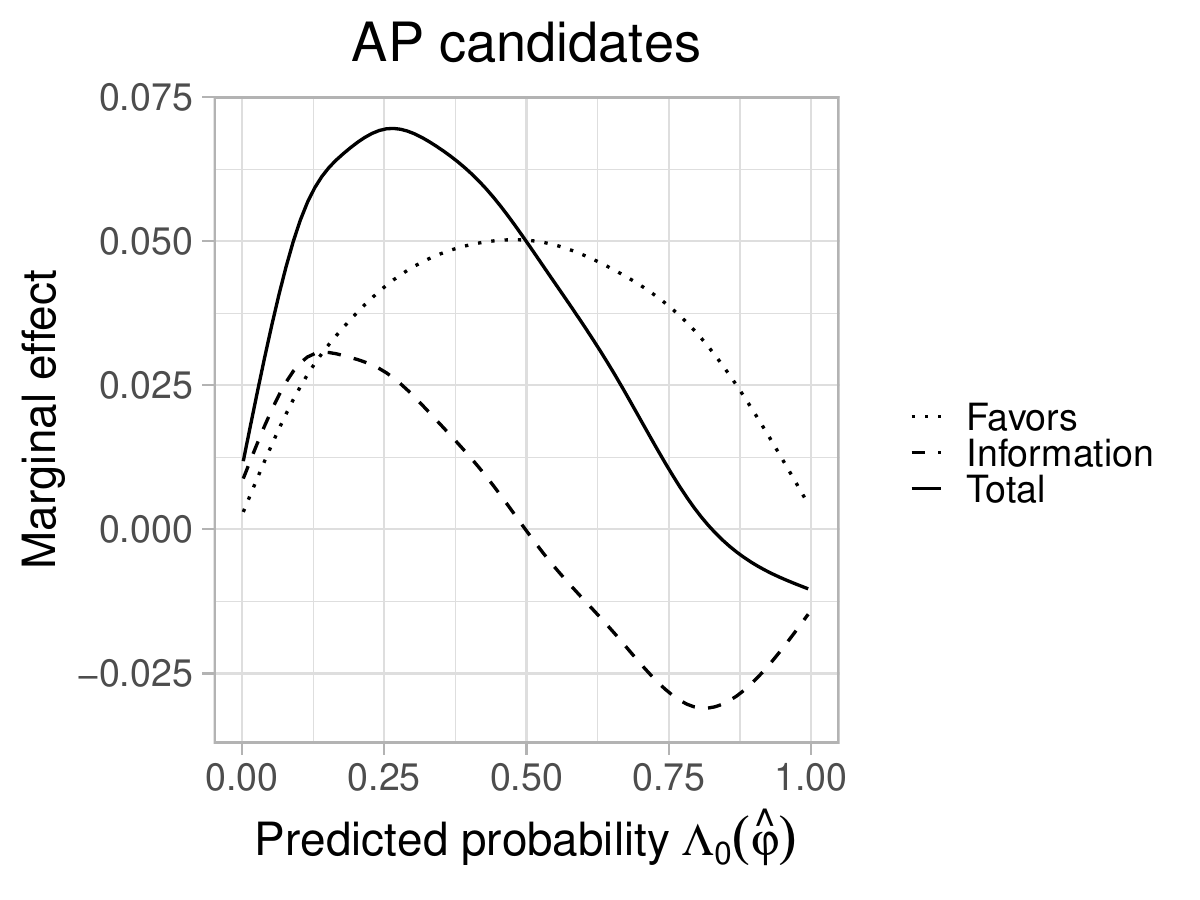} %
 		\includegraphics[scale =0.36]{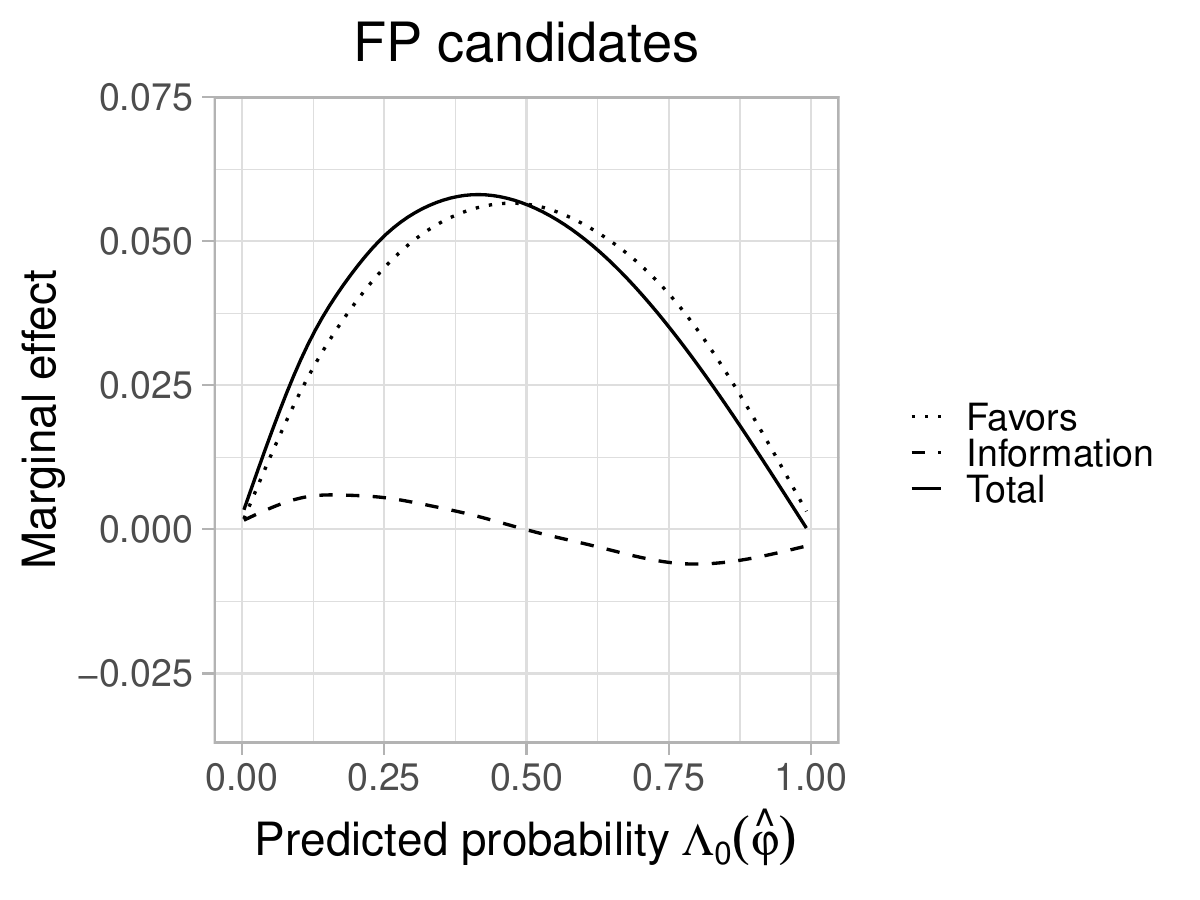}
 	\end{center}
 	\par
 	\begin{spacing}{0.7}
 	{\footnotesize \textit{Notes}: The plots are constructed using the estimates from Model \eqref{eq:knownmodel} reported in Table \ref{tbl:HetLogitBinaryParametricJoint}. Subsamples are indicated above each plot.}
 \end{spacing}
 \end{figure}

Table \ref{tbl:AvgEffects_Binary_Parametric_Italy_1} and Figure \ref{fig:MarginalEffects_Binary_Sieve_Italy} report results for the data on Italy. Candidates with no connection now have about 4 chances out of 10 to be promoted, a much higher baseline probability than in Spain. Being connected, by chance, to the jury leads to a $10 \%$ increase in promotion probability, on average, for both AP and FP candidates.  The information effect is present only for AP candidates, and, when considering the entire sample, it is relatively small compared to the favoritism effect. This masks an important heterogeneity, however. For candidates with low chances, the information effect is positive and represents about half of the overall average impact of connections. For candidates with high chances, the information effect is negative and of a similar magnitude as for candidates with low chances.

\begin{table}[!ht]
	\begin{center}
		\caption{Marginal effects, parametric model, binary connections, Italy} 
		\label{tbl:AvgEffects_Binary_Parametric_Italy_1}
		\small
  	\vspace{-10pt}
		\begin{tabular}{rccccccc}
			\hline
			\hline\\
			[-1.8ex]
			&\multicolumn{1}{c}{Baseline}& &\multicolumn{5}{c}{Marginal effects}\\
			\cline{2-2} \cline{4-8}\\
			[-1.8ex]
			&Predicted& &Total & Favors & Information &Information &Information \\
            && &&  & &Low chances& High chances\\
			\hline\\
			[-1.5ex] 

 AP & 0.42$^{***}$ & & 0.044$^{***}$ & 0.038$^{***}$ & 0.007$^{**}$ & 0.022$^{***}$ & -0.019$^{***}$ \\ 
   & (0.009) && (0.008) & (0.008) & (0.002) & (0.006) & (0.005) \\
 Observations&  25,291 && 25,291 & 25,291 & 25,291 & 16,000 & 9,291 \\
 \\
  FP & 0.408$^{***}$ & &0.043$^{***}$ & 0.041$^{***}$ & 0.002 & 0.004$^{*}$ & -0.004 \\ 
   & (0.007) & & (0.01) & (0.01) & (0.001) & (0.001) & (0.001) \\ 
  Observations&  13,264 && 13,264 & 13,264 & 13,264 & 8,717 & 4,547\\

			\hline
		\end{tabular}
	\end{center}
	\vspace*{-10pt}
	\begin{spacing}{0.7} 
	{\footnotesize \textit{Notes:}
		The average marginal effect of being connected calculated for candidates with at least one connection to eligible evaluators based on the estimates reported in Table \ref{tbl:HetLogitBinaryParametricJoint}. Bootstrapped standard errors clustered at the committee level are in parentheses. $^{*}$p$<$0.1;$^{**}$p$<$0.05; $^{***}$p$<$0.01.} 
	\end{spacing}
\end{table}





At the individual level, the overall impact of connections peaks at probability $0.26$ for AP candidates and at probability $0.41$ for FP candidates. For AP candidates with very high chances of being promoted, the overall impact of connections is negative, as the negative effect of information dominates the positive effect of favors.

\begin{figure}[H]
 	\caption{Marginal effects, parametric model, binary connections, China}
 	\label{fig:MarginalEffects_Binary_Parametric_China}
 	\begin{center}
 		\includegraphics[scale=0.36]{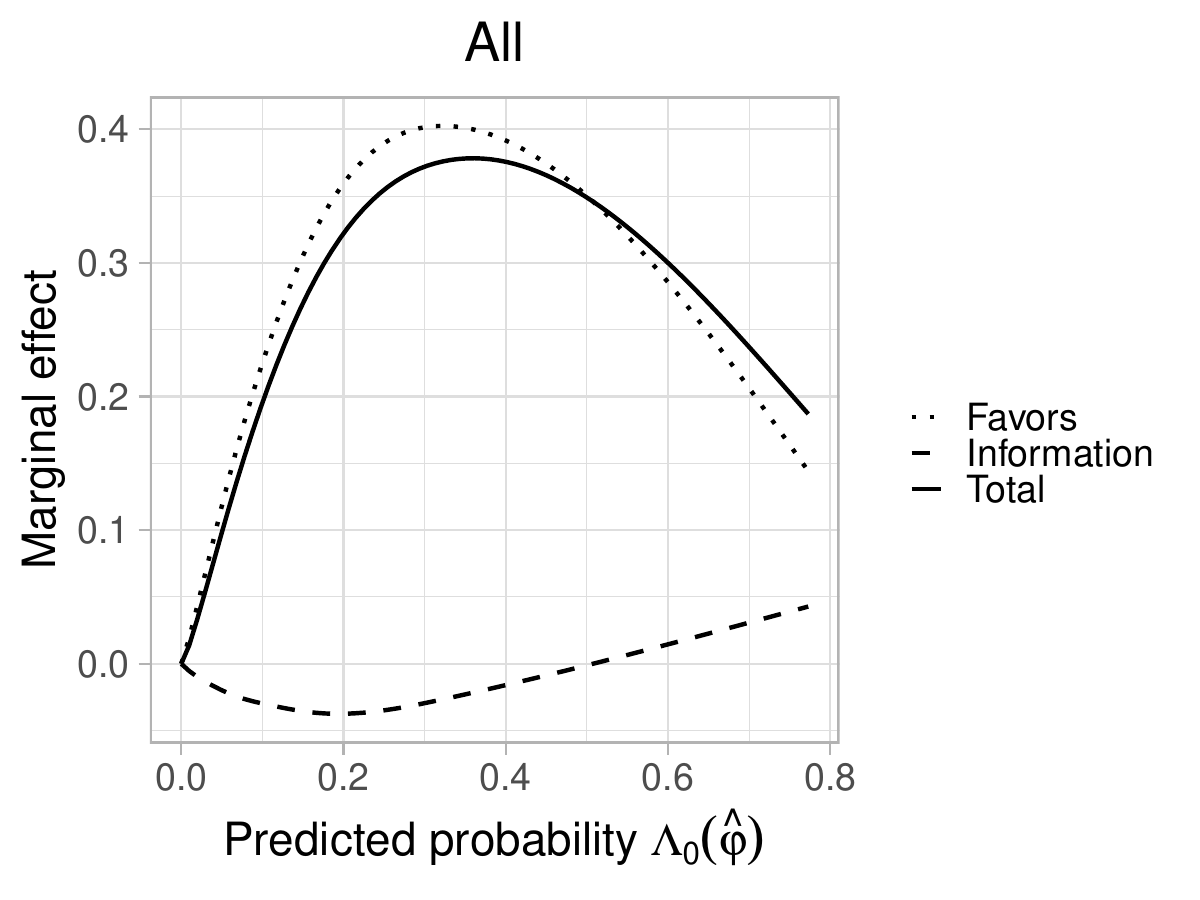}
 	\end{center}
 	\par
 	\begin{spacing}{0.7}
 	{\footnotesize \textit{Notes}: The plot is constructed using the estimates from Model \eqref{eq:knownmodel} reported in Table \ref{tbl:HetLogitBinaryParametricJoint}. The sample is indicated above the plot.}
 \end{spacing}
 \end{figure}

Table \ref{tbl:AvgEffects_Binary_Parametric_China} and Figure \ref{fig:MarginalEffects_Binary_Parametric_China} report results for the data on China. Unconnected provincial secretaries and governors only have, on average, a $5 \%$ chance to be promoted in any given year. Being connected to a member of the Politburo more than doubles these chances, leading to a $152 \%$ increase in promotion probability. The overall impact of connections peaks at probability $0.36$.

\begin{table}[!ht]
	\begin{center}
		\caption{Marginal effects, parametric model, binary connections, China} 
		\label{tbl:AvgEffects_Binary_Parametric_China}
		\small
  	\vspace{-10pt}
		\begin{tabular}{lccccccc}
			\hline
			\hline\\
			[-1.8ex]
			&\multicolumn{1}{c}{Baseline}& &\multicolumn{5}{c}{Marginal effects}\\
			\cline{2-2} \cline{4-8}\\
			[-1.8ex]
			&Predicted& &Total & Favors & Information &Information &Information \\
            && &&  & &Low chances& High chances\\
			\hline\\
			[-1.5ex] 

 All & 0.046$^{***}$ && 0.070$^{*}$ & 0.077$^{*}$ & -0.006 & -0.007 & 0.033 \\ 
   & (0.006) && (0.036) & (0.04) & (0.012) & (0.014) & (0.066) \\
   \\
  Observations & 966 && 966 & 966 & 966 & 944 & 22 \\


			\hline
		\end{tabular}
	\end{center}
	\vspace*{-10pt}
	\begin{spacing}{0.7}
	{\footnotesize 	\textit{Notes:} Average marginal effect of being connected based on the estimates reported in Table \ref{tbl:HetLogitBinaryParametricJoint}.  Standard errors are calculated at the province level using the Delta method. $^{*}$p$<$0.1;$^{**}$p$<$0.05; $^{***}$p$<$0.01.} 
\end{spacing}
\end{table}




Overall, these results indicate that, indeed, researchers may be able to identify favors from information effects in the impact of connections, with standard data collected at the time of promotion. Model (\ref{eq:knownmodel}) imposes strong assumptions, however. In what follows, we assess the validity of the previous results in two ways: by relaxing some of these assumptions and by testing restrictions implied by the model.

\subsection{Binary impact of connections, semiparametric model}\label{ss:binary_semiparametric}

We now relax the assumption that the unobservables of connected candidates have a logistic distribution. We estimate model (\ref{eq:FirstStage_General}) with polynomial approximations of functions $\varphi$ and $\psi$ and via lasso logit. In this Section, we reestimate the impacts of favors and information effects and compare the results with those obtained from the parametric model. In Section \ref{subsec:testing_restrictions}, we test restrictions implied by both models.

As discussed in Section \ref{subsec:econometric_framework}, the estimation of the bonus from favors requires inverting an estimated semiparametric function around value $1/2$. And this estimate is then used, in a second step, to estimate information effects. To recover informative estimates of favors and information effects, the data must then have enough observations for which the probability of being promoted if connected is around $0.5$. While in Italy this support condition turns out to be satisfied, in China and Spain we observe relatively few observations with predicted promotion probabilities near $0.5$. This appears to have little impact on the estimated bonus from favors ($B$) in China, but results in noisy and unstable estimates of $B$ in Spain.
 We next, therefore, focus on data on Italy and China, and provide a detailed discussion of this issue and of the results of the semiparametric regression on the Spanish data in Appendix A.

\begin{figure}[!ht]
 	\caption{Marginal effects, semiparametric model, binary connections, Italy}
 	\label{fig:MarginalEffects_Binary_Lasso_Italy}
 	\begin{center}
 		\includegraphics[scale=0.36]{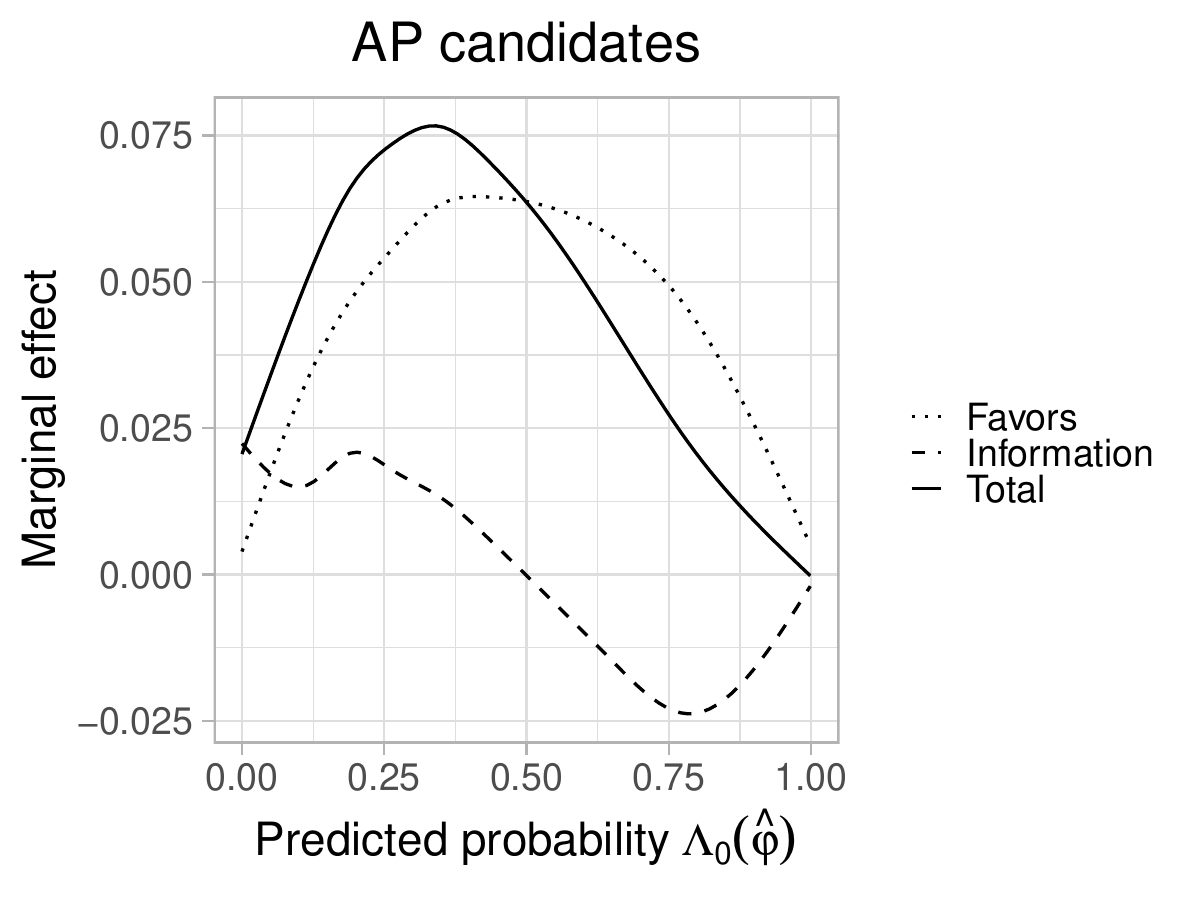} %
 		\includegraphics[scale =0.36]{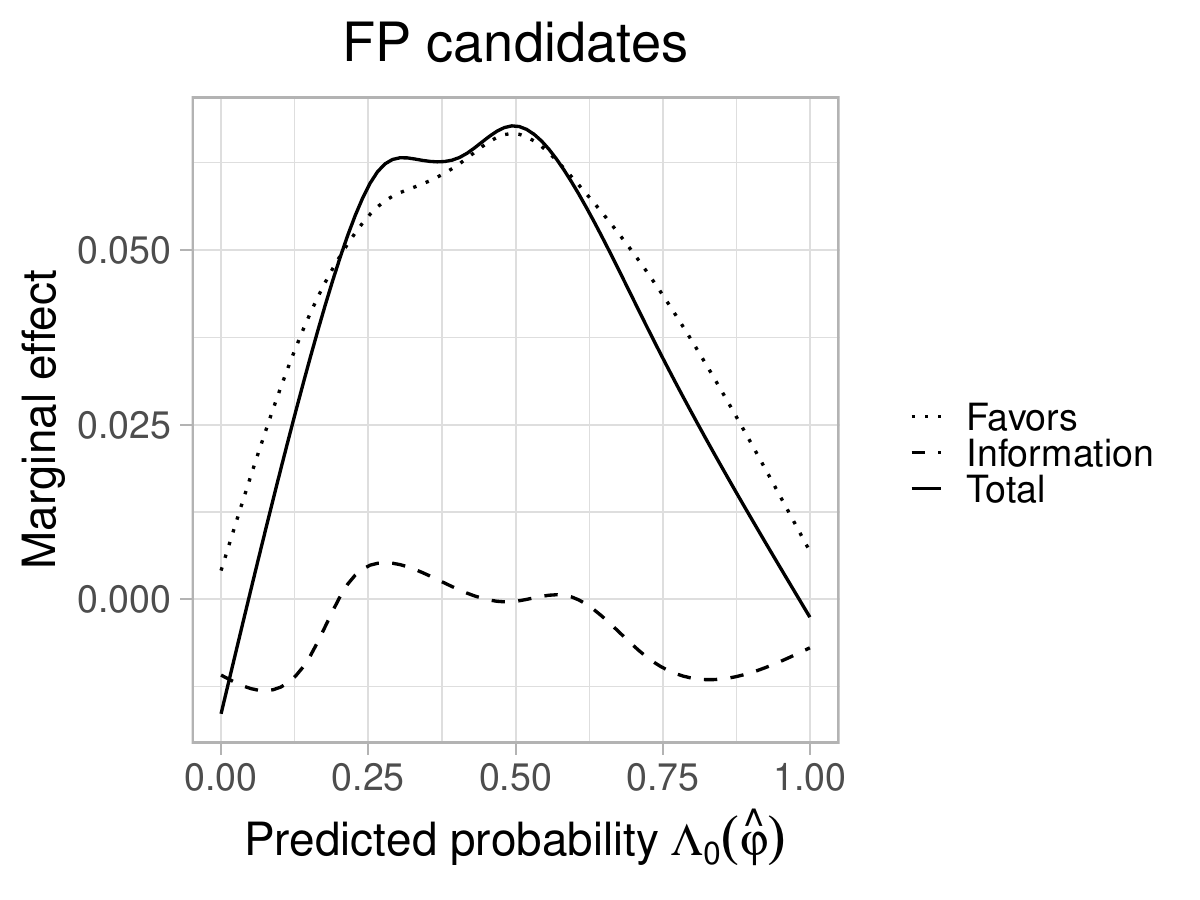}
 	\end{center}
 	\par
 	\begin{spacing}{0.7}
 	{\footnotesize \textit{Notes}: The plots are constructed using the estimates from Model \eqref{eq:FirstStage_General}. Subsamples are indicated above each plot.}
 \end{spacing}
 \end{figure}

Table \ref{tbl:AvgEffects_Binary_Semiparametric_Italy} and Figure \ref{fig:MarginalEffects_Binary_Lasso_Italy} present the results on academic promotions in Italy.\footnote{As an illustration, we note that the parametric model we use for Italy includes 458 regressors, while in the semiparametric model this number is 1352.} We see that the qualitative results are similar to those obtained in the parametric framework. We detect information effects at work for AP candidates in Italy and favors at work in all cases. Comparing the results reported in Table \ref{tbl:AvgEffects_Binary_Semiparametric_Italy}  and Table \ref{tbl:AvgEffects_Binary_Parametric_Italy_1}, we see that estimated average effect from favors is a bit larger in Table \ref{tbl:AvgEffects_Binary_Semiparametric_Italy}, equal to $0.050$ rather than $0.038$ for AP candidates ($B=0.287$ vs. $B=0.230$) and equal to $0.051$ rather than $0.041$ for FP candidates ($B=0.267$ vs $B=0.233$). Conversely, the estimated information effects are a bit lower. In particular, the average effect of information for AP candidates with low chances is equal to $0.014$ rather than $0.022$. This indicates that the parametric restrictions lead to overestimating information effects and underestimating favors.


\begin{table}[!ht]
 	\begin{center}
 		\caption{Marginal effects, semiparametric model, binary connections, Italy} 
 		\label{tbl:AvgEffects_Binary_Semiparametric_Italy}
 		\small
 		\begin{tabular}{lcccccccc}
			\hline
			\hline\\
			[-1.8ex]
			& & \multicolumn{1}{c}{Baseline} & &\multicolumn{5}{c}{Marginal effects}\\
			\cline{5-9}\\
			[-1.8ex]
			&$B$&Predicted& & Total & Favors & Information &Information &Information \\
            && &  & &&&Low chances& High chances\\
			\hline\\
			[-1.5ex] 

AP & 0.287$^{***}$ & 0.423$^{***}$ && 0.054$^{***}$ & 0.05$^{***}$ & 0.004$^{**}$ & 0.014$^{***}$ & -0.014$^{***}$ \\ 
   & (0.03) & (0.004) && (0.005) & (0.005) & (0.002) & (0.004) & (0.004) \\ 
  Observations & 25,291 & 25,291 && 25,291 & 25,291 & 25,291 & 15,999 & 9,292 \\ 
  \\
   FP & 0.266$^{***}$ & 0.410$^{***}$ && 0.050$^{***}$ & 0.051$^{***}$ & -0.001 & 0.0004 & -0.005 \\ 
   & (0.037) & (0.005) && (0.007) & (0.007) & (0.003) & (0.005) & (0.006) \\ 
  Observations & 13282 & 13282 && 13282 & 13282 & 13282 & 8943 & 4339 \\



			\hline
		\end{tabular}
 	\end{center}
 	\vspace*{-10pt}
  \par
 	\begin{spacing}{0.7}
 	{\footnotesize 	\textit{Notes:} Average marginal effect of being connected to the jury calculated for candidates with at least one connection to eligible evaluators. Bootstrapped standard errors clustered at the committee level are in parentheses.  $^{*}$p$<$0.1;$^{**}$p$<$0.05; $^{***}$p$<$0.01.} 
  \end{spacing}
 \end{table}

At the individual level, the overall impact of connections peaks at probability $0.33$ for AP candidates and at probability $0.51$ for FP candidates.  For AP candidates, Figure \ref{fig:MarginalEffects_Binary_Lasso_Italy} looks similar to Figure \ref{fig:MarginalEffects_Binary_Sieve_Italy}, with information effect negative for candidates with high chances of promotion. For FP candidates, while Figure \ref{fig:MarginalEffects_Binary_Lasso_Italy} looks somewhat different compared to Figure \ref{fig:MarginalEffects_Binary_Sieve_Italy}, suggesting that unobservables of connected FP candidates may not follow a logistic distribution.

\begin{figure}[H]
 	\caption{Marginal effects, semiparametric model, binary connections, China}
 	\label{fig:MarginalEffects_Binary_Lasso_China}
 	\begin{center}
 		\includegraphics[scale=0.36]{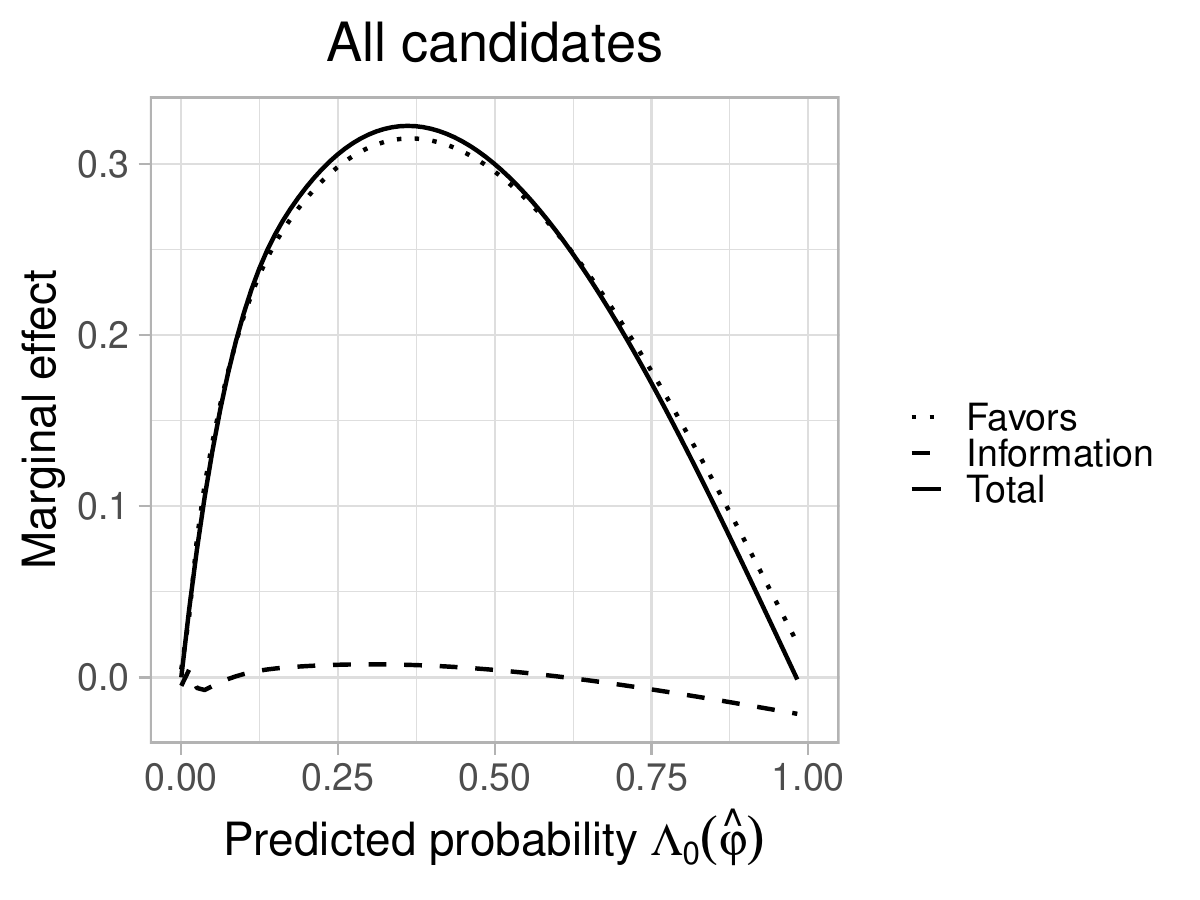} %
 	
 	\end{center}
 	\par
 	\begin{spacing}{0.7}
 	{\footnotesize \textit{Notes}:
    The plot is constructed using the estimates from Model \eqref{eq:FirstStage_General}. The sample is indicated above the plot.}
 \end{spacing}
 \end{figure}

Table \ref{tbl:AvgEffects_Binary_Semiparametric_China} and Figure \ref{fig:MarginalEffects_Binary_Lasso_China} present the results on political promotion in China. Results are, again, similar to those obtained from the parametric model. The estimated bias $B$ from favors is equal to $1.39$ rather than $1.46$ and is marginally statistically significant, while we do not detect evidence of information effects. The total effect peaks at a probability  of promotion equal to 
$0.42$, while the information effect peaks at probability $0.43$. 

\begin{table}[H]
 	\begin{center}
 		\caption{Marginal effects, semiparametric model, binary connections, China} 
 		\label{tbl:AvgEffects_Binary_Semiparametric_China}
 		\small
 		\begin{tabular}{lcccccccc}
			\hline
			\hline\\
			[-1.8ex]
			& & \multicolumn{1}{c}{Baseline} & &\multicolumn{5}{c}{Marginal effects}\\
			\cline{5-9}\\
			[-1.8ex]
			&$B$&Predicted& & Total & Favors & Information &Information &Information \\
            && &  & &&&Low chances& High chances\\
			\hline\\
			[-1.5ex] 

All & 1.394$^{*}$ & 0.063$^{***}$ && 0.062$^{**}$ & 0.062$^{**}$ & 0 & 0 & -0.009 \\ 
  & (0.791) & (0.003) && (0.029) & (0.03) & (0.006) & (0.006) & (0.016) \\ 
Observations & 966 & 966 && 966 & 966 & 966 & 938 & 28\\ 

			\hline
		\end{tabular}
\end{center}
\vspace*{-10pt}
  \par
 	\begin{spacing}{0.7}
 	{\footnotesize 	\textit{Notes:} Average marginal effect of being connected to the jury.  Year, term, and office $\times$ province fixed effects are
included. Bootstrapped standard errors clustered at the province level in parentheses. $^{*}$p$<$0.1;$^{**}$p$<$0.05; $^{***}$p$<$0.01.} 
  \end{spacing}
\end{table}





Overall, we obtain similar results when relaxing the assumption that the unobservables of connected candidates have a logistic distribution. In the next Section, we explore the testable implications of the two modeling approaches, parametric and semiparametric. 

\subsection{Testing the parametric and semiparametric models}\label{subsec:testing_restrictions}

In this Section, we formally test restrictions of the semiparametric and parametric frameworks, based on Corollary 1. Recall, within semiparametric model (\ref{eq:FirstStage_General}), the assumption of independence from observables implies that the conditional log-odds ratio for connected candidates, $e_1(\obs_i)$, is an increasing function of the conditional log-odds ratio for unconnected candidates, $e_0(\obs_i)$. And  this function is linear under the additional parametric restrictions of model (\ref{eq:knownmodel}), when unobservables follow logistic distributions. These are strong, nested testable restrictions, which we assess as follows.

\begin{figure}[H]
 	\caption{$e_1(e_0)$}
 	\label{fig:e1e0_joint}
 	\begin{center}
 		\includegraphics[scale=0.36]{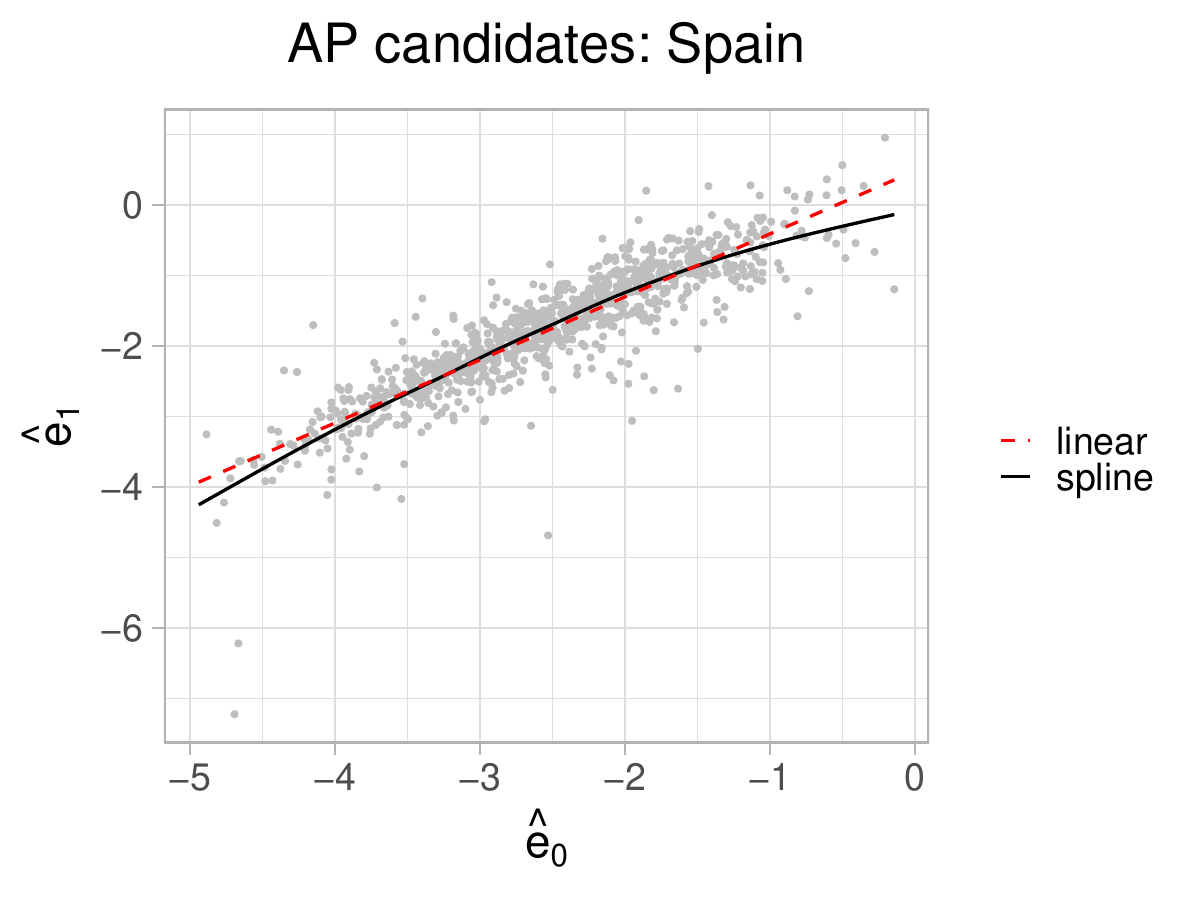} %
 	 		\includegraphics[scale=0.36]{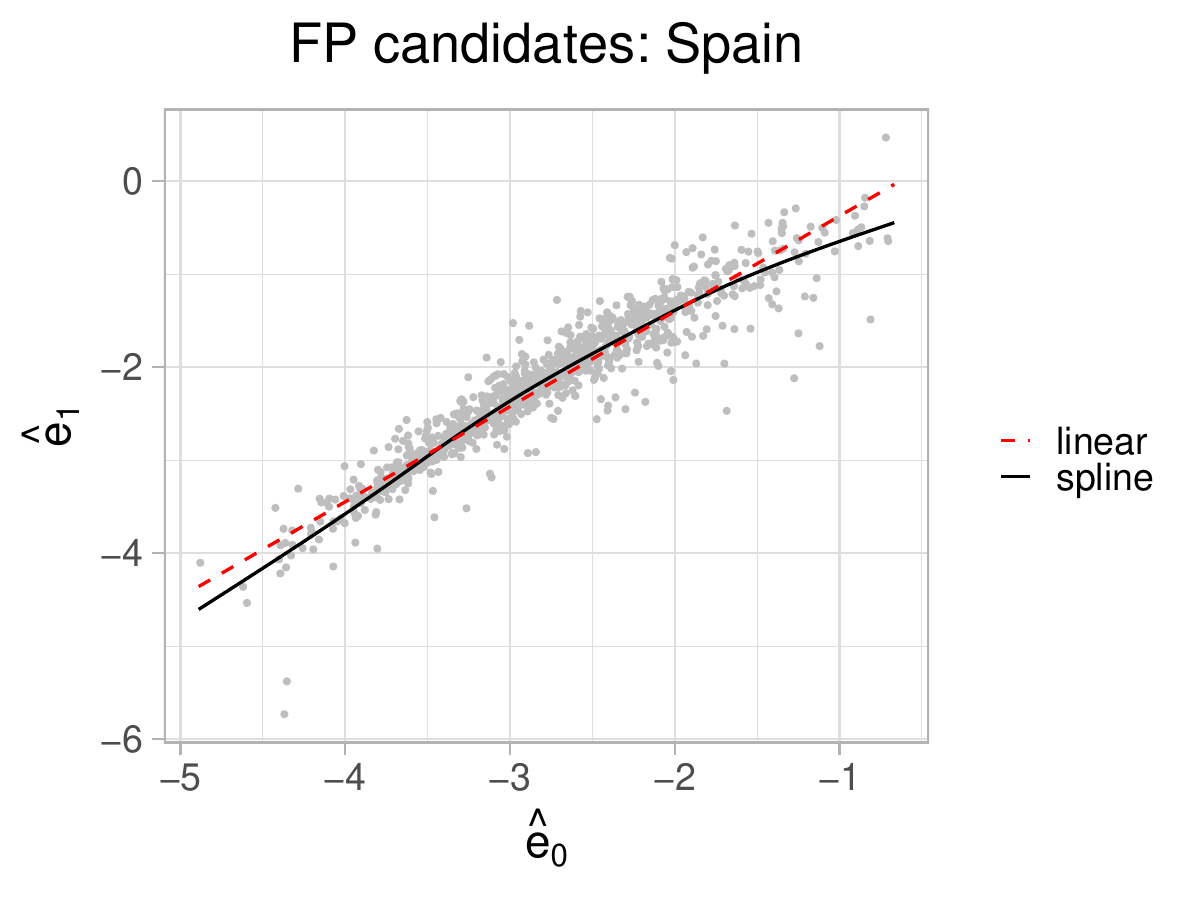} %
            	\includegraphics[scale=0.36]{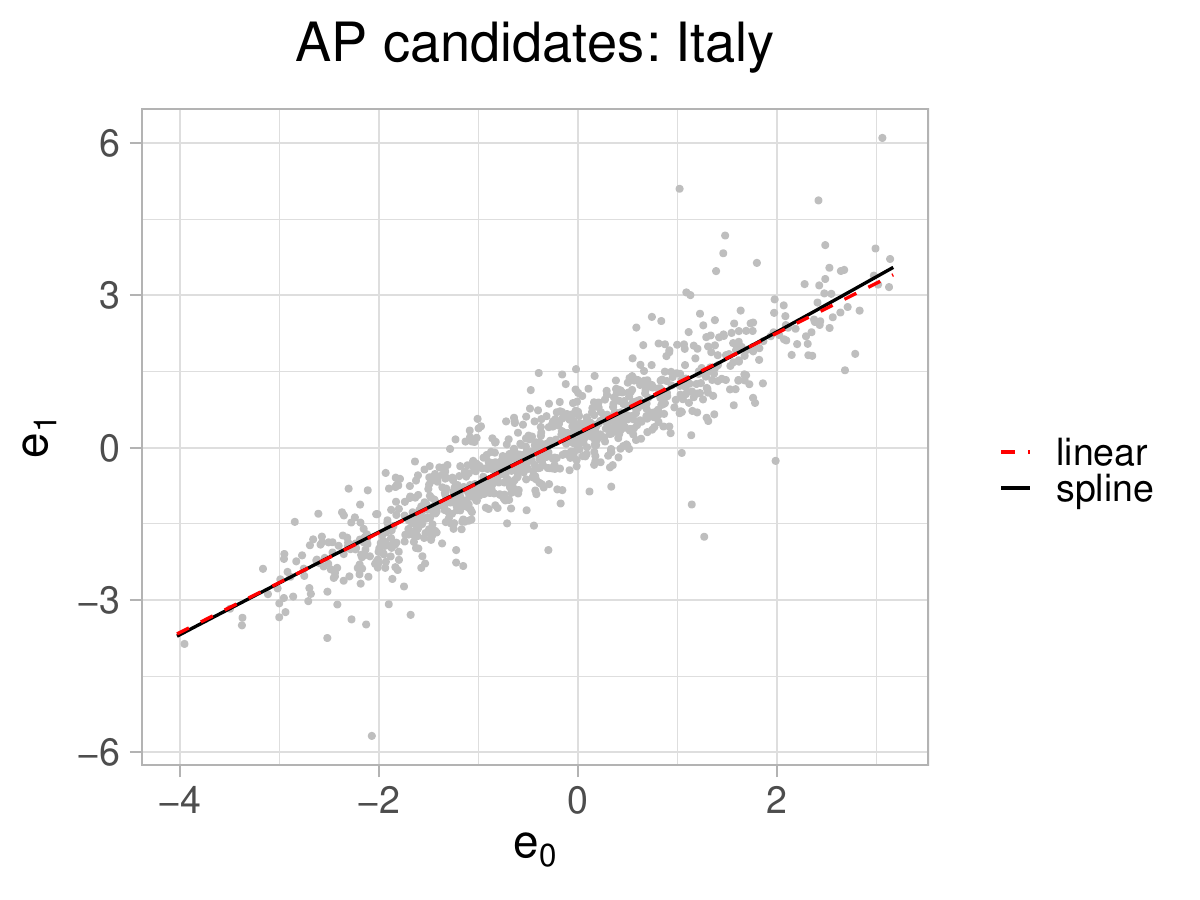} %
 	 		\includegraphics[scale=0.36]{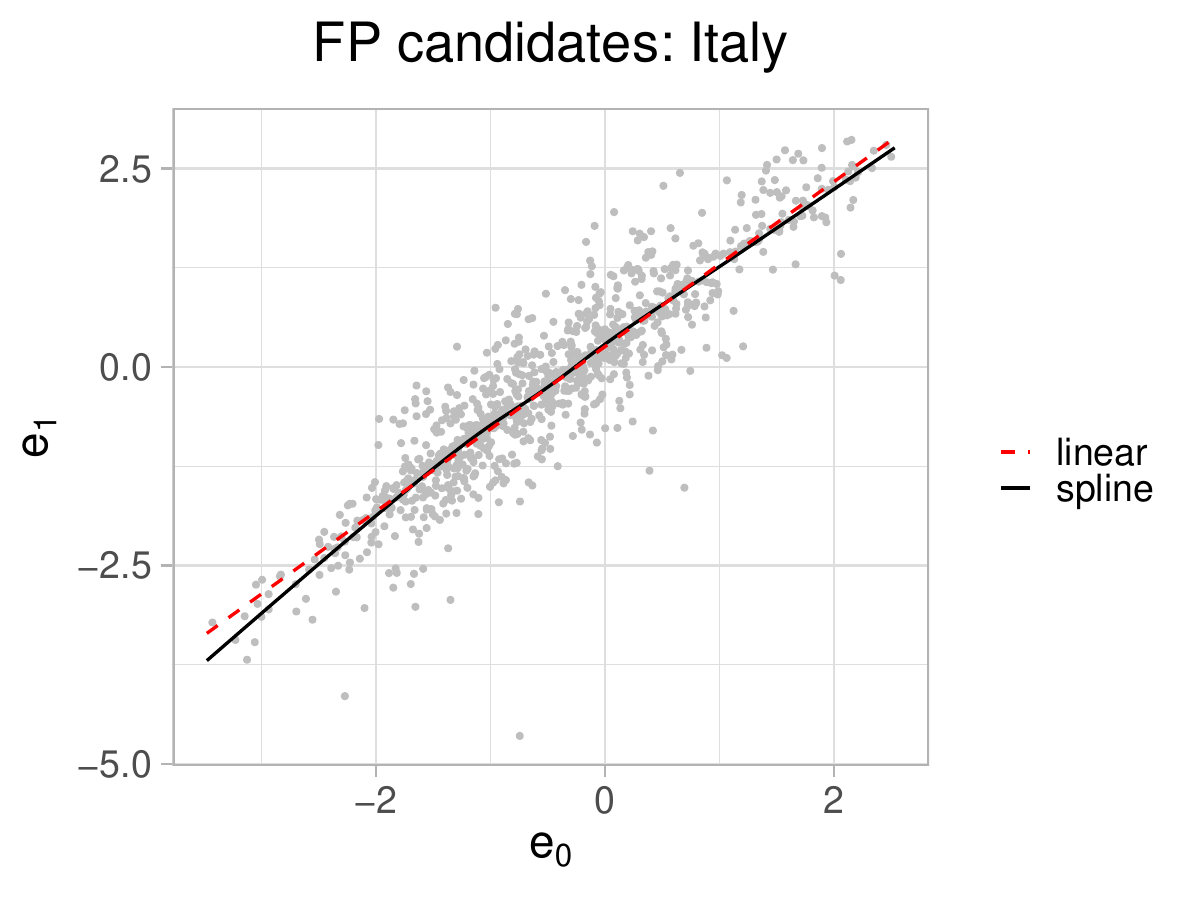} %
            	\includegraphics[scale=0.36]{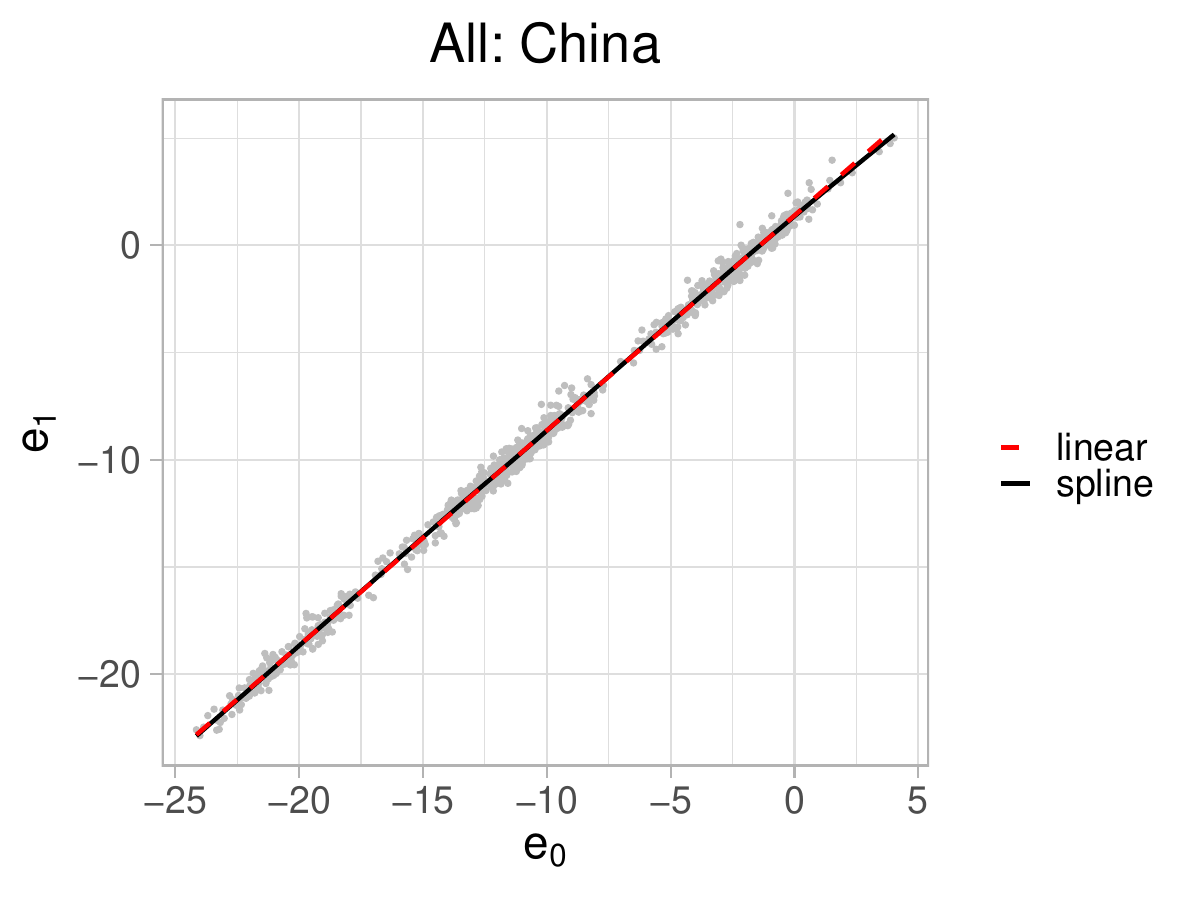} %
 	\end{center}
 	\par
 	\begin{spacing}{0.7}
 	{\footnotesize \textit{Notes}: Plots of the estimated function $e_1 (e_0)$.  Predicted $e_1$ and $e_0$ are obtained from  model (\ref{eq:FirstStage_General}). We plot linear regression (solid black line) vs. penalized spline regression (red dashed line). Each dot represents a par ($e_0$, $e_1$). For visibility we display only 1000  randomly chosen ($e_0$, $e_1$) pairs.}
 \end{spacing}
 \end{figure}

From the estimated model (\ref{eq:FirstStage_General}), we obtain estimated values of the two log-odds ratio for each candidate $i$. Importantly, our estimation does not impose independence from observables, and hence imposes no specific relationship between these two log-odds ratios.\footnote{We have: $e_0(\obs_i)=\varphi(\obs_i)$ and $e_1(\obs_i)=\varphi(\obs_i)+\psi(\obs_i)$ and the two functions $\varphi(.)$ and $\psi(.)$ are semiparametrically estimated without constraints.} Figure \ref{fig:e1e0_joint} plots these estimated ratios for the three datasets.\footnote{When plotting the relation $e_1(e_0)$ and conducting tests, we eliminate outlier observations of $e_0$ using Tukey's fences. This procedure eliminates around  $1\%$ observations across all samples. We do this since, in some tests, we rely on non-parametric estimation of $e_1(e_0)$, which can be quite sensitive to outliers.} Visually, the two restrictions appear to be well-supported by the data. We see that $e_1$ clearly tends to increase with $e_0$ in all samples we consider. Moreover, the relationship between $e_1$ and $e_0$ seems well-captured by a linear fit in the case of Italian and Chinese data, while for Spanish data, there seems to be some deviation at the tails of the empirical distribution of $e_0$.

We next provide formal statistical tests. 
First, we test the implication that $e_1$ is an increasing function of $e_0$. For each sample, we compute two rank correlation measures between $e_1$ and $e_0$, Spearman's $\rho$ and Kendall's $\tau$. We then perform the non-parametric monotonicity test proposed by \cite{silverman1981using} and implemented by \cite{bowman1998testing} (BJG).\footnote{The test fits local linear regressions and finds the minimal bandwidth required for the estimated function to be monotonic. This bandwidth is then used as a test statistic, and the p-value is computed via bootstrap. The test is implemented in R package \textit{sm} available at https://CRAN.R-project.org/package=sm.}

We report results in Table \ref{tbl:monotonicity}. We see that rank correlation between $e_1$ and $e_0$ is very high in each sample. Based on BJG's test, we cannot reject the hypothesis that the relation between $e_1$ and $e_0$ is monotonic.

\begin{table}[!ht]
 	\begin{center}
 		\caption{Does $e_1$ vary monotonously with $e_0$? }
 		\label{tbl:monotonicity}
 		\small
   	\vspace{-10pt}
		\begin{tabular}{rccc}
			\hline
			\hline\\
			[-1.8ex]
 & Spearman's $\rho$ & Kendall's $\tau$ & BJG \\ 
  \hline
  Spain & & &\\
  \hline\\[-1.8ex]
AP & 0.91$^{***}$  & 0.75$^{***}$  & 0.14 \\ 
  FP & 0.95$^{***}$  & 0.82$^{***}$  & 0 \\ 
  
 Italy & & &\\
  \hline\\[-1.8ex]
  AP  & 0.92$^{***}$ & 0.76$^{***}$ & 0 \\ 
  FP & 0.92$^{***}$ & 0.76$^{***}$ & 0 \\ 
   China & & &\\ 
   \hline\\ [-1.8ex]
 All  & 0.997$^{***}$ & 0.953$^{***}$ & 0\\  
 
   \hline
 		\end{tabular}
 	\end{center}
 	\vspace*{-10pt}
  \par
 	\begin{spacing}{0.7}
 	{\footnotesize 	\textit{Notes:} Column 1 reports Spearman's $\rho$. Column 2 reports Kendall's $\tau$. Column 3 reports the critical bandwidth obtained from BJG. BJG's test is implemented using `sm' package in R. The null hypothesis of BJG's test is that the curve is monotonic. For AP and FP candidates in Italy and  AP candidates in Spain, the curve is monotone at the smallest value of the bandwidth.} 
  \end{spacing}
 \end{table}
Second, we test the implication of the parametric model that $e_1$ is a linear function of $e_0$. We compare linear regressions of $e_1$ as a function of $e_0$ with various flexible estimations of the same relationship. We also compute likelihood ratio tests obtained by comparing linear regressions of $e_1$ on $e_0$ and of $e_1$ on $e_0$ and $e_0^2$. 

We report results in Table \ref{tbl:ComparingModels_e1e0}. The likelihood ratio test tells us that linearity is statistically rejected, except for China. From Figure \ref{fig:e1e0_joint}, we see that departures from linearity seem to be concentrated in the tails of the distribution of $e_0$. However, linearity  appears to provide a relatively good fit relative to other models, as seen by comparing  Mean Squared Errors (MSE).
 
\begin{table}[ht]
\begin{center}
\caption{Comparing different methods to estimate $e_1(e_0)$}\label{tbl:ComparingModels_e1e0}
\begin{tabular}{lcccccHcH}
  \hline
  \hline\\[-1.8ex] 
 & \multicolumn{2}{c}{Linear} & \multicolumn{2}{c}{Quadratic}  & LR & pval & Spline & Isotonic\\
 \hline\\[-1.8ex] 
 
 & MSE & AIC & MSE & AIC &  $\chi^2$ & pval& MSE & MSE  \\
 \\[-1.8ex] 
 \multicolumn{8}{l}{Spain}\\
 \hline\\ [-1.8ex] 
AP & 0.14 & 13180.47 & 0.133 & 12429.93 & 752.54$^{***}$ & 0 & 0.133 & 0.133 \\ 
  FP & 0.103 & 6882.03 & 0.097 & 6118.37 & 765.66$^{***}$ & 0 & 0.096 & 0.097 \\ 
  \\[-1.8ex] 
   \multicolumn{8}{l}{Italy}\\
   \hline\\[-1.8ex] 
   AP & 0.343 & 43300.49 & 0.343 & 43296.54 & 5.95$^{**}$  & 0.015 & 0.343 & 0.344 \\ 
  FP & 0.264 & 19187.57 & 0.262 & 19079.98 & 109.59$^{***}$  & 0 & 0.262 & 0.264 \\  \\
[-1.8ex] 
   \multicolumn{8}{l}{China}\\
   \hline\\[-1.8ex] 
  All & 0.147 & 893.42 & 0.147 & 893.94 & 1.48 & 0.223 & 0.147 & 0.113 \\
 \hline
   \hline
\end{tabular}
\end{center}
\vspace*{-10pt}
	\begin{spacing}{0.7}
	{\footnotesize 	\textit{Notes:} The first column indicates the sample. The next four columns of the table report the mean squared error (MSE) and the Akaike information criterion statistics (AIC) for linear and quadratic regression of $e_1$ on $e_0$, respectively. Sixth column reports the $\chi^2$ statistics of the likelihood ratio test comparing nested linear and quadratic models. The last column reports the MSE obtained from penalized spline regression of $e_1$ on $e_0$, respectively.  MSE is calculated using 10 fold cross validation. $^{*}$p$<$0.1;$^{**}$p$<$0.05; $^{***}$p$<$0.01.} 
\end{spacing}
\end{table}

In this Section, we assessed the relationship between the estimated log-odds ratios of being promoted when connected and when unconnected. Overall, we find strong support for the monotonicity property implied by the assumption of independence from observables. Linearity implied by the parametric restrictions is formally rejected, except on data on political promotions in China. Even when formally rejected, however, a linear relationship provides a good fit. This is consistent with the fact that estimates of favors and information effects obtained from the parametric and semiparametric frameworks are quantitatively quite similar.

\subsection{When the number and types of connections may matter}

We next relax the assumption that connections have a binary impact within the parametric framework of model (\ref{eq:knownmodel}). We now assume that the impacts of connections may depend on their number and types. For academic promotion in Spain, we follow \cite{bagues2017does} and distinguish between strong and weak ties. Denote by $n_{Si}$ the number of strong ties that candidate $i$ has with the jury and by $n_{Wi}$ the number of weak ties. For academic promotion in Italy and political promotion in China, we consider only one type of connection, labeled as a strong tie.\footnote{Recall, in the Spanish data, candidates have strong ties with PhD advisors, coauthors, and colleagues and weak ties to members of their PhD committee, to members of the PhD committees of their PhD students, and to fellow PhD committee members. In the Italian data, candidates have strong ties with coauthors and colleagues. In the Chinese data, provincial leaders have strong ties with members of the Politburo who were coworkers in the past.}

Table \ref{tbl:desc_number_links} reports the numbers of candidates with no connection, some connections, and at least two connections to the jury. In all samples, most connected candidates have one connection only. In the data on academic promotion in Spain, 28\% of candidates who have a strong connection to the jury, have more than one. This proportion is 25\% for weak ties. In the data on academic promotions in Italy, only 5\% of connected candidates have more than one connections to the jury. In China, 27\% of observations of connected candidates involve more than one connection to the jury.

We supplement model (\ref{eq:knownmodel}) with parametric assumptions on how favors and information effects from connections depend on their numbers and types. We consider the following linear and log-linear assumptions. For academic promotions in Spain, we assume that $B(n_{Si},n_{Wi})=\gamma_S n_{Si}+\gamma_W n_{Wi}$ and $ln(\sigma(n_{Si},n_{Wi}))=\delta_S n_{Si}+\delta_W n_{Wi}$. For academic promotions in Italy and political promotions in China, we set $B(n_{Si})=\gamma_S n_{Si}$ and $ln(\sigma(n_{Si}))=\delta_S n_{Si}$. These assumptions allow us to estimate the incremental impact of a strong or a weak tie on the effects of connections.

\begin{footnotesize}
\begin{table}[!htb]
    \caption{Parametric model,  accounting for number and types of connections}\label{tbl:HetLogitConnectionsParametricJoint}
    \vspace{-10pt}
    \begin{center}
    \begin{subtable}{.40\linewidth}
        \centering
        \caption*{\hspace{70pt} Spain}
        \small
        \begin{tabular}{@{\extracolsep{5pt}}lccc} 
        \hline \hline
        & (All) & (AP) & (FP)\\ 
        \hline \\[-1.8ex]

        $\gamma_S$ & 0.378$^{***}$ & 0.309$^{***}$ & 0.487$^{***}$ \\ 
  & (0.065) & (0.101) & (0.074) \\ 
  $\gamma_W$ & 0.062 & 0.006 & 0.160$^{*}$ \\ 
  & (0.095) & (0.216) & (0.083) \\  
  \\
   $\delta_S$ & 0.126$^{***}$ & 0.180$^{***}$ & 0.064 \\ 
  & (0.035) & (0.050) & (0.044) \\ 
   $\delta_W$ & 0.084$^{*}$ & 0.192$^{*}$ & 0.006 \\ 
  & (0.047) & (0.109) & (0.046) \\ 
        \\
        Obs. & 31,000 & 17,784 & 13,216 \\
        \hline 
        \end{tabular} 
    \end{subtable}%
    \hspace{0.8em} 
    \begin{subtable}{.40\linewidth}
        \centering
        \caption*{Italy}
        \small
        \begin{tabular}{ccc} 
        \hline \hline
        (All) & (AP) & (FP)\\ 
        \hline \\[-1.8ex] 

          0.226$^{***}$ & 0.235$^{***}$ & 0.231$^{***}$ \\ 
  (0.035) & (0.041) & (0.052) \\ 
        \\
        \\
        \\
        0.099$^{***}$ & 0.122$^{**}$ & 0.030 \\ 
       (0.037) & (0.048) & (0.046) \\ 
        \\
        & &\\
        & &\\ 
       69,020 & 47,426 & 21,594  \\ 
        \hline 
        \end{tabular} 
    \end{subtable}%
    \hspace{-10pt}
    \begin{subtable}{.10\linewidth}
        \centering
        \caption*{China}
        \small
        \begin{tabular}{c} 
        \hline \hline
        (All)\\ 
        \hline \\[-1.8ex] 
        0.793$^{*}$ \\ 
        (0.410) \\ 
        \\
        \\
        \\
       $-$0.127 \\ 
    (0.190) \\ 
        \\
        \\
         \\
         966\\ 
        \hline 
        \end{tabular} 
    \end{subtable}
    \end{center}
    \vspace*{-10pt}
    \begin{spacing}{0.7}
    {\footnotesize\textit{Notes}: Heteroscedastic logit estimates. The column name indicates the sample. Standard errors, in parentheses, are clustered at the committee level for Spain and Italy, and the province level for China. We include exam-specific controls for Spain, exam fixed effects for Italy, and year, term, and office$\times$province fixed effects for China. $^{*}$p$<$0.1; $^{**}$p$<$0.05; $^{***}$p$<$0.01.}
    \end{spacing} 
\end{table}
\end{footnotesize}

Table \ref{tbl:HetLogitConnectionsParametricJoint} provides results of regression estimates based on polynomial approximations of the grade function and on heteroscedastic logit. We see that an additional strong tie increases the bias from favors for all candidates in both Spain and Italy. It also increases information effects for AP candidates in both countries. It has no significant impact on information effects for FP candidates in Italy, and a small and marginally significant impact on information effects for FP candidates in Spain. Similarly, we find some evidence that connections yield favors in political promotions in China and no evidence that they generate information effects. Overall, results on the impact of strong ties are thus qualitatively similar to the results obtained when considering a binary impact of connections.

For academic promotions in Spain, we can also assess the impact of weak ties on promotion. For AP candidates, this is limited by the fact that candidates have relatively few weak ties, leading to imprecise estimates. We detect no impact of an additional weak tie on favors and an impact of $0.192$ on the information effect. This is quantitatively quite large - and indeed larger than the impact of a strong tie - but is only statistically significant at the $10\%$ level. 
For FP candidates we find that an additional weak tie leads to a strong and significant positive impact on favors, equal to about one third of the impact of an additional strong tie. By contrast, weak ties have no significant impact on information effects for FP candidates. These results indicate that the academic community in Spain at the time was functioning within a system of generalized favor exchange.

Overall, these results are consistent with those obtained under the assumption that connections have a binary impact, with information effects at work for AP candidates only and favors operating for all candidates. On the Spanish data, estimated impacts of strong versus weak ties depict a coherent picture. An additional weak tie yields a lower bias from favors than an additional strong tie, and information gains for AP candidates of a similar magnitude. The positive relationship between the number of links and the strength of the information effect is consistent with juries receiving more numerous or more precise signals about candidates who have more connections.

\section{Endogenous Connections}\label{sec:endogeneous_connections}

Our analysis, so far, has relied on the assumption that connections are conditionally random. We now relax this assumption. We show how to extend our parametric framework to endogenous connections. More precisely, we adapt a control function approach proposed by \cite{wooldridge2014quasi, wooldridge2015control}. We model the formation of connections and incorporate generalized residuals from the equation describing connection formation into the heteroscedastic logit used to identify favors from information effects. We then apply this extended model to the data on Italy, where 8\% of the initially selected evaluators resign from the jury and where some observable characteristics differ between connected and unconnected candidates even when controlling for the expected number of connections (Table \ref{tbl:RandomAssignementItaly}). Our estimates of favors and information effects are very similar to those obtained under the assumption of conditionally random connections. We test, and reject, connection endogeneity, validating the analysis of the Italian data presented in the previous sections.

Let us describe our methodology first. When connections are not conditionally random, distributions of unobservables $u_i$ and $v_i$ may differ between connected and unconnected candidates. We consider the binary variant of our model, where favors and information effects only depend on whether candidates are connected to the jury. In this case, the outcome (promotion) is binary and the endogenous variable of interest (connectedness) is also binary.\footnote{When both the outcome and the endogenous variable are binary, the corresponding nonparametric model is not identified, as noted by \citet{chesher2003identification} and \citet{wooldridge2014quasi}, justifying parametric asssumptions.} To address connection endogeneity, we adapt an approach proposed by \cite{wooldridge2014quasi, wooldridge2015control}.\footnote{As pioneered by  \cite{heckman1976common}, the general idea is to model selection and to exploit information from the selection stage to address endogeneity at the outcome stage. In contexts where continuous measures of connectedness are available (e.g. strength of ties), we could adapt the control function approach developed by \cite{blundell2004endogeneity}, and incorporate residuals of the connectedness equation directly as regressors in our framework.} Assume that there exists an instrument $q_i$ that affects connectedness but has no direct impact on promotion. We model connection formation as:
\begin{equation}\label{eq:selection_first_stage}
   c_i=1 \Leftrightarrow \pi(\obs_i,q_i)+\eta_i>0
\end{equation}
\noindent where $\eta_i \Perp (\obs_i,q_i)$ follows a normal distribution with mean $0$ and variance $1$ and $\pi(\cdot)$ is a function to be estimated.  In our application below, and following \cite{bagues2017does}, we set $q_i$ equal to the number of connections to the initially drawn committee. This number is random conditional on the expected number of connections with the pool of potential evaluators, and hence may affect promotion only through its impact on connectedness to the actual jury.

Define, then, the generalized residual of the model of connection formation as $r_i=\E(\eta_i | c_i,\obs_i,q_i)$. Furthermore, denote by $\varphi$ and $\Phi$ the pdf and cdf of the standard normal distribution and by $\lambda$ the inverse Mills ratio, i.e., $\lambda(x)=\frac{\varphi(x)}{\Phi(x)}$. In the probit framework, the generalized residual is equal to\footnote{To see why, note that $\E(\eta_i | c_i=1,\obs_i,q_i)=\E(\eta_i | \eta_i>-\pi(\obs_i,q_i))=\frac{\varphi(-\pi(\obs_i,q_i))}{1-\Phi(-\pi(\obs_i,q_i))}=\lambda(\pi(\obs_i,q_i))$ while $\E(\eta_i | c_i=0,\obs_i,q_i)=\E(\eta_i | \eta_i<-\pi(\obs_i,q_i))=-\lambda(-\pi(\obs_i,q_i))$.}
\begin{equation*}
  r_i=c_i\lambda(\pi(\obs_i,q_i))-(1-c_i)\lambda(-\pi(\obs_i,q_i)).
  \end{equation*}
Note that $r_i$ is mean-independent of observables, $\E(r_i|\obs_i,q_i)=0$.\footnote{We have: $\E(r_i|\obs_i,q_i)=\mathbb{P}(c_i=0)\E(r_i|\obs_i,q_i,c_i=0)+\mathbb{P}(c_i=1)\E(r_i|\obs_i,q_i,c_i=1)=-\Phi(-\pi(\obs_i,q_i))\lambda(-\pi(\obs_i,q_i))+\Phi(\pi(\obs_i,q_i))\lambda(\pi(\obs_i,q_i))=0$.} Our main assumption is then that connection endogeneity is fully controlled when conditioning on $r_i$. In our context, this means that $c_i \Perp (u_i,v_i) | (\obs_i,r_i)$.  We recover the equality in distributions of $u_i$ and $v_i$ when conditioning on the generalized residual.

To identify favors from information effects, we then treat $r_i$ as an additional observable and apply Theorem 1.  Let $w_i=v_i + \E(u_i|\theta_i,\obs_i,r_i,v_i)$. The assumption of independence from observables becomes $(u_i,\theta_i,v_i) \Perp (\obs_i,r_i)$. 
Under full support, independence from observables, and a homogeneous bonus from favors, then $B$ and the distribution of $w_i$ are identified if the distribution of $v_i$ is known. The corresponding parametric version of the model can then be written as:
\begin{equation*}
    \mathbb{P}(y_i=1|c_i=0,\obs_i,r_i)=\Lambda_0(\varphi(\obs_i,r_i))
\end{equation*}
\begin{equation*}
\mathbb{P}(y=1|c_i=1,\obs_i,r_i)=\Lambda_0\left(\frac{\varphi(\obs_i,r_i)+B}{\sigma}\right)
\end{equation*}

To estimate the model, we begin by estimating the model of connection formation, equation \eqref{eq:selection_first_stage}, imposing $\pi(\obs_i,q_i)=\gucon+\zeta q_i$. We use probit and report estimates of $\zeta$ in Table~\ref{tbl:selection_first_stage_AP_FP}. In both subsamples of AP and FP candidates, $\zeta$ is large and precisely estimated, confirming that $q_i$ is a relevant determinant of connectedness.

Generalized residual $r_i$ can be incorporated into $\varphi(\obs_i,r_i)$ in different ways. We report results for two natural and nested specifications. The first specification, reported in column (2) of Table \ref{tbl:selection_parametric} extends the baseline specification of $\gucon$ used in Table~\ref{tbl:HetLogitBinaryParametricJoint} by adding $r_i$ as an additional control, thus setting $\varphi(\obs_i,r_i) = \gucon + \xi^1 r_i$. The second specification, reported in column (3) of Table \ref{tbl:selection_parametric}, treats $r_i$ as any other variable $x_i$ included in the list of arguments of $\varphi (\cdot)$. This specification therefore augments $\varphi(\obs_i, r_i)$ by including interactions between $r_i$ and all variables in $\obs_i$ as well as quadratic term $r_i^2$. We estimate both specifications separately for the AP candidate sample and the FP candidate sample, and the corresponding results are presented in Table \ref{tbl:selection_parametric}. To facilitate comparison, we also report in column (1) in that table the corresponding estimates of model \eqref{eq:knownmodel}, under the assumption of conditionally random connections, first reported in Table \ref{tbl:HetLogitBinaryParametricJoint}. 

\begin{table}[!ht] 
\small
\vspace{-15pt}
\begin{center} 
  \caption{Parametric model, binary and endogenous connections, Italy } 
  \label{tbl:selection_parametric} 
\begin{tabular}{@{\extracolsep{5pt}}lccc} 
\\[-1.8ex]\hline 
\hline \\[-1.8ex] 
\\[-1.8ex] & (1) & (2) & (3)\\ 
\hline \\[-1.8ex] 
AP candidates &&&\\
\addlinespace

$r_i^2$ ($\xi^2$)  &  &  & 0.005 \\ 
  &  &  & (0.023) \\ 
  $r_i$ ($\xi^1$) &  & $-$0.019 & $-$0.087 \\ 
  &  & (0.041) & (0.244) \\ 
  B & 0.230$^{***}$ & 0.238$^{***}$ & 0.238$^{***}$ \\ 
  & (0.044) & (0.047) & (0.047) \\ 
  $\delta$ & 0.137$^{***}$ & 0.137$^{***}$ & 0.130$^{**}$ \\ 
   & (0.051) & (0.051) & (0.056) \\
   \addlinespace
LR test (1) vs (2) &  $\chi^2=0.15$ (p=0.697) &  &  \\ 
LR test (1) vs (3) & $\chi^2 =13.82$ (p=0.463) &  &  \\ 
\hline 
\addlinespace
  FP candidates  &&&\\
  \addlinespace

   $r_i^2$ ($\xi_m^2$) &  &  & $-$0.021 \\ 
  &  &  & (0.038) \\ 
  $r_i$ ($\xi_m^1$)&  & 0.097$^{*}$ & $-$0.077 \\ 
  &  & (0.056) & (0.391) \\ 
  $B$ & 0.233$^{***}$ & 0.194$^{***}$ & 0.193$^{***}$ \\ 
  & (0.057) & (0.061) & (0.062) \\ 
$\delta$& 0.027 & 0.029 & 0.026 \\ 
  & (0.052) & (0.052) & (0.054) \\ 
  
  \addlinespace
 LR test (1) vs (2) & $\chi^2 =2.44$ (p=0.119) &  &  \\ 
LR test (1) vs (3) & $\chi^2 = 18.71$ (p=0.176) &  &  \\ 
\hline 
\end{tabular} 
\end{center}
	\vspace*{-10pt}
{\footnotesize 	\textit{Notes:} Column (1) reports the estimates of model \eqref{eq:knownmodel}. Column (2) reports estimates of model \eqref{eq:knownmodel} with generalized residual $r_i$ as an added control. Column (3) reports the estimates of model \eqref{eq:knownmodel} with $r_i$, $r_i^2$, as well as the interaction between $r_i$ and other control variables as added controls. 
All specifications include the full set of exam fixed effects. Standard errors clustered at the committee level are in parentheses. The LR test is the likelihood ratio test comparing the indicated nested models. We report $\chi^2$ statistics and corresponding p-values.  $^{*}$p$<$0.1; $^{**}$p$<$0.05; $^{***}$p$<$0.01. }
\end{table}

For AP candidates, the estimates of $B$ and $\delta$ in Table~\ref{tbl:selection_parametric} are stable across specifications, with only small changes. The estimate of $B$ rises modestly from $0.230$ in column~(1) to $0.238$ in column~(3), while $\delta$ falls from $0.137$ to $0.130$. Following the approach in \citet{wooldridge2014quasi}, the null hypothesis that $c_i$ is exogenous corresponds to a zero coefficient on the generalized residual ($\xi^1=0$). In column (2), we do not reject this null. Likelihood ratio tests comparing the nested specifications in columns (1) and (2), as well as those in columns (1) and (3), indicate that adding $r_i$ does not significantly improve model fit.

For FP candidates, the estimate of $B$ in Table~\ref{tbl:selection_parametric} decreases from $0.233$ in column~(1) to $0.193$ in column~(3), whereas $\delta$ remains stable and statistically insignificant when $r_i$ is included. The generalized residual is marginally significant in column~(2), but likelihood ratio tests again indicate that neither specification in columns~(2) or~(3) improves the fit relative to column~(1).

Table~\ref{tbl:AvgEffects_Binary_Selection_Italy} reports the implied average marginal effects. For AP candidates, the estimated effects are nearly identical to those reported in Table~\ref{tbl:AvgEffects_Binary_Parametric_Italy_1}. For FP candidates, the average effect of being connected falls slightly from $0.043$ to $0.036$. This attenuation is entirely driven by a reduction in the estimated favoritism effect, which falls from $0.041$ reported in Table \ref{tbl:AvgEffects_Binary_Parametric_Italy_1}  to $0.034$, while the information effect is essentially the same as reported in Table \ref{tbl:AvgEffects_Binary_Parametric_Italy_1}.

\begin{table}[!ht]
	\begin{center}
		\caption{Marginal effects, parametric model, binary and endogenous connections, Italy} 
		\label{tbl:AvgEffects_Binary_Selection_Italy}
		\small
  	\vspace{-10pt}
		\begin{tabular}{rccccccc}
			\hline
			\hline\\
			[-1.8ex]
			&\multicolumn{1}{c}{Baseline}& &\multicolumn{5}{c}{Marginal effects}\\
			\cline{2-2} \cline{4-8}\\
			[-1.8ex]
			&Predicted& &Total & Favors & Information &Information &Information \\
            && &&  & &Low chances& High chances\\
			\hline\\
			[-1.5ex] 
\multicolumn{8}{c}{{Panel A: column (2) of  Table \ref{tbl:selection_parametric} }}\\
\textbf{} & & & & & & &  \\       
AP & 0.42$^{***}$ && 0.046$^{***}$ & 0.039$^{***}$ & 0.007$^{**}$ & 0.022$^{***}$ & -0.019$^{***}$ \\ 
   & (0.01) && (0.008) & (0.008) & (0.002) & (0.006) & (0.005) \\ 
  Observations & 25291 && 25291 & 25291 & 25291 & 16010 & 9281 \\ 
  \\
  FP & 0.41$^{***}$ && 0.036$^{***}$ & 0.034$^{***}$ & 0.002 & 0.005$^{*}$ & -0.004 \\ 
   & (0.004) && (0.011) & (0.011) & (0) & (0) & (0) \\ 
  Observations & 13264 && 13264 & 13264 & 13264 & 8680 & 4584 \\ 
\\
\multicolumn{8}{c}{{Panel B: column (3) of  Table \ref{tbl:selection_parametric}}}\\
\textbf{} & & & & & & &  \\   

 AP& 0.42$^{***}$ && 0.045$^{***}$ & 0.039$^{***}$ & 0.006$^{**}$ & 0.021$^{***}$ & -0.018$^{***}$ \\ 
   & (0.006) && (0.008) & (0.008) & (0.001) & (0.005) & (0.004) \\ 
Observations & 25291 && 25291 & 25291 & 25291 & 16010 & 9281 \\ 
\\
  FP & 0.41$^{***}$ && 0.036$^{***}$ & 0.034$^{***}$ & 0.002 & 0.004$^{*}$ & -0.004 \\ 
  & (0.004) && (0.011) & (0.011) & (0.001) & (0.001) & (0.002) \\ 
 Observations & 13264 && 13264 & 13264 & 13264 & 8682 & 4582 \\ 
 
  \hline
		\end{tabular}
	\end{center}
	\vspace*{-10pt}
	\begin{spacing}{0.7} 
	{\footnotesize \textit{Notes:}
		The average marginal effect of being connected is computed for candidates with at least one connection to eligible evaluators. 
        Panel A reports the average marginal effects for the model estimated in column (2) of Table \ref{tbl:selection_parametric}. Panel B reports the average marginal effects for the model estimated in column (3) of Table \ref{tbl:selection_parametric}. Bootstrapped standard errors clustered at the committee level are in parentheses.$^{*}$p$<$0.1;$^{**}$p$<$0.05; $^{***}$p$<$0.01.} 
	\end{spacing}
\end{table}

To sum up, we propose an approach to account for connection endogeneity in our framework. This approach corrects for differences in the distributions of $u_i$ or $v_i$ between connected and unconnected candidates that operate through the generalized residual $r_i$.

\section{Discussion and Conclusion}\label{sec:discussion}

In this article, we propose a new method to identify the respective roles of favors and information in the impact of connections, building on earlier work on discrimination. The method exploits distinct implications of the two effects on the relationship between candidates’ observables and success. We show that better information on connected candidates generates excess variance in latent errors. Once differences in variance are accounted for, differences in estimated promotion thresholds can identify the bias due to favors. We characterize the conditions under which both effects are identified and operationalize these ideas econometrically within a semiparametric framework. We then reanalyze the data assembled and studied in \cite{zinovyeva2015role, jia2015political}, and \cite{bagues2017does}. These three articles provide estimates of the causal impact of connections. Based on the same data, we analyze the mechanisms behind this impact. We notably provide decompositions of the overall impact of connections into separate favor and information components. Testable implications of the identifying assumptions are strongly supported in the data. Finally, we show how to account for connection endogeneity by incorporating a control function approach in our framework. 

As with any identification strategy, our method relies on a number of assumptions. We next discuss ways through which it can be externally validated. We also discuss how the identification of favors and information effects is affected when the assumptions of a risk neutral jury and a deterministic bias are relaxed. We conclude by highlighting the policy relevance of the method. 

\textbf{External validation} In our analysis, we derive testable restrictions of our framework and test these restrictions on the different datasets, exploiting data collected at time of promotion only. Access to data on post-promotion performance opens up other ways to validate our approach. In particular, we can assess differences in performance between connected and unconnected promoted candidates with the same estimated grade. The size and magnitude of these differences is then informative about the underlying mechanisms. Better information on connected candidates predict a higher performance post-promotion, while favors predict a lower performance post-promotion.\footnote{These predictions are valid if post-promotion performance depends on candidates’ true ability and is only affected by the presence of connected evaluators in the jury composition through its effect on promotion, see \citet[p.283]{zinovyeva2015role}.} The overall effect then depends on the relative strengths of the two channels.
 
We carry out this exercise for academic promotions in Spain, the only dataset containing information on post-promotion performance, in Appendix A. We perform individual-level regressions of post-promotion outcomes over the estimated grade at time of promotion and a connection indicator on the sample of promoted candidates. We see that promoted connected FP candidates perform worse in the five years post-promotion than promoted unconnected FP candidates, controlling for grade, in terms of publications, citations and citation-weighted publications. We do not detect differences on the number of PhD students and PhD committees. This is consistent with the fact that we only detect favors at work in this subsample, as reported in Table \ref{tbl:AvgEffects_Binary_Parametric_Spain}.

Interestingly, the effect of connection on post-promotion performance is also negative, but lower in magnitude, for AP candidates for the same three outcomes.\footnote{Magnitude of the estimates are comparable across subsamples since post-promotion outcomes are standardized within exams.} This is consistent with the facts that: (1) we detect both favors and information effects at work in this subsample, and (2) we find that favors quantitatively dominate information effects in their impacts on promotion, as reported in Table \ref{tbl:AvgEffects_Binary_Parametric_Spain}. Overall, the analysis of differences in post-promotion outcomes between connected and unconnected promoted candidates is fully consistent with our estimates of favors and information effects estimated from data at time of promotion.

\textbf{Risk averse jury} Throughout our analysis, we have maintained the assumption that the jury is risk neutral and only cares about the expected ability of the candidates. This appears to be a natural benchmark in the context of academic promotions in Spain and Italy, where the jury simply evaluates the suitability of candidates for advancement and bears no direct cost of promoting unsuitable candidates. In other settings, where the people who decide about promotion also bear the costs of incorrect decisions, risk aversion may be relevant. 

We develop a simple extension of our theoretical framework in Appendix A, under assumptions of mean-variance preferences and i.i.d. unobservables. In this setup, uncertainty on a candidate's true ability yields a grade penalty proportional to variance and that increases with risk aversion. The estimate of the difference in promotion thresholds now combines two effects: the bias due to favoritism and the difference in grade penalties coming from the difference in uncertainty between connected and unconnected candidates. Because of the extra information conveyed by connections, uncertainty on connected candidates is always lower than on unconnected candidates. The difference in grade penalties is thus always positive and the estimated difference in promotion thresholds now provides an upper bound for the bias from favors. Separately identifying risk aversion, favors, and information effects in the impact of connection provides an interesting, and challenging, direction for for future research.

\textbf{Stochastic bias} In our approach so far, we have assumed that the bias from favors is either homogeneous (Theorem \ref{th:IdentificationBenchmark}) or depends in a deterministic way on individual observables, such as gender (Theorem \ref{th:IdentificationExtended}). When the bias is not deterministic, but stochastic, this adds extra variance to the latent errors of connected candidates unrelated to information effects. In this case, the estimate of this extra variance should be interpreted as a upper bound of information effects, while the difference in promotion thresholds now identifies the expected bias.

\textbf{Policy implications} Finally, our approach has policy implications. Favoritism is a form of discrimination, which is illegal in many contexts, such as academic hiring and public procurement, and unethical in others, such as academic publications. Our method could serve as a diagnostic device to detect favoritism. Given data on candidates, connections and outcomes, it is straightforward to implement and essentially costless. It could notably be used to help target costly investigative techniques, such as audit studies.

 \newpage

\bibliography{LiteratureFavorInfo.bib}
 
 \newpage

 \section*{Appendix A: Additional Results}
\renewcommand{\thetable}{A.\arabic{table}}
\setcounter{table}{0}
\subsection*{A.1. Risk aversion}

Observe, first, that in the benchmark model of Section 2, the jury can be thought of as maximizing the expectation of $\Pi$ conditional on available information $\obs,\mathbf{c},\mathbf{v},\boldsymbol{\theta}$, where $\Pi=\sum_i \pi_i$ with $\pi_i=y_i(a_i+Bc_i)$ and $\pi_i$ is the jury's payoff of promoting candidate $i$ ($y_i=1$). Recall that $a_i=\varphi(\obs_i)+u_i+v_i$.

To think about risk aversion, consider a simple setup where the jury has mean-variance preferences, as in, for instance,  \citet{galeotti2021cross}, and denote by $\lambda$ the risk aversion parameter. Assume, also, that both $u_i$ and $u_i|\theta_i$ are i.i.d. and denote by $\sigma^2_u=\mathbb{V}(u_i)$ and $\sigma_c^2=\mathbb{V}(u_i|\theta_i)$. We maintain the independence assumption, i.e., $(u_i, \theta_i) \Perp (\obs_i, v_i)$.

With mean-variance preferences, the jury wants to maximize $\mathbb{E}(\Pi)-\lambda\mathbb{V}(\Pi)$. By independence, this is equal to $\sum_i \left[\mathbb{E}(\pi_i)-\lambda\mathbb{V}(\pi_i) \right]$. The utility of promoting unconnected candidate $i$ is
\begin{equation*}
    \E(a_i|\obs_i, v_i,c_i=0) - \lambda \V\left(a_i|\obs_i, v_i,c_i=0 \right)  =  \gucon - \lambda \sigma_u^2+ v_i  ,
\end{equation*}
By contrast, the utility of promoting connected candidate $i$ is
\begin{equation*}
    B+\E(a_i|\obs_i, v_i,\theta_i,c_i=1) - \lambda \V\left(a_i|\obs_i, v_i, \theta_i,c_i=1 \right)  = B+ \gucon - \lambda \sigma_c^2 + w_i  ,
\end{equation*}
where $w_i=v_i+\mathbb{E}(u_i|\theta_i)$.

Under mean-variance preferences, uncertainty in observables $u_i$ or $u_i|\theta_i$ get transformed into a grade penalty proportional to variance. By the law of total variance, $\mathbb{V}(u_i)=\mathbb{E}(\mathbb{V}(u_i|\theta_i))+\mathbb{V}(\mathbb{E}(u_i|\theta_i))$. This is equivalent to $\sigma_u^2=\sigma_c^2+\mathbb{V}(\mathbb{E}(u_i|\theta_i))$. This implies that $\sigma_c^2<\sigma_u^2$. Unobservable uncertainty is always lower for connected candidates, due to the additional signal from connections. By independence, $\mathbb{V}(w_i)=\mathbb{V}(v_i)+\mathbb{V}(\mathbb{E}(u_i|\theta_i))=\sigma^2 \mathbb{V}(v_i)$ with $\sigma>1$.

In the corresponding parametric framework, $v_i$ and $w_i$ follow logistic distributions with different variances. This yields
\begin{align*}
    \mathbb{P}(y_i=1|\obs_i,c_i=0)&=\Lambda_0(\varphi(\obs_i)-\lambda \sigma_u^2) \\
    \mathbb{P} (y_i=1|\obs_i,c_i=1)&=\Lambda_0\left(\frac{\varphi(\obs_i)+B-\lambda \sigma_c^2}{\sigma} \right)
\end{align*}

Applying the identification arguments of Theorem 1, we see that both $\sigma$, the extra variance coming from the additional signal, and the difference in promotion thresholds $B-\lambda \sigma_c^2+\lambda \sigma_u^2=B+\lambda(\sigma_u^2-\sigma_c^2)$ are identified. Difference in promotion thresholds is now always larger than or equal to the bias. 

\subsection*{A.2. Post-promotion outcomes}

We consider the sample of promoted candidates from the data of \cite{zinovyeva2015role}. We look at 5 outcomes post-promotion: number of publications, number of citations, total AIS, number of PhD students and number of PhD committees in the five years following the exam, see Section IV.D in the original paper. We run the following linear regressions:
\begin{equation*}
    y_{i}^p = \alpha_0 +  \alpha_1 c_{i} +  \alpha_2 grade_{i} +  \alpha_3 AP_{i} +  \alpha_4 AP_{i}  c_{i}+ \alpha_5 AP_i grade_i +  \epsilon_{i},
\end{equation*}
\noindent where $y_{i}^p$ is a post-promotion outcome, $c_i$ is a binary variable equal to $1$ if candidate $i$ is connected to the jury and $0$ otherwise, $grade_i$ is estimated using parametric model (7) described in Section 5 and estimated in Section 6, $AP_i$ is a dummy equal to $1$ for AP candidates and $0$ for FP candidates.

Results are reported in Table \ref{tbl:future_productivity_APvsFP}. We see that connected promoted FP candidates have less publications, less citations and less citation-weighted publications than unconnected promoted FP candidates ($\alpha_1<0$). This is also true for promoted AP candidates ($\alpha_1+\alpha_4<0$). However this difference is lower in magnitude for AP candidates than for FP candidates ($\alpha_4>0$). By contrast we do not detect differences in number of PhD students or number of PhD committees between connected and unconnected promoted candidates at either level.

We highlight two key differences between these regressions and the regressions underlying Table 6 in \cite{zinovyeva2015role}. First, we estimate different effects of connection for AP and FP candidates. Second, and consistently with our framework, we only control for observable characteristics at time of promotion through the estimated grade.

\begin{table}[!htbp] 
  \caption{Future productivity of promoted candidates} 
  \label{tbl:future_productivity_APvsFP} 
  \small
  \vspace{-10pt}
  \begin{center}
\begin{tabular}{@{\extracolsep{5pt}}lccccc} 
\hline 
\hline \\[-1.8ex] 
 & \multicolumn{5}{c}{\textit{Dependent variable:}} \\ 
\cline{2-6} 
 & Publications & Total AIS & Citations & PhD students & PhD committees \\ 
\\[-1.8ex] & (1) & (2) & (3) & (4) & (5)\\ 
\hline \\[-1.8ex] 
 grade ($\alpha_2$) & 0.279$^{***}$ & 0.305$^{***}$ & 0.262$^{***}$ & 0.246$^{***}$ & 0.216$^{***}$ \\ 
  & (0.038) & (0.039) & (0.038) & (0.034) & (0.037) \\ 
connected ($\alpha_1$) & $-$0.236$^{***}$ & $-$0.267$^{***}$ & $-$0.292$^{***}$ & $-$0.017 & 0.048 \\ 
  & (0.070) & (0.073) & (0.074) & (0.068) & (0.070) \\ 
AP ($\alpha_3$)& $-$0.200 & $-$0.345$^{**}$ & $-$0.368$^{***}$ & $-$0.337$^{***}$ & $-$0.443$^{***}$ \\ 
  & (0.127) & (0.136) & (0.138) & (0.122) & (0.123) \\ 
 grade:AP ($\alpha_5$) & 0.028 & $-$0.051 & $-$0.063 & $-$0.135$^{***}$ & $-$0.131$^{***}$ \\ 
  & (0.046) & (0.047) & (0.048) & (0.044) & (0.046) \\ 
 connected:AP ($\alpha_4$)& 0.151$^{*}$ & 0.150$^{*}$ & 0.164$^{*}$ & $-$0.008 & $-$0.067 \\ 
  & (0.084) & (0.087) & (0.085) & (0.084) & (0.086) \\ 
\hline \\[-1.8ex] 
Observations& 3523 & 3523 & 3523 & 3523 & 3523 \\


F stat ($\alpha_1 + \alpha
_4 =0$) & 3.437$^{*}$ & 6.046$^{**}$ & 8.913$^{***}$ & 0.254 & 0.142\\

\hline 
\end{tabular} 
\end{center}
\vspace{-10pt}
{\footnotesize \textit{Notes:} 
The sample consists
of promoted candidates in Spain. Standard errors clustered at the committee level are reported in parentheses. Row F stat reports F statistics of test $\alpha_1 + \alpha_4 =0$. $^{*}$p$<$0.1; $^{**}$p$<$0.05; $^{***}$p$<$0.01}
\end{table}

\pagebreak

 \clearpage

 \section*{Online Appendix B: Additional tables}
 
 \renewcommand{\thetable}{B.\arabic{table}}
\setcounter{table}{0}

 \begin{table}[!h] 
	\begin{center}
	\caption{Descriptive statistics, Spain, observables} 
	\label{tbl:DescriptiveObservables}
	\small
	\begin{tabular}{lccccccc}
		\hline
		\hline
		\\[-1.8ex]
		   & All & AP & FP & Eng. & H\&L & Sci. & Soc. Sci. \\ 
		 \hline		
         Promoted & 0.11 & 0.12 & 0.11 & 0.12 & 0.12 & 0.11 & 0.12 \\ 
           & (0.32) & (0.32) & (0.31) & (0.33) & (0.32) & (0.31) & (0.33) \\ \\
		 Female & 0.34 & 0.40 & 0.27 & 0.21 & 0.45 & 0.30 & 0.39 \\ 
		  & (0.47) & (0.49) & (0.44) & (0.41) & (0.50) & (0.46) & (0.49) \\ 
		 Age & 41.21 & 37.49 & 46.14 & 38.74 & 41.86 & 41.97 & 40.39 \\ 
		 & (7.59) & (6.41) & (6.06) & (7.07) & (7.62) & (7.55) & (7.54) \\ 
		 PhD in Spain & 0.78 & 0.83 & 0.70 & 0.83 & 0.77 & 0.76 & 0.78 \\ 
	     & (0.42) & (0.37) & (0.46) & (0.38) & (0.42) & (0.43) & (0.42) \\ 
		 Past Experience & 0.81 & 0.73 & 0.91 & 0.85 & 0.63 & 0.89 & 0.88 \\ 
		 & (1.27) & (1.27) & (1.26) & (1.36) & (0.94) & (1.40) & (1.30) \\
		 \\ 
		 Publications & 12.84 &  8.12 & 19.09 &  7.76 & 11.45 & 16.99 &  9.22 \\ 
		 & (18.31) & (14.06) & (21.18) & (12.88) & (11.39) & (24.10) & (11.61) \\ 
		 AIS & 0.72 & 0.70 & 0.74 & 0.52 & - & 0.80 & 0.62 \\ 
		 & (0.53) & (0.57) & (0.48) & (0.37) & - & (0.51) & (0.75) \\ 
		 \\
		 PhD Students & 1.00 & 0.24 & 2.00 & 0.83 & 0.61 & 1.45 & 0.66 \\ 
		 & (2.11) & (0.88) & (2.75) & (1.61) & (1.63) & (2.60) & (1.58) \\ 
		 PhD Committees & 3.61 & 0.88 & 7.23 & 2.40 & 3.04 & 4.81 & 2.67 \\ 
		 & (6.76) & (2.55) & (8.65) & (4.42) & (5.99) & (8.21) & (4.99) \\
		 \\
		 Observations & 31,243 & 17,799 & 13,444 &  4,783 &  9,005 & 12,858 &  4,597 \\ 
		 	\hline
		\end{tabular}
	\end{center}
	\vspace*{-10pt}
	 \begin{spacing}{1}
		{\footnotesize \textit{Notes: }Average values of observable characteristics. Standard deviations are in parentheses. FP and AP denote Full Professor and Associate Professor exams, respectively. Eng., H\&L, Sci., and Soc. Sci. are abbreviations for Engineering, Humanities and Law, Sciences, and Social Sciences, which are 4 broad scientific areas in our sample. AIS is the sum of international publications weighted by corresponding Article Influence Scores.}
	 \end{spacing}
\end{table}
  \clearpage
 \begin{table}[!h]  
	\begin{center}
	\caption{Descriptive statistics, Spain, connections} 
	\label{tbl:DescriptiveConnections} 
	\small
	\setlength{\tabcolsep}{5pt}
	\begin{tabular}{lccccccc}
		\hline
		\hline
		\\[-1.8ex]
		 & All & AP & FP  & Eng. & H\&L & Sci. & Soc. Sci. \\ 
		\hline
	    \multicolumn{1}{l}{Strong connections} & 31.71 & 29.08 & 35.18 & 37.78 & 27.65 & 31.13 & 34.94\\ 
	   \hline\\
	   [-1.8ex]
		Advisor  & 3.17 & 2.97 & 3.43 & 4.60 & 3.29 & 2.43 & 3.50  \\ 
		Coauthor  & 5.44 & 3.26 & 8.32 & 6.10 & 2.84 & 7.44 & 4.24 \\ 
		Colleague & 29.71 & 27.74 & 32.31 & 36.02 & 26.15 & 28.50 & 33.46 \\
			\\ [-1.8ex]
		\hline		
	 \multicolumn{1}{l}{Weak connections}	& 18.79 &  7.33 & 33.97 & 17.06 & 23.63 & 16.43 & 17.71 \\ 
		\hline\\
		[-1.8ex]
		Own PhD committee member & 7.08 & 5.31 & 9.43 &  8.05 & 10.22 &  4.39 &  7.48 \\ 
	    His PhD students' committee member   & 4.45 & 0.69 & 9.42 & 4.70 & 4.81 & 4.27 & 3.96 \\ 
		Fellow PhD committee member & 11.65 &  1.82 & 24.66 &  8.84 & 13.90 & 11.59 & 10.31 \\ 
		\hline
	\end{tabular}
	\end{center}
	\vspace*{-10pt}
{\footnotesize 	\textit{Notes: }The percentage of candidates with at least one connection of indicated type to the jury.}
\end{table}


   \clearpage

  \begin{table}[!h] 
	\begin{center}
	\caption{Descriptive statistics, Italy, observables} 
	\label{tbl:DescriptiveObservablesItaly}
	\small
	\begin{tabular}{lccccc}
		\hline
		\hline
		\\[-1.8ex]
		   & All & FP & AP & Non-STEMM & STEMM   \\ 
		 \hline
         Promoted & 0.37 & 0.36 & 0.37 & 0.34 & 0.39 \\ 
        & (0.48) & (0.48) & (0.48) & (0.47) & (0.49) \\ \\ 
		 Female & 0.38 & 0.31 & 0.41 & 0.42 & 0.35 \\ 
    & (0.49) & (0.46) & (0.49) & (0.49) & (0.48) \\ 
  Age & 44.48 & 48.67 & 42.57 & 44.18 & 44.68 \\ 
   & (7.9) & (7.59) & (7.28) & (7.95) & (7.86) \\ 
   \\
  Publications & 64.21 & 89.47 & 52.71 & 40.02 & 80.36 \\ 
  & (66.8) & (82.76) & (54.33) & (38.87) & (76.02) \\ 
  AIS & 38.95 & 56.79 & 30.83 & 1.48 & 63.96 \\ 
   & (83.67) & (108.53) & (67.94) & (7.76) & (100.36) \\ 
  A articles & 1.63 & 2.28 & 1.34 & 4.08 & 0 \\ 
   & (4.8) & (6.1) & (4.03) & (6.89) & (0) \\ 
  Articles & 36.98 & 53.37 & 29.52 & 15.02 & 51.64 \\ 
   & (51.2) & (65.18) & (41.27) & (21.72) & (59.33) \\ 
  Other publications & 8.76 & 10.38 & 8.02 & 7.77 & 9.42 \\ 
& (22.75) & (27.18) & (20.37) & (15.04) & (26.66) \\ 
  Pages in CV & 15.67 & 20.17 & 13.63 & 8.31 & 20.59 \\ 
  & (66.63) & (79.23) & (59.91) & (5.53) & (85.56) \\ 
  Coauthors per paper & 5.99 & 6.22 & 5.89 & 1.5 & 8.99 \\ 
   & (18.03) & (19.01) & (17.56) & (0.92) & (22.78) \\ 
  First author (per paper) & 0.22 & 0.22 & 0.22 & 0.13 & 0.28 \\ 
   & (0.2) & (0.19) & (0.2) & (0.18) & (0.18) \\ 
  Last author (per paper) & 0.12 & 0.15 & 0.11 & 0.08 & 0.15 \\ 
   & (0.16) & (0.17) & (0.15) & (0.15) & (0.16) \\ 
 \\
 Observations & 69020 & 21594 & 47426 & 27625 & 41395 \\
		 	\hline
		\end{tabular}
	\end{center}
	\vspace*{-10pt}
	 \begin{spacing}{1}
		{\footnotesize \textit{Notes: }Average values of observable characteristics. Standard deviations are in parentheses. FP and AP denote Full Professor and Associate Professor exams, respectively. STEMM is the abbreviation for Science, Technology, Engineering, Mathematics, and Medicine. \textit{A articles} refers to articles published in a list of journals indicated by the Ministry. There is no such list for STEMM fields.}
	 \end{spacing}
\end{table}
\newpage
\clearpage
  \begin{table}[!h]  
	\begin{center}
	\caption{Descriptive statistics, Italy, connections} 
	\label{tbl:DescriptiveConnectionsItaly} 
	\small
	\setlength{\tabcolsep}{5pt}
	\begin{tabular}{lccccc}
		\hline
		\hline
		\\[-1.8ex]
		 & All & FP & AP & Non-STEMM & STEMM  \\ 
		\hline
	    \multicolumn{1}{l}{Strong connections}  &15.97 & 18.76 & 14.70 & 11.27 & 19.11\\ 
	   \hline\\
	   [-1.8ex]
		Coauthor  & 6.96 & 8.61 & 6.22 & 1.85 & 10.38 \\ 
		Colleague & 11.62 & 13.44 & 10.80 & 10.25 & 12.54 \\
			\\ [-1.8ex]
		\hline		
	 \multicolumn{1}{l}{Same ssd} & 72.90 & 72.78 & 72.99 & 74.93 & 71.52\\ 
		\hline
	\end{tabular}
	\end{center}
	\vspace*{-10pt}
{\footnotesize 	\textit{Notes: }The percentage of candidates with at least one connection of indicated type to the jury. }
\end{table}
  \clearpage
\begin{table}[!h] 
	\begin{center}
	\caption{Descriptive statistics, China, observables and connections} 
	\label{tbl:DescriptiveObservablesChina}
	\small
  \begin{tabular}{lcccc}		\hline
		\hline
		\\[-1.8ex]
        & Candidate-year & All spells & Governor spells & Secretary spells \\ 
    \hline
 
Promoted & 0.07 & 0.26 & 0.34 & 0.18 \\ 
   & (0.26) & (0.44) & (0.48) & (0.39) \\ 
   \\

  Age & 56.24 & 56.28 & 55.36 & 57.18 \\ 
   & (3.75) & (3.93) & (3.75) & (3.9) \\ 
  College graduate (binary) & 0.82 & 0.82 & 0.82 & 0.82 \\ 
   & (0.39) & (0.39) & (0.39) & (0.39) \\ 
  Served in center (binary) & 0.36 & 0.4 & 0.36 & 0.43 \\ 
  & (0.48) & (0.49) & (0.48) & (0.5) \\ 
  Home (binary) & 0.26 & 0.23 & 0.31 & 0.15 \\ 
 & (0.44) & (0.42) & (0.47) & (0.35) \\ 
  Growth & 0.11 & 0.11 & 0.11 & 0.11 \\ 
   & (0.02) & (0.02) & (0.03) & (0.02) \\ 
  \\
    Connected & 0.21 & 0.25 & 0.23 & 0.27 \\ 
   & (0.41) & (0.43) & (0.43) & (0.45) \\ 
  \\
Observations& 966 & 258 & 128 & 130 \\ 
		 	\hline
		\end{tabular}
	\end{center}
	\vspace*{-10pt}
	 \begin{spacing}{1}
		{\footnotesize \textit{Notes: }Average values of observable characteristics for candidate-year (column 1), and by spells leading the provinces (columns 2--4). Standard deviations are in parentheses.   \textit{Served in center} equals 1 if the candidate previously held a position in the central government. \textit{Home} equals 1 if the candidate originates from the province currently governed. \textit{Growth} denotes the province’s GDP growth rate.  }
	 \end{spacing}
\end{table}
\newpage
\clearpage
\begin{table}[H] 
\begin{center}
\caption{Descriptive statisticts, number and types of connections}\label{tbl:desc_number_links}
\begin{tabular}{lccccccc}
\hline
  \hline\\[-1.8ex] 
  & \multicolumn{3}{c}{Strong} & & \multicolumn{3}{c}{Weak} \\
  & $n_{Si} =0$ & $n_{Si}  > 0$ & $n_{Si} >1$ & & $n_{Wi} =0$ & $n_{Wi} >0$ & $n_{Wi} >1$ \\
    \cmidrule(lr){2-4} \cmidrule(lr){6-8}
    Spain (AP) & 12623 & 5176 & 1385 &  & 16495 & 1304 &  92  \\ 
  Spain (FP) & 8714 & 4730 & 1355 &  & 8877 & 4567 & 1380 \\
  \\
  Italy (AP) & 40453 & 6973 & 258 &  & - & - & - \\ 
  Italy (FP) & 17543 & 4051 & 302 &  & - & - & - \\ 
  \\
 China &761 & 205 &  55 &  & - & - & -\\ 
 \hline
\end{tabular}
\end{center}
\vspace*{-10pt}
	\begin{spacing}{0.7}
	{\footnotesize 	\textit{Notes:} Number of observations with zero connections ($n_{Xi} =0$), at least one connection ($n_{Xi} >0$), and more than one connection ($n_{Xi} >1$), with $X \in \{S, W\}$. In Spain, we distinguish between strong ($S$) and weak ($W$) connections. In Italy and China, we consider only one type of connection, which we label as strong ($S$) in this table. }
\end{spacing}
\end{table}
\clearpage
\begin{table}[!ht]
	\begin{center}
	\caption{Balance tests, Spain} 
	\label{tbl:BalanceTest}
	\small
	\begin{tabular}{lccccc} 
	  \hline 
		\hline \\[-1.8ex] 
		& AIS & Publications & PhD & PhD  & Past\\
		& & &students& committees&experience \\ 
		\hline \\[-1.8ex] 
		& \multicolumn{5}{c}{\textit{Without controls for the expected number of connections}} \\ 
		\cline{2-6} 
		Strong &  0.009 & 0.010 & $-$0.017$^{***}$ & $-$0.012$^{*}$ & 0.009 \\ 
		& (0.007) & (0.007) & (0.006) & (0.007) & (0.008) \\ 
		Weak & 0.007 & 0.051$^{***}$ & 0.180$^{***}$ & 0.298$^{***}$ & 0.027$^{***}$   \\ 
		&(0.007) & (0.008) & (0.010) & (0.011) & (0.007)\\ 
		\hline 
		\\[-1.8ex]
		& \multicolumn{5}{c}{\textit{Including controls for the expected number of connections}} \\ 
		\cline{2-6} 
		Strong & $-$0.001 & $-$0.011 & 0.002 & $-$0.006 & $-$0.004  \\ 
		& (0.010) & (0.011) & (0.010) & (0.010) & (0.012)  \\ 
		Weak &   $-$0.005 & $-$0.008 & 0.013 & 0.010 & 0.003 \\ 
		&  (0.011) & (0.013) & (0.016) & (0.016) & (0.012) \\ 
		\\[-1ex]
		
		Observations & 31,243  & 31,243  & 31,243  & 31,243 & 31,243  \\ 
		\hline 
	\end{tabular} 
		\end{center}
		\vspace*{-10pt}
	{\footnotesize 	\textit{Notes: }
    Regressions of each observable (columns) on the number of strong and weak connections to the jury (rows). Standard errors, clustered at the committee level, are in parentheses. Upper panel excludes controls for expected connections; lower panel includes them. $^{*}$p$<$0.1; $^{**}$p$<$0.05; $^{***}$p$<$0.01. }  

\end{table}

\begin{landscape}
\small
\begin{table}[!htbp] 
  \begin{center}
  \caption{Balance tests, Italy} 
  \label{tbl:RandomAssignementItaly} 

\begin{tabular}{lccccccccc} 
\\[-1.8ex]\hline 
\hline 

 & AIS & Publications & Articles & A & Other  & Pages  & Coauthors & First & Last \\
  & &  &  & articles & publications & in CV & Coauthors & authored & authored \\
\hline \\[-1.8ex] 
& \multicolumn{9}{c}{\textit{Without controls for the expected number of connections}} \\
\cline{2-10} 
Strong& 0.024$^{**}$ & 0.099$^{***}$ & 0.085$^{***}$ & 0.028$^{***}$ & $-$0.022$^{**}$ & 0.088$^{***}$ & 0.019$^{**}$ & $-$0.021$^{**}$ & 0.028$^{***}$ \\ 
  & (0.009) & (0.010) & (0.010) & (0.006) & (0.010) & (0.010) & (0.010) & (0.010) & (0.010) \\  
\hline
& \multicolumn{9}{c}{\textit{With controls for the expected number of connections}} \\
\cline{2-10} 
 Strong & $-$0.023$^{**}$ & $-$0.003 & $-$0.007 & $-$0.014$^{*}$ & $-$0.011 & $-$0.024$^{**}$ & $-$0.025$^{**}$ & 0.011 & $-$0.008 \\ 
  & (0.012) & (0.012) & (0.012) & (0.007) & (0.012) & (0.012) & (0.012) & (0.012) & (0.012) \\ 
 \hline \\[-1.8ex] 
Observations & 69,020 & 69,020 & 69,020 & 69,020 & 69,020 & 69,020 & 69,020 & 69,020 & 69,020 \\ 
\hline 
\end{tabular} 
\end{center}
		\vspace*{-10pt}
	{\footnotesize 	\textit{Notes: }
    Regressions of each observable (columns) on the number of strong connections to the jury. Standard errors, clustered at the committee level, are in parentheses. Upper panel excludes controls for expected connections; lower panel includes them. $^{*}$p$<$0.1; $^{**}$p$<$0.05; $^{***}$p$<$0.01. }  
\end{table} 
\end{landscape}
\clearpage
\begin{table}[!ht]

		\caption{Parametric model, binary connections, approximation of $\gucon$, Spain} \label{tbl:SelectingGrade_BinaryParametricSpain}
		\small
					\vspace{-15pt}
					\begin{center}
	\begin{tabular}{lccccccc}
		\hline \hline
		& Model 1 & Model 2 & Model 3 & Model 4 & Model 5 & Model 6 & Model 7 \\ 
		\hline

      Bias ($B$)& 0.654$^{***}$ & 0.459$^{***}$ & 0.476$^{***}$ & 0.505$^{***}$ & 0.532$^{***}$ & 0.392$^{***}$ & 0.521$^{***}$ \\ 
  & (0.105) & (0.105) & (0.124) & (0.112) & (0.100) & (0.112) & (0.097) \\ 
 Information ($\delta$) & 0.017 & 0.127$^{**}$ & 0.149$^{**}$ & 0.118$^{**}$ & 0.100$^{**}$ & 0.189$^{***}$ & 0.123$^{**}$ \\ 
  & (0.055) & (0.050) & (0.060) & (0.056) & (0.051) & (0.053) & (0.049) \\


        \\
        Observations & 31,000  &31,000 & 31,000 & 31,000 &31,000  & 31,000 &31,000 \\
		Log Loss & 0.325 & 0.319 & 0.323 & 0.319 & 0.316 & 0.328 & 0.330 \\ 
  Log Lik. & -10044.29 & -9837.60 & -9901.92 & -9740.88 & -9666.01 & -9545.86 & -9372.69 \\ 
  df & 16 & 27 & 94& 105 & 116 & 391 & 534 \\ 
		\hline
	\end{tabular}
\end{center}
	\vspace*{-10pt}
{\footnotesize 	\textit{Notes:}We run the regressions on the full sample of both AP and FP candidates. Model 1: Linear; Model 2: Linear + quadratic terms; Model 3: Linear and double interactions but no quadratic terms; Model 4: Quadratic; Model 5: Second order polynomial with cubes and no triple interactions; Model 6: Second order polynomial with triple interactions but no cubes; Model 7: Third order polynomial. Log Loss is the average log loss obtained from 10 fold cross validation.} 
\end{table}

\clearpage

\begin{table}[!ht]

		\caption{Parametric model, binary connections, approximation of $\gucon$, Italy} \label{tbl:SelectingGrade_BinaryParametricItaly}
		\small
					\vspace{-15pt}
					\begin{center}
	\begin{tabular}{lccccccc}
		\hline \hline
		& Model 1 & Model 2 & Model 3 & Model 4 \\
		\hline
  

  Bias ($B$) & 0.202$^{***}$ & 0.216$^{***}$ & 0.233$^{***}$ & 0.219$^{***}$ \\ 
  & (0.040) & (0.037) & (0.038) & (0.038) \\ 
  Information ($\delta$) & 0.121$^{***}$ & 0.084$^{**}$ & 0.100$^{***}$ & 0.109$^{***}$ \\ 
  & (0.043) & (0.039) & (0.035) & (0.040) \\ 
  
  \\    
  Observations & 69,020   &69,020   & 69,020  & 69,020 \\ 
  Log Loss & 0.555 & 0.520 & 0.527 & 0.516 \\ 
  Log Lik. & -37895.41 & -35431.39 & -35791.26 & -35056.14 \\ 
  df & 382 & 393 & 448 & 459 \\
		\hline
	\end{tabular}
\end{center}
	\vspace*{-10pt}
{\footnotesize 	\textit{Notes:} We run the regressions on the full sample of both AP and FP candidates. Model 1: Linear; Model 2: Linear + quadratic terms; Model 3: Linear and double interactions but no quadratic terms; Model 4: Quadratic. Log Loss is the average log loss obtained from 10 fold cross validation. For specifications including a more complex model for $\gucon$, the estimation algorithm does not converge.} 
\end{table}

\clearpage
\begin{table}[!ht]
	
		\caption{Parametric model, binary connections, robustness to the inclusion of FE, Spain} \label{tbl:RobustnessFE_ParametricSpain}
  \small
	\vspace{-15pt}
  \begin{center}
	\begin{tabular}{lccc}
		\hline \hline
		& Exam Variables & Exam variables + Limited FE & Full FE \\ 
		\hline
            Bias ($B$)& 0.532$^{***}$ & 0.527$^{***}$ & 0.195 \\ 
         & (0.100) & (0.101) & (0.129) \\ 
      Information ($\delta$) & 0.100$^{**}$ & 0.107$^{**}$ & 0.310$^{***}$ \\ 
       & (0.051) & (0.051) & (0.054) \\ 
            \\
         Observations & 31,000  &31,000  & 31,000 \\    
		AIC & 19564.02 & 19641.15 & 21232.22 \\ 
  Log Lik. & -9666.01 & -9633.58 & -9601.11 \\ 
  df & 116 & 187 & 1015 \\ 
  LR &  & 64.87 & 64.94 \\ 
		\hline
	\end{tabular}
 \end{center}
 	\vspace*{-10pt}

{\footnotesize 	\textit{Notes:}  Heteroscedastic logit estimates using the preferred approximation of $\gucon$ selected based on Table \ref{tbl:SelectingGrade_BinaryParametricSpain}.  We run the regressions on the full sample of both AP and FP candidates. The specification in the first column includes only exam-specific variables. The specification in the second column additionally includes \textit{broad research area} $\times$ \textit{exam type} $\times$ \textit{exam wave} fixed effects. There are 4 broad research areas, two types of exams, and 9 exam waves.  The third column includes the full set of exam fixed effects.  Row AIC reports the value of the Akaike information criterion for the corresponding model. The minimizing value of the likelihood function is reported in the row Log Lik. Row df reports the degrees of freedom of the estimated model. Row LR reports the value of the Likelihood Ratio $\chi^2$ statistics of comparison of the more complex model with a simpler model in the preceding column. }
\end{table}

\clearpage
\begin{table}[!ht]
\caption{Parametric model, binary connections, robustness to the inclusion of FE, Italy} 
\label{tbl:RobustnessFE_ParametricItaly}
\small
\vspace{-15pt}
\begin{center}
\begin{tabular}{lcccc}
  \hline \hline
 & Exam  \& & Exam  \& Committee var. & Exam var. & Full FE \\ 
 & Committee  var.&  + discipline FE & + committee FE& \\ 
  \hline

Bias ($B$) & 0.220$^{***}$ & 0.222$^{***}$ & 0.215$^{***}$ & 0.219$^{***}$ \\ 
  & (0.037) & (0.037) & (0.037) & (0.038) \\ 
  Information ($\delta$) & 0.116$^{***}$ & 0.114$^{***}$ & 0.121$^{***}$ & 0.109$^{***}$ \\ 
  & (0.042) & (0.042) & (0.040) & (0.040) \\ 
  \\
  Observations & 69,020   &69,020  & 69,020  & 69,020 \\ 
  AIC & 71223.11 & 71096.72 & 70995.47 & 71030.28 \\ 
Log Lik. & -35503.55 & -35425.36 & -35218.74 & -35056.14 \\ 
  df & 108 & 123 & 279 & 459\\ 
  LR &  & $156.39^{***}$ & $413.25^{***}$ & $325.19^{***}$ \\ 
  
   \hline
\end{tabular}
\end{center}
\vspace*{-10pt}
{\footnotesize 	\textit{Notes:} Heteroscedastic logit estimates using the preferred approximation of $\gucon$ selected based on Table \ref{tbl:SelectingGrade_BinaryParametricItaly}. We run the regressions on the full sample of both AP and FP candidates.  The
specification in the first column includes exam-specific and committee-specific variables. The specification in the second column additionally includes research discipline fixed effects (16 disciplines). The specification in the third column includes exam-specific variables and committee-fixed effects (184 committees). The fourth column
includes the full set of exam fixed effects (367 exams). Row AIC reports the value of the Akaike information criterion for the corresponding model. The minimizing value of the likelihood function is reported in the row Log Lik. Row df reports the degrees of freedom of the estimated model. Row LR reports the value of the Likelihood Ratio  $\chi^2$ statistics of comparison of the more complex model with a simpler model in the preceding column.}
\end{table}

\clearpage

\begin{table}[!ht]
	
		\caption{Parametric model, binary connections, robustness to the inclusion of FE, China} \label{tbl:RobustnessFE_ParametricChina}
  \small
	\vspace{-15pt}
  \begin{center}
	\begin{tabular}{lccc}
		\hline \hline
		& Year FE & Year FE + Term FE & Year FE + Term FE \\
        &  &  & + Office$\times$Province FE\\
		\hline
             Bias ($B$) & 1.650$^{***}$ & 1.644$^{***}$ & 1.460$^{**}$ \\ 
  & (0.311) & (0.242) & (0.696) \\ 
  Information ($\delta$) & $-$0.756$^{**}$ & $-$0.972$^{***}$ & $-$0.164 \\ 
  & (0.372) & (0.372) & (0.355) \\ 
  \\
  Observations & 966 &966&966\\
  AIC & 459.31 & 449.58 & 461.81 \\ 
  Log Lik. & -209.66 & -193.79 & -138.90 \\ 
  df & 20 & 31 & 92 \\ 
  LR &  & 31.73 $^{***}$ & 109.78$^{***}$ \\ 
		\hline
	\end{tabular}
 \end{center}
 	\vspace*{-10pt}

{\footnotesize 	\textit{Notes:}  Heteroscedastic logit estimates. The specification in the first column includes year fixed effects. The specification in the second column additionally includes  term fixed effects. The third column additionally includes office $\times$ year fixed effects.  Row AIC reports the value of the Akaike information criterion for the corresponding model. The minimising value of the likelihood function is reported in the row Log Lik. Row df reports the degrees of freedom of the estimated model. Row LR reports the value of the Likelihood Ratio $\chi^2$ statistics of comparison of the more complex model with a simpler model in the preceding column. }
\end{table}



\clearpage

\clearpage
\begin{table}[!htbp] \begin{center}
  \caption{Parametric model, binary connections, selection, first stage, Italy} 
  \label{tbl:selection_first_stage_AP_FP} 
\begin{tabular}{@{\extracolsep{5pt}}lcc} 
\\[-1.8ex]\hline 
\hline \\[-1.8ex]  
\\[-1.8ex] & (AP) & (FP)\\ 
\hline \\[-1.8ex] 
 $\zeta$& 3.619$^{***}$ & 3.523$^{***}$ \\ 
  & (0.169) & (0.157) \\ 
  \\
  Observations & 47,426 & 21,594  \\
 \hline \\[-1.8ex] 
\end{tabular} 
\end{center}
\vspace*{-10pt}
\begin{spacing}{0.7}
{\footnotesize\textit{Notes}:  The probit estimate of equation \eqref{eq:selection_first_stage}.   Standard errors clustered at the committee level are in parentheses. All specifications include exam-fixed effects. $^{*}$p$<$0.1;$^{**}$p$<$0.05; $^{***}$p$<$0.01. }
\end{spacing}
\end{table}

\newpage

	


 \renewcommand{\thetable}{C.\arabic{table}}
\setcounter{table}{0}
\subsection*{Online Appendix C: Semiparametric model, Spain}

When we estimate the semiparametric model on the Spanish data, the total effects of connections are nearly identical to those from the parametric specification in Table~2 (0.076 vs. 0.075 for AP candidates and 0.050 vs. 0.051 for FP candidates). The Spanish sample, however, exhibits major limitation in the support in the region required to identify the  parameter $B$. As a consequence, the data do not allow a precise decomposition of the total effect into information and favoritism.

Identification of $B$ requires estimating equation~\eqref{eq:SecondStage_General} nonparametrically 
and inverting it at point where the predicted probability of promotion is $1/2$. The Spanish data has a few observations near this point:  only 27observations fall in the $(0.5, 0.55)$ interval for AP candidates and only 12 for FP candidates. For both samples, all of these observations are classified as outliers  when applying Tukey's fences criteria. 
As a result, the estimates of $B$ for both AP and FP candidates are 
statistically insignificant and have large standard errors. Moreover, the point  estimates of $B$ are sensitive to dropping a small number of observations  with high predicted baseline promotion probabilities, due to sensitivity of the nonparametric estimation of \eqref{eq:SecondStage_General} to outliers. We therefore conclude that the support limitations in the Spanish data prevent a reliable separation of the information and favoritism channels using our semiparametric approach. For 
completeness, we report the corresponding estimates in 
Table~\ref{tbl:AvgEffects_Binary_Semiparametric_Spain} 

\begin{table}[H]
 	\begin{center}
 		\caption{Marginal effects, semiparametric model, binary connections, Spain} 
 		\label{tbl:AvgEffects_Binary_Semiparametric_Spain}
 		\small
 		\begin{tabular}{lcccccccc}
			\hline
			\hline\\
			[-1.8ex]
			& & \multicolumn{1}{c}{Baseline} & &\multicolumn{5}{c}{Marginal effects}\\
			\cline{5-9}\\
			[-1.8ex]
			&$B$&Predicted& & Total & Favors & Information &Information &Information \\
            && &  & &&&Low chances& High chances\\
			\hline\\
			[-1.5ex]

 AP & -0.284 & 0.097$^{***}$ && 0.076$^{***}$ & -0.033 & 0.108 & 0.109 & -0.071$^{***}$ \\ 
   & (0.978) & (0.005) && (0.008) & (0.079) & (0.077) & (0.078) & (0.022) \\ 
  Observations & 15,492 & 15,492 && 15,492 & 15,492 & 15,492 & 15,410 & 82 \\  \\
  FP& 0 & 0.077$^{***}$ && 0.05$^{***}$ & 0 & 0.05 & 0.05 & -0.058$^{***}$ \\ 
   & (0.503) & (0.005) && (0.007) & (0.049) & (0.047) & (0.047) & (0.023) \\ 
  Observations & 12,717 & 12,717&& 12,717 & 12,717 & 12,717 & 12,678 & 39 \\
			\hline
		\end{tabular}
 	\end{center}
 	\vspace*{-10pt}
  \par
 	\begin{spacing}{0.7}
 	{\footnotesize 	\textit{Notes:}
    Average marginal effect of being connected to the jury calculated for candidates with at least one
connection to eligible evaluators. Bootstrapped standard errors clustered at the committee level are in parentheses. $^{*}$p$<$0.1;$^{**}$p$<$0.05; $^{***}$p$<$0.01.} 
  \end{spacing}
 \end{table}

\pagebreak
\newpage

\end{document}